\RequirePackage{xcolor}
\RequirePackage{ifpdf}
\documentclass[letterpaper]{JHEP3}
\usepackage{amsmath}
\usepackage{epsfig}
\usepackage{subfigure}
\usepackage{enumitem}
\pdfoutput=1

\newcommand{\roughly}[1]{\mathrel{\raise.3ex\hbox{$#1$\kern-0.85em
\lower1ex\hbox{$\sim$}}}}
\newcommand{\lsim}{\roughly<}
\newcommand{\gsim}{\roughly>}

\def\nn{\nonumber}

\newcommand{\be}{\begin{equation}}
\newcommand{\bee}{\begin{equation}}
\newcommand{\ee}{\end{equation}}
\newcommand{\beea}{\begin{eqnarray}}
\newcommand{\eea}{\end{eqnarray}}
\newcommand{\bea}{\begin{eqnarray}}

\def\nott#1{\setbox0=\hbox{$#1$}                
   \dimen0=\wd0                                 
   \setbox1=\hbox{/} \dimen1=\wd1               
   \ifdim\dimen0>\dimen1                        
      \rlap{\hbox to \dimen0{\hfil/\hfil}}      
      #1                                        
   \else                                        
      \rlap{\hbox to \dimen1{\hfil$#1$\hfil}}   
      /                                         
   \fi}                                         %
\def\Dsl{\nott{D}}
\def\cDsl{\nott{\cD}}

\def\uxsl{\hbox{/\kern-.4000em$u$}}
\def\uxslsm{\hbox{\smaller/\kern-.5600em$u$}}
\def\pxpsl{\hbox{/\kern-.5000em$p$}}
\def\epssl{\hbox{/\kern-.5600em$\epsilon$}}
\def\delsl{\hbox{/\kern-.7000em$\nabla$}}
\def\lxpsl{\hbox{/\kern-.5600em$l$}}
\def\kxpsl{\hbox{/\kern-.5600em$k$}}
\def\qxpsl{\hbox{/\kern-.3900em$q$}}

\def\pref#1{(\ref{#1})}
\def\exd{{\rm d}}
\def\ol#1{{\overline{#1}}}

\def\QCD{{\scriptscriptstyle QCD}}
\def\EF{{\scriptscriptstyle EF}}

\def\cA{{\cal A}}

\def\cD{{\cal D}}
\def\cC{{\cal C}}
\def\cE{{\cal E}}
\def\cF{{\cal F}}
\def\cG{{\cal G}}
\def\cH{{\cal H}}
\def\cJ{{\cal J}}
\def\cK{{\cal K}}
\def\cL{{\cal L}}

\def\cN{{\cal N}}
\def\cO{{\cal O}}
\def\cP{{\cal P}}

\def\cR{{\cal R}}
\def\cS{{\cal S}}
\def\cT{{\cal T}}

\def\cV{{\cal V}}
\def\cW{{\cal W}}
\def\cX{{\cal X}}
\def\cY{{\cal Y}}
\def\cZ{{\cal Z}}

\def\bfx{{\bf x}}

\def\mfa{{\mathfrak a}}
\def\mfb{{\mathfrak b}}
\def\mfc{{\mathfrak c}}

\def\mfe{{\mathfrak e}}
\def\mff{{\mathfrak f}}
\def\mfg{{\mathfrak g}}
\def\mfh{{\mathfrak h}}
\def\mfk{{\mathfrak k}}
\def\mfm{{\mathfrak m}}
\def\mfn{{\mathfrak n}}

\def\mfw{{\mathfrak w}}
\def\mfz{{\mathfrak z}}
\def\mfD{{\mathfrak D}}
\def\mfF{{\mathfrak F}}
\def\mfH{{\mathfrak{H}}}

\def\mfK{{\mathfrak{K}}}
\def\mfL{{\mathfrak L}}

\def\ssA{{\scriptscriptstyle A}}
\def\ssB{{\scriptscriptstyle B}}
\def\ssC{{\scriptscriptstyle C}}
\def\ssD{{\scriptscriptstyle D}}

\def\ssF{{\scriptscriptstyle F}}

\def\ssH{{\scriptscriptstyle H}}
\def\ssI{{\scriptscriptstyle I}}
\def\ssJ{{\scriptscriptstyle J}}

\def\ssL{{\scriptscriptstyle L}}
\def\ssM{{\scriptscriptstyle M}}
\def\ssN{{\scriptscriptstyle N}}
\def\ssP{{\scriptscriptstyle P}}

\def\ssR{{\scriptscriptstyle R}}
\def\ssS{{\scriptscriptstyle S}}
\def\ssT{{\scriptscriptstyle T}}

\def\ssW{{\scriptscriptstyle W}}
\def\ssX{{\scriptscriptstyle X}}
\def\ssY{{\scriptscriptstyle Y}}
\def\ssZ{{\scriptscriptstyle Z}}

\def\TEV{{\scriptscriptstyle TeV}}

\def\KK{{\scriptscriptstyle KK}}
\def\KL{{\scriptscriptstyle KL}}
\def\KS{{\scriptscriptstyle KS}}
\def\SM{{\scriptscriptstyle SM}}
\def\EH{{\scriptscriptstyle EH}}
\def\EF{{\scriptscriptstyle EF}}

\def\EW{{\scriptscriptstyle EW}}

\def\PPN{{\scriptscriptstyle PPN}}

\title{Yoga Dark Energy: Natural Relaxation  and Other Dark Implications of a Supersymmetric Gravity Sector}

\author{C.P.~Burgess,${}^{1,2,3}$ Danielle Dineen${}^1$ and F.~Quevedo${}^{4}$\\

{\it 
${}^1$ Department of Physics \& Astronomy, McMaster University,\\ 
\hspace{1 cm} 1280 Main Street West, Hamilton ON, Canada.\\
${}^2$ Perimeter Institute for Theoretical Physics,\\ 
\hspace{1 cm} 31 Caroline Street North, Waterloo ON, Canada.\\
${}^3$ CERN, Theoretical Physics Department, Gen\`eve 23, Switzerland.\\
${}^4$ DAMTP, Cambridge University, Wilberforce Road,  Cambridge, United Kingdom.
}
}

\preprint{CERN-TH-2021-192}

\date{\today}

\abstract{We construct a class of 4D `yoga' (naturally relaxed) models for which the gravitational response of heavy-particle vacuum energies is strongly suppressed. The models contain three ingredients: (i) a relaxation mechanism driven by a scalar field (the `relaxon'), (ii) a very supersymmetric gravity sector coupled to the Standard Model  in which supersymmetry is non-linearly realised, and (iii) an accidental approximate scale invariance expressed through the presence of a low-energy dilaton supermultiplet. All three are common in higher-dimensional and string constructions and although none suffices on its own, taken together they can dramatically suppress the net vacuum-energy density.  The dilaton's {\it vev}~$\tau$ determines the weak scale $M_\ssW \sim M_p/\sqrt\tau$. We compute the potential for $\tau$ and find it can be stabilized in a local de Sitter minimum at sufficiently large field values to explain the size of the electroweak hierarchy, doing so using input parameters no larger than O(60) because the relevant part of the scalar potential arises as a rational function of $\ln\tau$. The de Sitter vacuum energy at the minimum is order $c\,  M_\ssW^8 \propto 1/\tau^4$, with a coefficient $c \ll \cO(M_\ssW^{-4})$. We discuss ways to achieve $c \sim 1/M_p^4$ as required by observations. Scale invariance implies the dilaton couples to matter like a Brans-Dicke scalar with coupling large enough to be naively ruled out by solar-system tests of gravity. Yet because it comes paired with an axion it can evade fifth-force bounds through the novel screening mechanism described in {\tt ArXiV:2110.10352}. Cosmological axio-dilaton evolution predicts a natural quintessence model for Dark Energy, whose evolution might realize recent proposals to resolve the Hubble tension, and whose axion contributes to Dark Matter. We summarize inflationary implications and some remaining challenges, including the unusual supersymmetry breaking regime used and the potential for UV completions of our approach.}

\dedicated{Dedicated to the memory of Steven Weinberg: a physicist's Standard Model.}

\begin{document}

\section{Introduction}

The cosmological constant problem \cite{CCWeinberg, Burgess:2013ara, CCReviews} seems hopeless. Despite years of effort and much model-building \cite{DEReviews} no technically natural mechanism has been found that reconciles the large vacuum fluctuations associated with known particles with the small gravitational response to the vacuum revealed by the evidence for Dark Energy \cite{SNDE, PlanckDE}. Indeed, it is widely believed that a symmetry-based or relaxation-type \cite{RelMechs,Graham:2019bfu} mechanism does not exist, and this point of view has driven much of the community towards anthropic \cite{CCWeinberg, CCAnthropic} and/or landscape arguments (see for instance \cite{Landscape, Li:2020rzo}). Although these might ultimately prove to be the way Nature works, a proper assessment of their likelihood suffers from the absence of compelling-yet-natural alternatives with which to compare.

We here propose a class of models that we hope can provide such a point of comparison. These models are designed to address the low-energy\footnote{By this we mean we focus on how the vacuum energies of known particles ({\it e.g.}~the electron) can avoid gravitating; arguably the hardest part of the problem because its solution involves changing the properties of well-measured particles at experimentally accessible energies. We leave open questions of UV completion (except where these introduce new constraints -- see \S\ref{ssec:UVcomplete}), but do so knowing that there are multiple ways (including supersymmetry) to suppress the vacuum energy of undiscovered particles above the weak scale. For other recent proposals for a technically natural vacuum energy see for instance \cite{Kaloper:2013zca,Kaloper:2015jra, Copeland:2021czt}.} (and so hardest) part of the cosmological constant problem and are built on the interplay of three separate ingredients, all of which seem to play important roles: 

\begin{enumerate}[label=(\roman*)]
\item A very supersymmetric gravity sector, for which supermultiplets are split by much less than for Standard Model fields (see \cite{LowESugra} for a discussion of some other implications and the naturality of this assumption).

\item A relaxation mechanism in which a `relaxon' scalar field\footnote{There is also no `i' in `relaxon' (as opposed to {\it relaxion} \cite{Relaxion}) since for us this field need not be an axion.} dynamically reduces the leading non-gravitational vacuum energy. 

\item Accidental approximate scale invariance, including the implied low-energy dilaton, $\tau$, such as is known to be a generic property of low-energy string vacua \cite{Burgess:2020qsc} and higher-dimensional supergravities more generally \cite{Salam:1989fm, Burgess:2011rv, SUGRAscaleinv, BMvNNQ, GJZ}. 
\end{enumerate}

Although each of these ingredients has a plausible UV pedigree, we here avoid unnecessary UV baggage (like extra dimensions) and instead let all three stand on their own within a simple 4D context, with a view to better understanding the underlying mechanisms that could be at work. Indeed, this kind of phenomenological approach lends itself to the cosmological constant problem, which is at heart a low-energy problem rather than a high-energy one (for the reasons given in footnote 1). (We do examine UV completions more explicitly in \S\ref{ssec:UVcomplete}, with a view to understanding the independent new constraints that having a UV provenance for these ingredients can introduce.) 

The presence of the dilaton introduces a $\tau$-dependence to particle masses, so we first start with a general EFT at low energies and ask how the $\tau$-dependence associated with the vacuum energy, $\delta V \sim m^4(\tau)$, can be dynamically suppressed. We then push the EFT into the UV to see how far it can go, and ask how it extends to energies near the weak scale, but well below the masses of any putative superpartners for Standard Model fields (which therefore do not appear to be supersymmetric\footnote{Despite supersymmetry playing an important role one of our first predictions therefore is a successful one: the {\it absence} of Standard-Model superpartners at the LHC.} at all). We focus on whether our three ingredients suffice to adequately suppress the gravitational response as the Standard Model fields themselves are integrated out. They appear to do so, subject to a few provisos discussed below. 

Because the model we propose has a number of moving parts it is instructive here to summarize the underlying reasons why it works. The first ingredient -- the assumption that gravity is described by $\cN = 1$ supergravity down to very low energies -- is important largely because of the auxiliary fields, $F^\ssA$, that its linear realization requires to be in the low-energy scalar potential. Although these fields do not propagate,\footnote{Concrete evidence for the importance of keeping track of non-propagating fields in low-energy EFTs comes from Quantum Hall systems, although the fields in these examples are usually topological gauge potentials \cite{QHE, EFTBook}. From this point of view it is suggestive that in known UV completions 4D auxiliary fields like $F$ start life as 4-form fields that often convey topological information from higher dimensions, both in string theory \cite{LuisForm} and extra-dimensional gravity more generally \cite{MyForm}.} they are required in order to linearly realize supersymmetry. Crucially, their presence changes the way that UV physics can enter into the low-energy potential; because supersymmetry-breaking masses necessarily themselves involve $F$ the contribution of virtual heavy nonsupersymmetric states to the low-energy potential tends to be $\delta V \propto M^2 F + \hbox{h.c.}$ (where $M$ is the UV scale) rather than directly as an $F$-independent term like $\delta V \propto M^4$ \cite{LowESugra}. Even though $M^4$ eventually arises once $F$ is integrated out, the form involving $F$ shows that the most UV-sensitive effective couplings have a reduced dimension. 

The second important consequence of having low-energy auxiliary fields is the structure that their elimination imposes on the scalar potential, which comes as the usual sum and differences of squares: $V = V_\ssF + V_\ssD$ with
\be \label{VFdef}
  V_\ssF = e^{K/M_p^2} \left[ K^{\bar A B} \ol{D_\ssA W} D_\ssB W - \frac{3|W|^2}{M_p^2} \right]   \quad \hbox{with} \quad
  D_\ssA W := W_\ssA + \frac{K_\ssA W}{M_p^2} \,,
\ee
and
\be \label{VDdef}
  V_\ssD = \frac12 \, \mfF^{\alpha\beta} \mfD_\alpha \mfD_\beta  \,,
\ee
familiar from $\cN=1$ supergravity, where $K$ is the supersymmetric K\"ahler potential, $W$ is the holomorphic superpotential, $\mfF^{\alpha\beta}$ is the inverse of the real part of the holomorphic gauge kinetic function, $\mff_{\alpha\beta}$, and $\mfD_\alpha$ are the `moment maps' for the gauge symmetries \cite{FreedmanVanProeyen} (whose detailed form is not needed here). Subscripts on $K$ and $W$ denote differentiation with respect to any complex scalars $Z^\ssA$. This implies in particular that the dominant `globally supersymmetric' term (the terms unsuppressed by $1/M_p$) arise as a square, 
\be \label{Vglobpos}
   V_{\rm glob} = K^{\ol\ssA\ssB} \ol{W_\ssA} W_\ssB \,, 
\ee
and so can vanish at its minimum very naturally. 

The above observation is only useful if \pref{VFdef} can also be used when supergravity is coupled to systems like the Standard Model, for which the matter does {\it not} come in $\cN=1$ supermultiplets and for which the potential usually need not be positive. The generality of the above form ultimately follows from the generality of the rules for nonlinearly realizing supersymmetry described in \cite{Komargodski:2009rz}, together with its coupling to supergravity \cite{Bergshoeff:2015tra, DallAgata:2015zxp, Schillo:2015ssx}. Ref.~\cite{LowESugra} explores more concretely why generic potentials are consistent with the above supergravity form, using this general framework. When supersymmetry is nonlinearly realized (as it must be in such theories) there is always a low-energy superfield $X$ that is nilpotent, $X^2 = 0$, since this is what is required to represent the goldstino \cite{Komargodski:2009rz}, and it is typically true that $W_\ssX \neq 0$ for this field. For systems where global supersymmetry breaks badly in the UV, for example, the positivity of \pref{Vglobpos} is consistent with the non-supersymmetric low-energy scalar potential $U$ not being positive because $W_\ssX \simeq \mu^2 + U/(2\mu^2) + \cdots$ and so $|W_\ssX|^2 \simeq \mu^4 + U + \cdots$, since constant terms in the potential are irrelevant in global supersymmetry. Supergravity complicates things because gravity couples to all sources of energy, but also the gravity sector introduces new auxiliary fields. In what follows we imagine that $X$ is the only supermultiplet to descend from the UV sector\footnote{The generality of the emergence of $X$ in the far infrared carrying the main supersymmetry-breaking order parameter (even if supersymmetry should be partly broken in a more complicated way, including by $D$-terms in the UV, say) is argued in \cite{Komargodski:2009rz}} with nonzero derivative for $W$, 
so that $V_{\rm glob} \propto |W_\ssX|^2$. 

The relaxation mechanism is now built around the structure of the scalar potential described above. A (nonsupersymmetric) relaxon field $\phi$ is introduced, whose mass is assumed to be a bit smaller than the electron mass (so that it survives to appear in the low-energy theory below the lightest known dangerous Standard Model field). This scalar appears in particular in $W_\ssX$, and so long as a configuration exists for which $W_\ssX = 0$ then this will be a minimum for $V_{\rm glob}$. (A very similar mechanism is also commonly at work in supersymmetric gauge theories, where charged scalars automatically seek the zero of the positive $D$-term potential given in \pref{VDdef}.) The relaxon field likes in this way to zero out the biggest (order $M_p^0$) contribution in \pref{VFdef}, causing $W_\ssX$ to be Planck suppressed once gravitational interactions are included. We return below to why it remains consistent to use the formalism of nonlinearly realized supersymmetry when $W_\ssX$ is suppressed in this way.

Such a mechanism still leaves order $M_p^{-2}$ contributions to \pref{VFdef}, and because these are not positive definite they cannot as simply be removed using the same kind of relaxon mechanism. Here is where accidental scale invariance finally plays a role. Motivated by the accidental scaling symmetries known to be common in the low-energy limit of higher-dimensional supergravity, we propose that the theory comes to us with an action that is expanded in inverse powers of a large scalar field $\tau \gg 1$, 
\be
  S = S_0 + S_1 + S_2 + S_3 + \cdots \,,
\ee
with  each term in this expansion scaling homogeneously in the sense that $S_n \to \lambda^{1 - ns} S_n$ when $g_{\mu\nu} \to \lambda g_{\mu\nu}$ and $\tau \to \lambda^s \tau$ for constant $\lambda$.  This is as would be expected if $S_0 \to \lambda S_0$ and each successive term scales with an additional power of $1/\tau$ relative to the previous one. The scaling of $S_0$ is chosen to be consistent with the scaling of the 4D Einstein-Hilbert action, $S_\EH \propto M_p^2\int \exd^4x\, \sqrt{-g}\; \cR$, when written in Einstein frame. 

Within a supergravity framework we imagine $\tau$ being combined with an axion, $\mfa$, into a complex axio-dilaton field $\cT = \frac12(\tau + i \mfa)$ that, together with a spin-half field $\xi$, forms a proper\footnote{By so doing we assume that the mass splittings in this dilaton supermultiplet are small enough that all members remain in the low-energy theory, unlike for the Standard Model sector. We verify below that the gravitational coupling of these fields do keep their splittings to be similar to those in the gravity sector.} supermultiplet, $T$. Invariance under the axion shift symmetry $\mfa \to \mfa + c$ ensures $K$ depends only on $\tau = T + \ol T$ and that $W$ is $T$-independent, and the above condition of accidental approximate scale invariance says $K$ admits the expansion
\be \label{ScalingK}
   e^{-K/(3M_p^2)} = \tau F - k + \frac{h}{\tau} + \cO(1/\tau^2)
\ee
where $F$ is possibly a scale-invariant function of other fields, and none of $F$, $k$, $h$ and so on can depend on powers of $\tau$. They can be functions of any other fields besides $T$ (and, as it turns out \cite{Aghababaie:2002be}, potentially also on logarithms of $\tau$, as we shall see). 

Now comes the final bit of magic. The scale invariance of the leading $K = - 3 M_p^2 \ln(\tau F)$ term in \pref{ScalingK} suffices to prove \cite{Burgess:2008ir} that it is automatically of `no-scale' form \cite{NoScale}, for which $K$ satisfies the identity 
\be \label{noscaleform0}
   K^{\ol\ssA\ssB}K_{\ol\ssA} K_\ssB = 3 M_p^2 \,.
\ee
This guarantees the flatness of the potential along the directions in field space that do not appear in $W$ (such as $T$). But even though the action is {\it not} scale invariant in the same way when the first subdominant term in \pref{ScalingK} is kept, so $K = - 3M_p^2 \ln(\tau F - k)$, it happens that \pref{noscaleform0} remains true \cite{Burgess:2020qsc} provided only that $k$ does not depend on $T$. This type of accidental preservation of the no-scale structure beyond leading order in a large field expansion was first noticed in certain string compactifications \cite{Berg:2005ja, Cicoli:2007xp}, where it is called an `extended no-scale structure' and fits within the general approach described in \cite{Barbieri:1985wq}. The interplay between scale invariance and supersymmetry is more than the sum of its parts \cite{Burgess:2020qsc}: the flat potential for $\tau$ gets lifted at one higher order in $1/\tau$ than would naively be expected.\footnote{For aficionados: this is how we evade (really, co-exist with) Weinberg's no-go theorem \cite{CCWeinberg}. Although the theorem says flat directions built on scale invariance must be lifted, it does not say by how much and so does not preclude supersymmetry making the lifting smaller than would otherwise be generic.}

For the present purposes, what is nice about this last observation is that it means that the $1/M_p^2$ contributions to the potential also vanish, even after the relaxon has been integrated out, leaving the final dominant result at order $V \propto 1/M_p^4$. This is the start of the explanation for why the vacuum energy turns out to be of order $V \simeq m_{\rm vac}^4$ where $m_{\rm vac} \sim M_\TEV^2/M_p$ and $M_\TEV$ is of order the TeV scale. 

Because of the underlying scale invariance and the expansion in powers of $1/\tau$, the powers of $1/M_p$ in $V$ turn out to go along with powers of $1/\tau$ leading to a result for the potential that has size $V \sim M^8/(\tau M_p)^4$, where $M$ is the generic UV scale appearing everywhere in $K$ and $W$ on dimensional grounds. The upshot is that the generic $|W_\ssX|^2/\tau^2$ part of the potential -- including in particular any $M_\TEV^4$ contributions due to SM particles with masses $M_\TEV \propto 1/\sqrt\tau$ -- is cancelled, leaving a low-energy potential that depends on other parameters. Yet both the weak scale and the vacuum-energy scale are predicted to depend on $\tau$ in a manner consistent with $V \propto M_\TEV^4$.  

The next question becomes: why should the field $\tau$ be stabilized at such large values? \S\ref{ssec:DilationStabilization} shows that radiative corrections generically imply the function $k$ can depend on $\ln \tau$ \cite{Aghababaie:2002be}, and mild assumptions about this dependence give a potential for $\tau$ that {\it is} stabilized at very large values. Because these functions depend only logarithmically on $\tau$ minima can arise at astronomically large values while only dialing in hierarchies amongst the parameters in $k$ that are of order $\ln\tau$. Furthermore, standard renormalization-group (RG) methods allow this minimum to be reliably explored without losing control over the underlying radiative corrections.   

With this full picture in mind we can return to the question, deferred above, as to why a nonlinearly realized treatment of the Standard Model fields can be consistent even though the relaxon adjusts to ensure that $W_\ssX$ vanishes. These two conditions might normally be thought to contradict one another because it is the auxiliary field, $F^\ssX$, for the goldstino multiplet $X$, that is the measure of the size of supersymmetry breaking in the unseen sector that badly breaks supersymmetry (and thereby gives superpartners to the Standard Model large masses). In particular, the formulation of nonlinearly realized supersymmetry assumes $F^\ssX$ is a UV scale and works as an expansion in powers of $1/F^\ssX$. But in global supersymmetry the field equations usually predict that $F^\ssX$ is given by
\be\label{GlobalFW}
  F^\ssX \propto K^{\ol\ssX\ssX} \ol{W_\ssX} \,, 
\ee
and so large $F^\ssX$ should be inconsistent with small or vanishing $W_\ssX$. 

We argue that there are two reasons why the above framework is nonetheless consistent. First, in supergravity $F^\ssX$ is instead determined by 
\be \label{LocalFW}
   F^\ssX \propto K^{\ol\ssX\ssX} \ol{D_\ssX W} = K^{\ol\ssX\ssX} \left[\ol{W_\ssX} + \frac{K_{\ol\ssX} \ol{W}}{M_p^2} \right]
\ee
rather than by \pref{GlobalFW}, and so need not vanish even if $W_\ssX$ does. Second, relaxation actually implies that $W_\ssX$ is Planck-suppressed rather than strictly zero. Both of these can be consistent with a large-$F^\ssX$ expansion, if the Planck-suppressed terms in \pref{LocalFW} are sufficiently big. 

Ultimately the suppression of the vacuum energy relative to the weak scale depends on the size of $\tau$, with $\tau \sim 10^{26}$ proving to be consistent with the two observed hierarchies $M_\TEV \sim M_p/\sqrt\tau$ and $V \sim (M_\TEV^2/M_p)^4$. However -- as discussed in \S\ref{ssec:UVcomplete} -- additional constraints on how large $\tau$ can be arise once its UV origins are more explicit. The same UV frameworks also provide extra sources of suppression (such as warping), making the final solution likely involve a cocktail of suppressions, possibly along the lines described in \S\ref{ssec:UVcomplete}.

Explicit details of the above construction are given in later sections, but an immediate consequence of the scale invariance and any successful suppression of the cosmological constant is that the dilaton field $\tau$ must be very light, with a mass of order the present-day Hubble scale.\footnote{The size of the dilaton mass in this model is a special case of a general result \cite{AndyCostasnMe} that a gravitationally coupled scalar field, $\cL \sim M_p^2 (\partial \theta)^2 + V(\theta)$, whose potential at its minimum successfully gives the observed dark energy density, generically predicts a mass for $\theta$ that is of order the present-day Hubble scale, $H$. It does so because the generic condition $V''(\theta) \sim V(\theta) \sim \cO(1)$ at the minimum implies a mass $m^2 \sim V/M_p^2$, which is order $H^2$ whenever the scalar potential dominates the universal energy density.} It follows that it must be cosmologically active up to the current epoch, and so {\it predicts} Dark Energy must be described by a specific type of near-scale-invariant quintessence theory \cite{quintessence, ScaleAnomalyCC}, but (remarkably) one for which both the cosmological constant and the quintessence-field mass would be technically natural. 

But it gets better than this. A gravitationally coupled scalar as light as the Hubble scale should stick out in tests of gravity like social skills at a physics meeting. Indeed, the low-energy lagrangian relevant to astrophysics is explored in \S\ref{sec:Phenomenology} where it is shown that the underlying scale invariance forces the dilaton $\tau$ to couple to Standard Model matter as does a Brans Dicke scalar \cite{BransDicke} (at leading order in $1/\tau$ -- a {\it great} approximation when $\tau \sim 10^{26}$). And it does so with a coupling that is apparently too large to have escaped detection in precision tests of gravity in the solar system and elsewhere \cite{TestsofGR}. A more careful look, however, shows that its supersymmetric partner (the axion) can save the day, and does so because of the target-space axion-dilaton interactions also automatically predicted by the model. As explored in more detail in \cite{ADScreening}, the axion-dilaton interactions have the effect of making matter-dilaton couplings largely generate external axion fields (rather than dilaton fields), which are much less effective at altering test-particle motions within the solar system and so can escape detection. We call this mechanism `axion homeopathy' because it can work for extremely small direct axion-matter couplings, provided only that these are nonzero. 

Because the phenomenology of the axio-dilaton field is so crucial to the viability of such models, \S\ref{sssec:DilatonCosmology} provides a preliminary discussion of axio-dilaton cosmology and checks that the most basic things work (though without doing justice to the entirety of the constraints that a viable model must ultimately pass -- further studies of structure formation and CMB properties within this framework are important to explore). However even if axio-dilaton phenomenology eventually poses challenges to the version of this approach we present here, we regard any such model-building problems within this general framework to be a good trade for progress on the (much harder) cosmological constant problem.  

There are also many other ways to test this picture, such as through tests of gravity and the changes predicted in cosmology during well-measured epochs (such as the variations in fundamental masses -- in Planck units -- that are predicted whenever the dilaton field $\tau$ varies in space and time). Intriguingly, some of these may actually help with the Hubble tension \cite{HubbleTension} by allowing particle masses (in particular $m_e$) all to differ by a common factor at recombination relative to their values today. Such a scaling potentially exploits the mechanism described in \cite{Sekiguchi:2020teg}, and we briefly check that the basic requirements of this mechanism can be satisfied.

So what is the catch? For the long-distance physics (below the eV scale) relevant to astrophysics, we do not see a fundamental one yet and not for want of looking. The main provisos about which we worry are described in \S\ref{ssec:LoopsAndTN} and \S\ref{sec:Conclusions} below. They start with the observation that the large value for $\tau$ required to explain the hierarchies also implies a breakdown of EFT methods well below electroweak scales. It does so because $\tau \sim 10^{26}$ implies the axion decay constant is $f_a \sim M_p/\tau \sim 10$ eV. This need not in itself be a problem because supersymmetric extra dimensions \cite{SLED} could provide a plausible UV completion at these scales, while remaining consistent with SM degrees of freedom being four-dimensional as assumed here.

The worries come once the low-energy picture is embedded into such a UV completion, because new constraints on the value of $\tau$ can arise depending on precisely how this is done. For instance a natural choice in extra-dimensional models identifies $\tau$ with the extra-dimensional volume modulus, but this seems to require $\tau \lsim 10^{20}$. Either $\tau$ must arise differently in the UV completion or there must be additional sources of hierarchical suppression (or both). We explore some of the options in \S\ref{ssec:UVcomplete}. Other worries include ensuring the naturalness of having the relaxon be so light; exploring the detailed stability of the nonlinearly realized supergravity form in regions for which $W_\ssX$ is Planck-suppressed; and so on.

Our presentation is organized as follows. \S\ref{sec:Relaxation} describes the main mechanism in some detail, starting by explicitly writing out the EFT applicable at energies just below the electron mass. Since all of the dangerous Standard Model particles are integrated out at this point the EFT at these scales illuminates most clearly the interplay between scale-invariance and the relaxation mechanism. Standard Model fields are then reintroduced, allowing their couplings to low-energy states to be made more precise.

Knowledge of Standard Model couplings allows a more explicit assessment of naturalness issues, such as why low-energy gravitational physics is relatively insensitive to the loops of Standard Model fields. This is the topic of \S\ref{ssec:LoopsAndTN}. Since most of the hierarchies of scale in the model are set by the background value of the dilaton field, this section also shows how this field can be stabilized, along the lines described above. 

\S\ref{sec:Phenomenology} makes a down payment on the most pressing phenomenological challenges, including a derivation of the low-energy EFT relevant to astronomical and cosmological tests. These give the field equations used in \cite{ADScreening} to evade the constraints on dilaton-matter couplings coming from tests of General Relativity (GR) in the solar system (whose results are merely quoted here). \S\ref{sssec:DilatonCosmology}  provides a preliminary evaluation of the cosmological evolution of the axio-dilaton fields for comparison with some features of late-universe cosmology. This includes a brief discussion of potential relevance to the Hubble tension, mentioned above. 

Finally \S\ref{sec:Conclusions} summarizes some of the implications of our proposal; contrasts our approach with other discussions that build on the role of scale invariance. Along the way this section outlines several topics for future investigation -- such as possible implications for dark matter, inflation, baryogenesis and neutrino physics -- that we do not study here.

\section{The model}
\label{sec:Relaxation}

The goal in this section is to set up the lagrangian for a low-energy nonsupersymmetric world coupled to supergravity, for which there is a relaxation mechanism that dynamically removes the gravitational response of any vacuum energy obtained by integrating out Standard Model fields. When doing so we lean heavily on the description given in \cite{LowESugra} of the form this lagrangian must take \cite{Komargodski:2009rz, Bergshoeff:2015tra, DallAgata:2015zxp, Schillo:2015ssx}.

The idea is two-fold: first combine supersymmetry and scale invariance with a relaxation mechanism to make the leading part of the scalar potential small and then use the `extended no-scale' mechanism to minimize the damage to the scalar potential that inevitably must come once subdominant scale-invariance breaking effects are included.

\subsection{SUSY scaling and relaxation}
\label{ssec:QNSRelaxation}

We start with a minimal formulation, to which complications are later introduced as needed. We concentrate (in the first instance) on the effective theory that applies below the electron mass in order to focus on the more difficult low-energy part of the cosmological constant problem, putting aside until later a discussion of the theory further into the UV. We do not take the low-energy potential of this theory to be particularly small because loops of heavier Standard Model fields are assumed to contribute to it in an unsupressed way, giving contributions involving the power of $m$ (for a field of mass $m$) required on dimensional grounds (and so includes the dangerous  contributions $V \ni m^4$).  

\subsubsection{Low-energy particle content}

In this EFT some Standard Model fields remain: the photon and the neutrinos, but because these do not themselves contribute dangerous vacuum energies we do not include them explicitly in our initial description of the model's vacuum dynamics. (They are of course included in later sections about low-energy phenomenology.) Our focus is initially to see how the new degrees of freedom we propose in this section can suppress the contribution of UV-sensitive terms in the lagrangian to the spacetime curvature.  
  
To this end we postulate several new low-energy degrees of freedom: a very supersymmetric low-energy gravity sector (with both a massless graviton and very light gravitino); a dilaton-dilatino scalar supermultiplet $T = (\cT, \xi)$ that contains the low-energy pseudo-Goldstone boson for accidental scale invariance plus its supersymmetric partners. All of the above particles are assumed to be very light, and this assumption is checked {\it ex post facto} by computing their masses once the model is fully formulated. 

In addition to the above fields we also add a real scalar, $\phi$, not to be confused with the Standard Model's Higgs (which has been integrated out). The scalar $\phi$ nonlinearly realizes supersymmetry (as do the photon and left-handed neutrinos), since its superpartner is assumed to be heavy and so to have already been integrated out. The relaxaton field $\phi$ itself should be light enough to appear in the EFT below the electron mass, and were it not for this condition its role could be played by the Standard Model Higgs. We discuss in \S\ref{ssec:LoopsAndTN} the naturalness issues associated with $\phi$ being this light. 

According to the rules for nonlinearly realizing supersymmetry given in \cite{Komargodski:2009rz, Bergshoeff:2015tra, DallAgata:2015zxp, Schillo:2015ssx, LowESugra}, we are to build our EFT using the standard rules \cite{FreedmanVanProeyen, WB} for constructing a supergravity lagrangian using specific types of constrained superfields for each particle in the theory that does not have an explicit superpartner: 
\begin{itemize}
\item The UV supersymmetry breaking order parameter $F^\ssX$ and the spin-half goldstino field $G$ (ultimately eaten by the gravitino) turn out to be described by a chiral multiplet $X$ satisfying a nilpotent condition $X^2 = 0$ that allows its scalar component $\cX \in X$ to be expressed in terms of $G$ and $F^\ssX$. 
\item Scalar fields like the `relaxon' $\phi$ are represented by a superfield $\Phi$ that satisfies the constraint that the supercovariant derivative $\ol{\cD}(X \ol{\Phi})$ vanishes, which says $X \ol\Phi$ is left-chiral. This constraint removes both the fermionic and auxiliary field parts of $\Phi$ as independent variables, leaving only the scalar $\phi$ as a physical degree of freedom. The constraint also implies $X \cF(\Phi,\ol{\Phi})$ is left-chiral for more general functions $\cF$. A reality condition for $\phi$ gets represented in this framework as the constraint $X \ol{\Phi} = X \Phi$.
\end{itemize}

With these constraints the theory is specified at the two-derivative level by giving the K\"ahler potential $K$, superpotential $W$ and gauge kinetic function $\mff_{\alpha\beta}$ as functions of the above superfields. 

\subsubsection{Low-energy lagrangian}

Accidental scale invariance is incorporated by assuming the theory has an expansion in powers\footnote{Because the theory is organized as an expansion in powers of $1/\tau$ we take the field $\tau$ to be dimensionless (and so not canonically normalized).} of $1/\tau$, where $\tau := \cT + \ol\cT$. For supersymmetric theories this means $W$ and $e^{-K/(3M_p^2)}$ should be expanded in this way, although for $W$ this doesn't say much because we also assume $\mfa :=2\, \hbox{Im}\,\cT$ enjoys an axionic shift symmetry that prevents $\cT$ from appearing in $W$ at all.

Scale invariance for the leading term of the expansion of $e^{-K/(3M_p^2)}$ requires it to be a homogeneous function of $\cT$ and we are free to define $\cT$ so that this leading term is homogeneous degree one. Subsequent orders in $1/\tau$ break this scale invariance, as do loops built from the leading term (because the above assumptions ensure that $1/\tau$ plays the role of $\hbar$ in this part of the theory). We assume all such corrections arise as integer powers of $1/\tau$ (apart from logarithms -- more about which below).

The first terms in the expansion of the K\"ahler function therefore become
\bea\label{RelaxKW}
  &&K \simeq -3 M_p^2 \ln \cP \quad\hbox{with} \quad \cP(\tau,X,\ol X,\Phi,\ol\Phi) = \tau - k + \frac{h}{\tau} + \cdots \\
  &&\qquad\qquad \hbox{and} \quad W\simeq w_0(\Phi) +  X w_{\ssX}(\Phi, \ol \Phi)  \,,\nn
\eea
where the ellipses denote higher orders in $1/\tau$ and the functions $k$, $h$ and $W$ are otherwise chosen to be the most general consistent with the constraints $X^2 = X(\ol\Phi - \Phi) = 0$:
\be \label{kexpX}
  k = \frac{1}{M_p^2} \Bigl\{ \mfK(\Phi,\ol{\Phi},\ln \tau) + \Bigl[ X  \mfK_\ssX(\Phi,\ol{\Phi},\ln\tau) + \hbox{h.c.} \Bigr] +  \ol{X} X \mfK_{\ssX\ol\ssX}(\Phi,\ol{\Phi},\ln\tau) \Bigr\}\,,
\ee
and similarly for $h$ and higher-order terms (although these are not needed in what follows, except briefly in \S\ref{ssec:LoopsAndTN}). The series in $1/\tau$ is written explicitly so the functions $k$ and $h$ do not depend on powers of $\tau$, but they can in principle\footnote{As we see below a logarithmic dependence on $\tau$ can naturally arise once loop corrections are included.} depend on $\ln\tau$ as is indicated in \pref{kexpX}.

Since all factors of $M_p$ are explicit and $X$ has dimension mass, the function $\mfK$ has dimension (mass)${}^{2}$, while $\mfK_\ssX$ has dimension (mass) and $\mfK_{\ssX\ol\ssX}$ is dimensionless. (The $M_p$'s are chosen so that the $\phi$ kinetic term and the `global supersymmetry' term involving $|w_\ssX|^2$ in \pref{VFtauexp} are both independent of $M_p$.) The superpotential functions $w_n$ similarly have dimension (mass)${}^{3-n}$. Because of the constraints $X^2 = X (\ol{\Phi} - \Phi) = 0$ it is always possible to rescale $X \to \tilde X \cF(\Phi,\ol{\Phi})$ with $\cF$ chosen to set $\mfK_{\ssX\ol\ssX} = 1$ (provided $\mfK_{\ssX\ol\ssX}$ does not depend on $\ln\tau$).   

The scalar potential $V_\ssF$ obtained given these functions $K$ and $W$ is as given in \pref{VFdef}, repeated here for convenience
\be \label{VFdef2}
  V_\ssF = e^{K/M_p^2} \left[ K^{\bar A B} \ol{D_\ssA W} D_\ssB W - \frac{3|W|^2}{M_p^2} \right]   \,.
\ee
Despite appearances there is an important change here relative to ordinary supergravity: the absence of independent auxiliary fields in the constrained field $\Phi$ implies that the sums on the indices $A$ and $B$ only run over the fields  $z^\ssA := \{ T, X\}$ and not also over $\Phi$ \cite{DallAgata:2015zxp}. 

Working to leading nontrivial order in $1/\tau$ we drop $h/\tau$ and higher orders, so
\be\label{cPvsKnmfK}
  \cP \simeq \tau - k = \tau  - \frac{\mfK}{M_p^2}  
\ee
and the K\"ahler metric and its inverse become
\be  \label{KInv0}
   K_{\ssA \ol \ssB} \simeq \frac{3M_p^2}{\cP^2}  \left( \begin{array}{ccc}
  1  &&  -k_{\ol \ssX} \\  -k_ \ssX &&  \cP \, k_{ \ssX \ol  \ssX} + k_ \ssX k_{\ol  \ssX}
\end{array} \right) 
\quad \hbox{and} \quad 
   K^{\ol \ssB \ssA} \simeq \frac{\cP}{3M_p^2}  \left( \begin{array}{ccc}
  \cP + \mfk^2  &&  k^{ \ssX} \\   k^{\ol  \ssX} &&  k^{\ol  \ssX  \ssX}  
  \end{array}
\right) \,,
\ee
where $z^\ssA := \{ T, X\}$ and subscripts on $k$ as usual denote differentiation. Furthermore $\mfk^2 := k^{\ol \ssX \ssX} k_ \ssX k_{\ol \ssX}$, $k^ \ssX := k^{\ol \ssX \ssX} k_{\ol \ssX}$ and $k^{\ol \ssX} = k^{\ol \ssX \ssX} k_ \ssX$ with $k^{\ol \ssX \ssX} := 1/ k_{ \ssX \ol \ssX}$. 

At leading order the K\"ahler covariant derivatives for $T$ and $X$ (evaluated at $X = 0$) similarly become
\be \label{DTWnDXW}
   D_\ssT W = \frac{K_\ssT W}{M_p^2} =  -\frac{3w_0}{\cP} \quad \hbox{and} \quad
   D_\ssX W = W_\ssX + \frac{K_\ssX W}{M_p^2} = w_{\ssX} + \frac{3\mfK_{\ssX} w_0}{\cP M_p^2} \,.
\ee
The useful identity $K^{\ol \ssB\ssA} K_\ssA = - \bar z^{\ol\ssB}$ (that follows -- see {\it e.g.}~eq.~\pref{Kupper1} -- from the no-scale \cite{NoScale} property $K^{\ol\ssB\ssA} K_{\ol\ssB} K_\ssA = -3 M_p^2$) then shows that the Einstein-frame auxiliary fields $F^{\ol\ssB} = e^{K/(2M_p^2)} K^{\ol\ssB\ssA} D_\ssA W$ are 
\be \label{Fauxcov}
    F^{\ol\ssT} = -   e^{K/(2M_p^2)} \frac{\tau \, W}{M_p} \simeq  \frac{w_0}{\sqrt\tau M_p}  \quad\hbox{and} \quad
    F^{\ol\ssX} = e^{K/(2M_p^2)} W_\ssX \simeq \frac{w_\ssX}{\tau^{3/2}} \,,
\ee
where the approximate equalities use $\cP \simeq \tau$. 
 
Denoting (as above) $z^\ssA = \{T, X\}$ and neglecting subdominant powers of $1/\tau$ one finds the potential \pref{VFdef2} becomes (see eq.~\pref{VFBasicFormzz} of Appendix \ref{App:CaseNo2} for details)
\be \label{VFtauexp}
  V_\ssF  \simeq  \frac{1}{\cP^2} \left[ \frac13 \, \mfK^{\ol\ssX \ssX} \ol{w_{\ssX}} w_{\ssX} + \frac{\mfK^{\ol\ssX \ssX} \mfK_{\ssX \ol\ssT}}{M_p^2} \; w_0 \ol{w_{\ssX}}  + \frac{\mfK^{\ol\ssX \ssX} \mfK_{\ssT \ol\ssX}}{M_p^2}\; w_{\ssX} \ol{w_0} - \frac{3(\mfK_{\ssT\ol\ssT}  - \mfK^{\ol\ssX \ssX} \mfK_{\ssT \ol\ssX} \mfK_{\ssX \ol\ssT})}{  1  + 2 \mfK^{\ssX\ol\ssX}  \mfK_\ssX \mfK_{\ol\ssX}/M_p^2 } \; \frac{|w_0|^2}{M_p^4}
 \right] \,.  
\ee
The powers of $M_p$ are here shown explicitly for later convenience, and follow directly from the dimensions of $X$ and $T$ and from \pref{cPvsKnmfK}. The unusual $M_p^{-4}$ of the last term is an artefact of $T$ being dimensionless. 

Notice that only the first term survives if $\mfK$ is independent of $T$, and this happens because in this case \pref{cPvsKnmfK} becomes a no-scale model. Because $\mfK$ depends on $T$ only through $\ln \tau$ it follows that each derivative with respect to $T$ costs a power of $1/\tau$ and so the $w_0$--$w_\ssX$ mixing terms of \pref{VFtauexp} arise at order $1/\tau^3$ while the $|w_0|^2$ term first appears at order $1/\tau^4$. Contributions from the function $h$ in \pref{RelaxKW} involve at least one additional power of $1/\tau$ compared to those shown.

The kinetic terms for the physical scalars $z^\ssI := \{ \cT \,, \phi \}$ are given by the second derivatives of $K$ in the usual way: $\mfL_{{\rm kin\, scal}}= - \sqrt{-g}\, K_{\ssJ\ol\ssI} \partial_\mu z^\ssJ \partial^\mu \bar z^{\ol\ssI}$ \cite{DallAgata:2015zxp}. Evaluating at $X = 0$ and working to lowest order in $1/\tau$ gives
\bea \label{TkinL0}
  -\frac{\mfL_{{\rm kin\, scal}}}{\sqrt{-g}} &=&  K_{\ssT \ol\ssT} \,\partial^\mu \ol{\cT} \partial_\mu \cT + K_{\Phi \ol\Phi} \,\partial^\mu \ol\phi\, \partial_\mu \phi + \Bigl( K_{\Phi\ol\ssT}\, \partial^\mu \ol \cT \,\partial_\mu \phi + \hbox{h.c.} \Bigr)  \\
  &=& \left[ \frac{3M_p^2}{\cP^2} \, \partial^\mu \ol{\cT} \partial_\mu \cT +  \frac{3}{\cP}\, \mfK_{\Phi\ol{\Phi}} \, \partial^\mu \ol\phi \,\partial_\mu \phi  - \frac{3}{\cP^2}\,  \mfK_{\Phi} \Bigl( \partial^\mu \ol \cT \,\partial_\mu \phi  + \hbox{h.c.} \Bigr) \right] \left[ 1 + \cO\Bigl(\tau^{-1} \Bigr) \right] \,. \nn
\eea
The leading off-diagonal kinetic term for fluctuations can be removed through a field redefinition of the form $\delta\phi \to \delta\phi + A \,\delta\cT$ where $A = \mfK_\Phi/(\tau\mfK_{\Phi\ol\Phi})$ is a function of the background fields; while correcting the diagonal kinetic terms only by subdominant powers of $1/\tau$. 

\subsubsection{Relaxation mechanism}
\label{sssec:RelaxationScales}

When minimizing the potential $V_\ssF$ of \pref{VFtauexp} with respect to $\Phi$ it is the $|w_{\ssX}|^2$ term that should dominate for large $\tau$ because it is proportional to the fewest powers of $1/\tau$ (and the fewest powers of $1/M_p$). 

\subsubsection*{The case $\mfK_{\ssX\ol\ssT} = 0$}

For simplicity of explanation suppose first that $\mfK_{\ssX\ol\ssT} = 0$ (the more general case is handled below, with similar results). In this case all $w_\ssX$--\,$w_0$ cross terms vanish and the potential \pref{VFtauexp} simplifies to 
\be \label{VFtauexpyy}
  V_\ssF  \simeq \frac{1}{\cP^2} \left[ \frac13 \, \mfK^{\ol\ssX \ssX} | w_{\ssX}|^2  - \frac{3\mfK_{\ssT\ol\ssT}  }{  1  + 2 \mfK^{\ssX\ol\ssX}  \mfK_\ssX \mfK_{\ol\ssX}/M_p^2 } \; \frac{|w_0|^2}{M_p^4}
 \right] \,,
\ee
which the supergravity auxiliary field structure has ensured is a sum of squares,\footnote{Notice that the second term in this potential is positive when $\mfK_{\ssT\ol\ssT} < 0$ (as we assume henceforth) and this is allowed because $\mfK_{\ssT\ol\ssT}$ does {\it not} control the sign of the kinetic term for $\cT$ ({\it c.f.}~eq.~\pref{TkinL0}).} with the $|w_\ssX|^2$ term arising at order $1/\tau^2$ and the $|w_0|^2$ term arising suppressed by both $1/\tau^4$ and $1/M_p^4$. Of these it is the $|w_\ssX|^2$ that contains the usual dangerous contributions to the vacuum energy, since \cite{LowESugra} shows $w_\ssX$ receives contributions of order $m^2$ when integrating out non-supersymmetric particles of mass $m$.  

Suppose however that this function has a zero for some nonzero value of $\Phi$, such as\footnote{For reasons given in \S\ref{ssec:Loops} we assume a symmetry $\phi \to - \phi$ to prevent $W$ from depending linearly on $\phi$. Even though we take the simplest quadratic case, our discussion below  extends straightforwardly to any function with a non trivial zero.} if
\be \label{RelaxWx}
   w_{\ssX} = g( \ol{\Phi} \Phi - v^2 ) = g(\Phi^2 - v^2)
\ee
near some $\Phi = v$, where the last equality uses the constraint $X \ol{\Phi} = X \Phi$ to trade $\ol\Phi$ for $\Phi$ and the parameter $g$ is dimensionless. Because the $|w_\ssX|^2$ term is both nonnegative and the largest term in the potential, it is energetically favourable for $\phi$ to seek the zero of $w_{\ssX}$, thereby turning this term off and allowing the $|w_0|^2$ term to dominate. As noted above, this remaining term vanishes if $\mfK$ is independent of $\ln \tau$, and does so because in this case the supergravity becomes a no-scale model for which $T$ is a flat direction. 

The second $\phi$-derivative of the potential near this point is of order $V_{\Phi\ol{\Phi}} \sim |w_{\ssX\Phi}|^2/\tau^2 \sim (g v)^2/\tau^2$. Comparing this to the $\phi$ kinetic term, which -- assuming $\mfK_{\Phi\ol{\Phi}}$ to be order unity -- has the form $\cZ\, \partial_\mu \ol\phi \, \partial^\mu \phi$ with $\cZ \sim 1/\tau$ for large $\tau$, we see that the $\phi$ mass is of order 
\be \label{mphivstau}
  m_\phi^2 \sim \frac{V_{\Phi\ol{\Phi}}}{\cZ_{\Phi\ol{\Phi}}} \sim \frac{(gv)^2}{\tau} \,.
\ee

For comparison, the value of the potential at this minimum is given (because $w_{\ssX} = 0$ and $\mfK_{\ssX\ol\ssT} = 0$) by
\be \label{Vmintau4}
   V_{\rm min} \sim \frac{\mfK_{\ssT\ol\ssT} |w_0|^2}{\tau^2 M_p^4} \sim \frac{M^2 \mu_\ssW^6}{\tau^4 M_p^4}
\ee
which uses $\mfK = \mfK(\ln\tau)$ to conclude $\mfK_{\ssT\ol\ssT} \propto 1/\tau^2$ and otherwise assumes $\mfK \sim M^2$ and $w_0 \sim \mu_\ssW^3$ when $\Phi = v$, where $M$ and $\mu_\ssW$ are generic UV scales\footnote{As we see below these scales both turn out to be large, but are different. This difference is argued in \S\ref{ssec:LoopsAndTN} to be both natural and compatible with expectations coming from string compactifications.} to be determined (with $M$ eventually identified with the Planck scale $M_p$). In later sections we see that the $\tau$-dependence of ordinary (Standard Model) particles in this scenario is also generically 
\be \label{weakscalevsMtau}
   m_\TEV \sim \frac{M}{\sqrt\tau}
\ee
and so the $\tau$-dependence of this leading contribution to the potential suggests an origin for the successful numerology $V_{\rm min} \sim (m_\TEV^2/M_p)^{4}$ regardless of the value of $\tau$.

This entire picture using nonlinearly realized supersymmetry only makes sense if the $F$-term of the $X$ multiplet is large, presumably the scale of the masses of the superpartners of Standard Model fields, and one might worry that this is not possible if $W_\ssX = w_{\ssX}$ is being arranged to vanish. This would indeed be a legitimate worry to the extent that $F^\ssX \propto w_\ssX$ (as is true both in global supersymmetry and \pref{Fauxcov}). The conclusion $w_\ssX = 0$ turns out to be a consequence of the non-essential simplifying assumption $\mfK_{\ssX\ol\ssT} = 0$ made above, however, so before proceeding further we first pause to relax this assumption.

\subsubsection*{More general $\phi$-dependence and the case $\mfK_{\ssX\ol\ssT} \neq 0$}

To show what happens when $\mfK_{\ssX\ol\ssT} \neq 0$ we return to the potential \pref{VFtauexp}, reproduced here for convenience of reference
\be \label{VFtauexpagain}
  V_\ssF   \simeq  \frac{1}{\cP^2} \left[ \frac13 \, \mfK^{\ol\ssX \ssX} \ol{w_{\ssX}} w_{\ssX} + \frac{\mfK^{\ol\ssX \ssX} \mfK_{\ssX \ol\ssT}}{M_p^2} \; w_0 \ol{w_{\ssX}}  + \frac{\mfK^{\ol\ssX \ssX} \mfK_{\ssT \ol\ssX}}{M_p^2}\; w_{\ssX} \ol{w_0} - \frac{3(\mfK_{\ssT\ol\ssT}  - \mfK^{\ol\ssX \ssX} \mfK_{\ssT \ol\ssX} \mfK_{\ssX \ol\ssT})}{  1  + 2 \mfK^{\ssX\ol\ssX}  \mfK_\ssX \mfK_{\ol\ssX}/M_p^2 } \; \frac{|w_0|^2}{M_p^4}
 \right] \,,
\ee
and ask what the low energy potential for $\tau$ is after extremization with respect to $\phi$. This extremization is also relatively simple to do when $\mfK_{\ssX}$ (and so also $\mfK^{\ol\ssX \ssX}$ and $\mfK_{\ssT\ol\ssX}$) are independent of $\phi$ (though we also relax the assumption of $\phi$-independence below). 

In this case $\phi$ enters into $V_\ssF$ only through $w_\ssX$ and so extremizing $V_\ssF$ with respect to $\phi$ is equivalent to extremizing with respect to $w_\ssX$. Since the dependence of $V_\ssF$ on $w_\ssX$ is quadratic the extremization with respect to $w_\ssX$ is easily obtained by evaluating at the saddle point 
\be \label{wxsaddle}
   w_\ssX = - \frac{3 \mfK_{\ssX\ol\ssT} \,w_0}{M_p^2} \,,
\ee
showing that in the general case $w_\ssX$ does not vanish, but is both Planck-suppressed and down by a power of $1/\tau$ (because logarithmic dependence of $\mfK$ implies $\mfK_{\ssX\ol\ssT} \sim 1/\tau$). Using this in the potential leads to
\be \label{VFBasicFormzzmin22}
  V_\ssF  \simeq  -\frac{3|w_0|^2}{\tau^2 M_p^4} \left[  \mfK^{\ol\ssX \ssX} \mfK_{\ssX\ol \ssT} \mfK_{\ssT\ol\ssX}   + \frac{\mfK_{\ssT\ol\ssT}  - \mfK^{\ol\ssX \ssX} \mfK_{\ssT \ol\ssX} \mfK_{\ssX \ol\ssT}}{  1  +2 \mfK^{\ssX\ol\ssX}  \mfK_\ssX \mfK_{\ol\ssX}/M_p^2 }
 \right] =: \frac{U}{\tau^4}  \,.
\ee
Although the overall sign in this expression is negative the square bracket (and therefore also $V_\ssF$) can have either sign depending on the properties of $\mfK$.

This derivation also shows why the simplifying assumption that $\mfK^{\ol\ssX\ssX}$ and $\mfK_{\ssT\ol\ssX}$ be $\phi$-independent is not crucial. The main point is that the dominant $\phi$ dependence enters $V_\ssF$ at order $1/\tau^2$ and because this arises proportional to $|w_\ssX|^2$ it always favours configurations that make $w_\ssX$ vanish. Although the subdominant $1/\tau^3$ terms in $V_\ssF$ move the minimum away from $w_\ssX = 0$, its displacement from zero it at most of order $\delta w_\ssX \sim 1/\tau$, and this is equally true whether or not $\phi$ enters the potential through $\mfK$ or $W$. Because the dominant result is then quadratic in $\delta w_\ssX$ it very robustly contributes $V_\ssF \sim 1/\tau^4$ once $\phi$ is eliminated. Although the detailed form of $U(\ln\tau)$ in \pref{VFBasicFormzzmin22} can depend on UV specifics, what is important is that the leading result is always proportional to $1/\tau^4$.

\subsubsection{Scales and supersymmetry breaking}
\label{sssec:SUSYScales}

We return now to the question of how large $F^\ssX$ is, and the consistency of using nilpotent fields. The starting point is expression \pref{wxsaddle} for the size of $w_\ssX$ which implies $w_\ssX \sim M \mu_\ssW^3/(\tau M_p^2)$, where we again estimate $\mfK_{\ssX\ol\ssT}   \sim M/\tau$ and $w_0 \sim \mu_\ssW^3$ on dimensional grounds, with the power of $\tau$ coming from differentiating a function of $\ln\tau$. In this case the auxiliary field given in \pref{Fauxcov} becomes
\be \label{FXEF}
  F^{\ol\ssX}  \sim  e^{K/(2M_p^2)}  K^{\ol\ssX\ssX}  w_\ssX   \sim \frac{M \mu_\ssW^3}{ \tau^{3/2} M_p^2} \,.
\ee
As we see in more detail in \S\ref{ssec:LoopsAndTN}, the scale $\mu_\ssX$ defined by $F^\ssX = \mu^2_\ssX$ sets the mass scale for {\it e.g.}~scalar superpartners for SM particles, and so must be a UV scale (since it was the integrating out of these particles that necessitates use of the nilpotent formalism).

Eq.~\pref{FXEF} for $\mu_\ssX$, the expression $m_\TEV \sim M/\tau^{1/2}$ for the SM-particle masses (also justified in more detail below) and\footnote{We include here a factor $\epsilon^5$ in $V_{\rm min}$ where $\epsilon \lsim 1/60$ -- whose origins lie in the stabilization mechanism for $\tau$, as explained below eq.~\pref{VFBasicFormzzminzs1NAPPx} -- since this `order-unity' factor matters when inferring a size for $\tau$.} $V_{\rm min} = \epsilon^5 m_{\rm vac}^4$ for the vacuum energy determine the three input parameters $\mu_\ssW$, $M$ and $\tau$. The relaxon mass $m_\phi$ then fixes $gv$. In particular, successful phenomenology requires
\be \label{mvacest}
    V_{\rm min} \sim \frac{\epsilon^5 M^2 \mu_\ssW^6}{\tau^4 M_p^4}   \sim  \Bigl( 10^{-11} \; \hbox{GeV} \Bigr)^4   \,,
\ee
while SM scales are set (up to small dimensionless gauge and Yukawa couplings) by 
\be \label{Mconstrainedtau}
   m_\TEV \sim \frac{M}{\tau^{1/2}} \sim 10^3 \; \hbox{GeV}  \,,
\ee 
while superpartner masses must satisfy
\be \label{muXest}
   \mu_\ssX^2   \sim  \frac{M\mu_\ssW^3}{\tau^{3/2} M_p^2} \gsim  \Bigl( 10^4 \; \hbox{GeV} \Bigr)^2 \,.
\ee

These can be cast into two useful dimensionless combinations
\be \label{SPext}
    \frac{\mu_\ssX^2}{m_\TEV^2} \sim \frac{\mu_\ssW^3}{M M_p^2 \sqrt\tau}   \gsim 100 \,,
\ee
and
\be \label{mvacestmuX}
    \frac{V_{\rm min}}{\mu_\ssX^4} \sim \frac{\epsilon^5 }{\tau}   \lsim  \Bigl( 10^{-15}\Bigr)^4 \sim 10^{-60}  \,.
\ee
Eq.~\pref{mvacestmuX} shows we wish to have $\tau$ as large as possible, and \pref{Mconstrainedtau} argues that this requires $M$ also to be chosen as large as possible. If $M \sim M_p \sim 10^{18}$ GeV then \pref{Mconstrainedtau} implies $\tau \sim 10^{30}$ and so \pref{SPext} and \pref{mvacestmuX} then give $\epsilon \sim 10^{-5}$ and $\mu_\ssW \sim 4.6 \, M_p \,\tau^{1/6} \sim 10^5 M_p$:
\be \label{baseline1}
  M \sim M_p \,, \quad \mu_\ssW \sim 10^5 \, M_p \,, \quad \tau \sim 10^{30} \quad\hbox{and} \quad \epsilon \sim 10^{-5} \,.
\ee
Although having scales larger than $M_p$ might seem unusual, \S\ref{ssec:UVcomplete} argues why it can actually be expected for the scale $\mu_\ssW$ if the UV completion is a string compactification. 

The above estimates rely heavily on eq.~\pref{VFBasicFormzzmin22} for the scalar potential, and this (in particular the factors of $\epsilon$) relies on precisely how a minimum arises for $V$ that fixes the present-day value for $\tau$. These estimates use the stabilization mechanism described in detail in \S\ref{sssec:LogStab}. The sensitivity of the above numbers to these choices can be assessed by comparing this stabilization mechanism with the alternative mechanism sketched in \S\ref{sssec:AltStab} (which so far as we can see turns out not to improve on the numbers given above). 

Although SM superpartners dominantly acquire masses through their couplings to $F^\ssX$ (see \S\ref{ssec:MatterCouplings}) the same is not true of the gravitino, which responds to the total invariant order parameter
\be
   \cF := \Bigl( K_{\ssA\ol\ssB} F^\ssA \ol F^{\ol\ssB} \Bigr)^{1/2} = \Bigl[  e^{K/M_p^2} K^{\bar\ssA \ssB} \ol{D_\ssA W} \, D_\ssB W\Bigr]^{1/2} 
   \sim e^{K/(2M_p)}  \frac{|W|}{M_p}  \sim    \frac{\mu_\ssW^3}{\tau^{3/2} M_p} \,,
\ee
where the approximate equalities use \pref{Fauxcov}, the form for $K$ and the earlier estimate for $w_\ssX$. The corresponding gravitino mass is
\be \label{gravitinomass}
  m_{3/2} = e^{K/(2M_p^2)} \frac{|W|}{M_p^2} \sim \frac{\cF}{M_p} \sim  \frac{\mu_\ssW^3}{M_p^2 \tau^{3/2}}\,,
\ee
which for the benchmark numerical estimates of \pref{baseline1} gives 
\be \label{gravitinomassF}
 \cF \sim \frac{100 M_p^2}{\tau} \gsim  \Bigl( 10^{4} \; \hbox{GeV} \Bigr)^2 
 \quad \hbox{and} \quad
 m_{3/2} \sim \frac{100 M_p}{\tau}  \gsim 10^{-14} \; \hbox{GeV}  = 10^{-5} \; \hbox{eV} \,. 
\ee

\subsubsection{Relaxon mass}
\label{sssec:Relaxmass}

The mass of the relaxing field $\phi$ must be small enough that it is present in the low-energy EFT where the relaxation takes place. This requires it to be lighter than the lightest known particle whose vacuum energy is known to be a problem: the electron.\footnote{Notice that we are able to take the $\phi$ mass to be this large because technical naturalness does {\it not} care about quadratic divergences, and instead is concerned with the dependence of low-energy observables on physical quantities like renormalized masses $m$ (see {\it e.g.}~\cite{Burgess:2013ara} for more details).}  We therefore ask
\be  
    m_\phi  \sim \frac{gv}{\sqrt\tau} \quad\hbox{to be smaller than} \quad
     m_e \sim y_e m_\TEV \sim \frac{y_e M}{\sqrt\tau} \simeq 0.5  \; \hbox{MeV}  \,.
\ee
This assumes no other particles are present with masses below the $\phi$ mass but much larger than the cosmological constant scale. The bound $m_\phi \lsim m_e$ implies $gv \lsim y_e M$ and so for the benchmark numbers of \pref{baseline1}
\be \label{gvvalues}
  gv \lsim m_e \, \tau^{1/2}  \sim 5 \times 10^{12} \; \hbox{GeV} \ll M \,.
\ee

To further determine the size of $g$ requires knowing the relative sizes of $v$ and $M$. Since  \S\ref{ssec:LoopsAndTN} shows that loops involving a Standard Model particle with mass $m \sim M/\sqrt\tau$ contribute $\delta w_\ssX \sim M^2$, naturalness requires the constant term in \pref{RelaxWx} not to be small -- {\it i.e.}~$g v^2 \sim M^2$. Demanding $m_\phi \sim m_e$ therefore requires $g \sim y_e^2$.

\begin{figure}[t]
\begin{center}
\includegraphics[width=150mm,height=80mm]{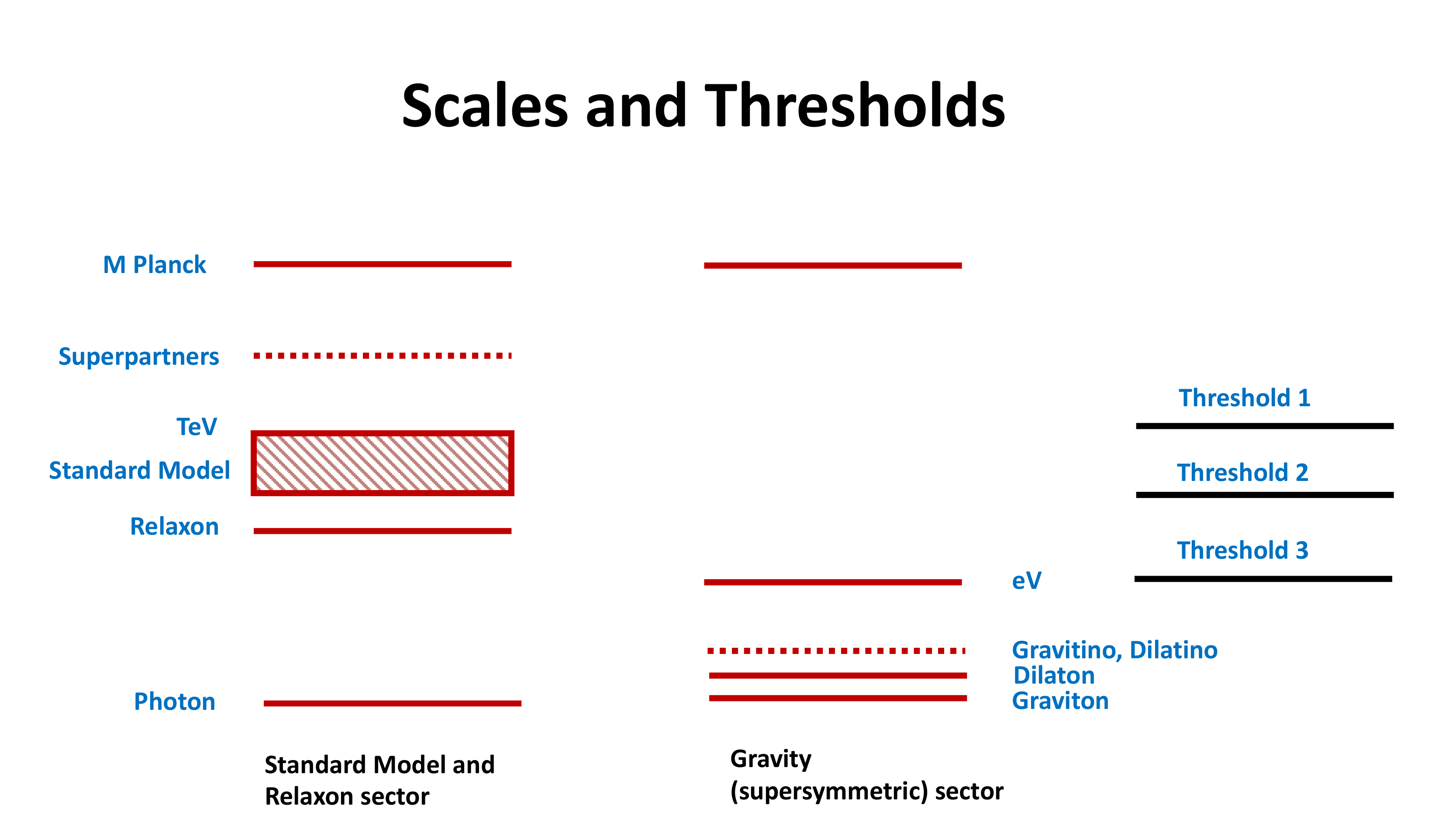} 
\caption{\footnotesize{Scales and thresholds differentiating the nonlinearly realized supersymmetric sector (including the Standard Model and the relaxon field) with the gravity sector (including the graviton and axio-dilaton multiplets) in which supersymmetry is linearly realized. The line marked `eV' represents the naive scale for the axion decay constant, where the gravity sector requires UV completion, possibly to a higher-dimensional supergravity. This scale is drawn above the `gravitino' scale, as happens when \pref{m32warped} is true and $b > a$, although these scales coincide in most of the scenarios we examine. Three different EFT thresholds are described in the main text: Threshold 1 denotes the point below which SM degrees of freedom must nonlinearly realize supersymmetry (because their superpartners are integrated out) while Threshold 2 describes the EFT for which the SM degrees of freedom are also integrated out but in which the relaxon must still be present. Threshold 3 represents the scale below which the gravity sector must be 4-dimensional and so for which our 4D EFT relevant to astrophysics and cosmology applies.} \label{Fig:Scales} }
\end{center}
\end{figure}

Figure \ref{Fig:Scales} illustrates the different scales separating the non-linearly realized supersymmetry breaking sector to the linearly realized supersymmetric gravity sector.

\subsection{Couplings to ordinary matter}
\label{ssec:MatterCouplings}

The remainder of this section fills in what this framework implies for the properties of the new fields and explores how they couple to ordinary SM fields. Along the way we justify several earlier assertions, such as that SM particles have masses proportional to $\tau^{-1/2}$.

We first focus on SM couplings, returning to Dark Sector properties in \S\ref{ssec:SUSYsecProps}. SM particles are again included using the nilpotent framework of \cite{Komargodski:2009rz, Bergshoeff:2015tra, DallAgata:2015zxp, Schillo:2015ssx}. This involves adding new constrained superfields for each individual SM particle, along the lines of \cite{LowESugra}, including:
\begin{itemize}
\item A chiral multiplet $Y$ satisfying the constraint $XY = 0$ for each Standard Model fermion. This constraint allows the scalar part $\cY$ to be expressed in terms of other fields. Although $Y^2$ is nonzero for fields satisfying this constraint, it happens that the constraint implies $Y^3 = 0$ pointwise. 
\item Standard Model gauge bosons are contained within left-chiral spinor multiplets $\cW$ (just as if they were supersymmetric) but their fermionic partners get projected out as independent fields by the constraint $X \cW = 0$. 
\item The Higgs is described by a superfield $H$ subject to the constraint $\ol\cD (X \ol{H}) = 0$, much as was done for the `relaxon' $\phi$ above. (If it weren't for the condition that the relaxaton field $\phi$ should be light enough to appear in the EFT below the electron mass, its role could be played by the Standard Model Higgs itself.)
\end{itemize}

With these ingredients one simply turns the same crank as in the previous section to compute the component lagrangian. (A simpler alternate derivation of matter couplings to the dilaton is also given in \S\ref{ssec:DilatonMatterPhenomenology}.) We illustrate the main points of this construction by summarizing the example of a single charged Dirac fermion, whose presence is encoded by adding two multiplets $Y_\pm$ (representing, say, the left-handed electron and positron). Both multiplets are subject to the constraint $X Y_\pm = 0$. The couplings of this field to the fields $T$, $X$ and $\Phi$ is obtained as before by including $Y_\pm$ in the K\"ahler function and superpotential:
\bea \label{KvsPhiXY}
  &&K(T,\ol T, X, \ol X, \Phi , \ol \Phi,Y_\pm, \ol Y_\pm)  \simeq  -3 M_p^2 \ln \left( \tau - k + \frac{h}{\tau}+\cdots \right) \nn\\
  &&\quad\hbox{and} \quad W(T,X,\Phi, Y_\pm)  \simeq  w_0(\Phi, Y_+ Y_-) +  X w_{\ssX}(\Phi,Y_+ Y_-) \,,
\eea
where charge conservation requires $W$ to depend only on the combination $Y_+Y_-$, and
\be
  k = \frac{1}{M_p^2} \Bigl\{ \mfK(\Phi,\ol{\Phi},Y_\pm , \ol Y_\pm,\ln\tau) + \Bigl[ X \mfK_{\ssX}(\Phi,\ol{\Phi}, \ol Y_\pm,\ln \tau) + \hbox{h.c.} \Bigr] +  \ol{X} X \; \mfK_{\ssX\ol\ssX}(\Phi,\ol{\Phi},\ln\tau)  \Bigr\} \,.
\ee

We assume the potential is chosen so that the auxiliary fields $F^{\ssY\pm}$ vanish in the vacuum, as required by unbroken electromagnetic gauge invariance. Since $XY_\pm = 0$ constrains the sclar part $\cY_\pm$ to be a function of the fermions, the field $Y_\pm$ is `pure fluctuation' and the lagrangian can be expanded in powers of $Y_\pm$. The leading parts of the K\"ahler function and superpotential are
\be  \label{wformsSM}
   w_0 = \mfw_0(\Phi) + \mfm(\Phi) \, Y_+ Y_- + \cdots \quad \hbox{and} \quad
   w_{\ssX} = \mfw_{\ssX}(\Phi) + \mfn(\Phi) \, Y_+Y_- + \cdots \,,
\ee
and $\Phi$ and $Y_\pm$ can be rescaled so that 
\bea \label{mfKvsPhiY}
  \mfK &=& \Phi \ol \Phi + Y_+ \ol Y_+ + Y_- \ol Y_- + \hbox{(cubic and higher)} \nn\\
   \mfK_\ssX &=& \kappa (\Phi, \ol \Phi) + \mfe (Y_+ \ol Y_+ + Y_- \ol Y_-) + \cdots \,,
\eea
and so on.  

Denoting, as before, $\cP \simeq \tau  - ({\mfK}/{M_p^2})$ one finds in addition to \pref{DTWnDXW} the new K\"ahler derivatives (evaluated at $\cX = 0$)
\be  \label{DTWnDXW1}
   D_\pm W = W_{\ssY_\pm} + \frac{K_{\ssY_\pm} W}{M_p^2} = \mfm \, Y_\mp + \frac{3\ol Y_\pm}{\cP M_p^2} (\mfw_0 + \mfm \, Y_+ Y_-)\,,
\ee
which automatically vanishes when fermions are set to zero. This precludes this fermion from mixing with the gravitino and also keeps these new supercovariant derivatives from contributing to the potential $V_\ssF$, which therefore again has the form given in \pref{VFtauexp}. The discussion in the previous section of the minima of this potential goes through as before, leading (at large $\tau$) again to \pref{VFtauexpyy} -- or \pref{VFBasicFormzzmin22}. 

Denoting the fields collectively as $z^\ssA := \{T, z^a\}$ with $z^a := \{ X, Y_+,Y_-\}$, the leading parts of the K\"ahler metric and its inverse evaluated at $X=Y_\pm = 0$ generalize \pref{KInv0} to
\be  \label{KInv}
   K_{B \ol A} = \frac{3M_p^2}{\cP^2}  \left( \begin{array}{ccc}
  1  &&  -k_{\bar a} \\  -k_b &&  \cP \, k_{b \bar a} + k_b k_{\bar a}
\end{array} \right) 
\quad \hbox{and so} \quad 
   K^{\bar A B} = \frac{\cP}{3M_p^2}  \left( \begin{array}{ccc}
  \cP + \mfk^2  &&  k^{b} \\   k^{\bar a} &&  k^{\bar a b}  
  \end{array}
\right) \,,
\ee
where $\mfk^2 := k^{\bar a b}k_{\bar a} k_b$. Notice that these expressions show that the matter fields do not alter the kinetic terms for the scalars $\phi$ and $\cT$ -- leaving these still given by \pref{TkinL0}.

When $\mfK_{\ssX\ol\ssX} \simeq \mfK_{\ssY_\pm \ol\ssY_\pm} \simeq 1$ these formulae imply
\be \label{usefulKDs}
   K_{\ssX \ol \ssX} \simeq
   K_{\ssY_+ \ol \ssY_+} = K_{\ssY_- \ol \ssY_-} \simeq \frac{3}{\tau}  \quad 
   K_{\ssY_+ \ssY_- \ol \ssX} \simeq \frac{3 \mfe}{\tau} \,,
\ee
when evaluated at $\cX = \cY_\pm = 0$ and to leading order in $1/\tau$. The leading part of the kinetic terms for the fermion $\psi_\pm \in Y_\pm$ are given by
\be \label{psikinL}
   \cL_{{\rm kin}\,\psi} \simeq - \frac{3}{\cP} \Bigl( \ol\psi \gamma_\ssL \Dsl \psi +  \ol \psi \gamma_\ssR \Dsl \psi \Bigr) \simeq - \frac{3}{\tau} \; \ol\psi \Dsl \psi \,,
\ee
though these coefficients are also $\phi$-dependent at subdominant orders in $1/\tau$.

The fermion mass term is similarly given by the supergravity expression
\be
  \cL_{{\rm mass}\, \psi}  = - \frac12 \, m_{\ssA\ssB} \ol \psi^\ssA \gamma_\ssL \psi^\ssB + \hbox{h.c.}
\ee
where
\be
   m_{\ssA\ssB} = e^{K/(2M_p^2)} D_\ssA D_\ssB W = e^{K/(2M_p^2)}\left[  \left(\partial_\ssA + \frac{K_\ssA}{M_p^2} \right) D_\ssB W - \Gamma^\ssC_{\ssA\ssB} D_\ssC W \right]
\ee
and the target-space connection is $\Gamma^\ssC_{\ssA\ssB} = g^{\ssC\ol\ssD} \partial_\ssA g_{\ssB\ol\ssD} =  K^{\ssC \ol \ssD} K_{\ssA\ssB\ol\ssD}$. Our interest is in $m_{+-} := m_{\ssY_+ \ssY_-}$ for which we can use $W_{\ssY_\pm} = K_{\ssY_\pm} = 0$ leaving the only nonzero terms
\be \label{Ymass}
   m_{+-} = e^{K/(2M_p^2)}   W_{\ssY_+\ssY_-} -  K_{\ssY_+\ssY_- \ol\ssX} \ol F^{\ol\ssX} 
   \simeq \frac{\mfm}{\cP^{3/2}}  -    \frac{3\mfe}{\cP}   \left( \frac{w_\ssX}{\tau^{3/2}}\right)  \,,
\ee
which uses \pref{wformsSM} and evaluates $F^{\ol\ssX}$ using \pref{Fauxcov}. Expression \pref{Ymass} shows that the $F^\ssX$-dependent term is subdominant in powers of $1/'\tau$ and so can be dropped, leading to 
\be
  \cL_{{\rm mass}\, \psi}  \simeq -  \frac{\mfm}{\cP^{3/2}} \Bigl( \ol \psi_+ \gamma_\ssL \psi_-  + \hbox{h.c.} \Bigr) \simeq -   \frac{\mfm}{\tau^{3/2}}\; \ol   \psi   \psi \,,
\ee
where the last equality assumes $\mfm$ is real. Comparing with the kinetic term reveals the $\tau$-dependence of the physical mass to be
\be \label{SMmassvstau}
    m_\psi \simeq \frac{ \mfm}{3\sqrt\cP} \sim \frac{ \mfm}{\sqrt\tau} \,,
\ee
as claimed earlier. 

When the dust settles what is important is that the leading dependence of $m_\psi$ on powers of $\tau$ comes completely from the factor $e^{K/(2M_p^2)}$ in \pref{Ymass}, and as a result they can be traced to the Weyl rescalings $\tilde g_{\mu\nu} = e^{K/(3M_p^2)} g_{\mu\nu}$ required to put the lagrangian into Einstein frame. This is important for several reasons. First it ensures this $\tau$-dependence is universal, and so is the same for all SM particles. This is because the $\tau$ dependence ultimately enters only through dimensionful parameters, and so can only do so through the quadratic term in the Higgs potential once formulated in an $SU_c(3) \times SU_\ssL(2) \times U_\ssY(1)$ invariant way. This ensures that dimensionless physical quantities like Yukawa couplings and Higgs self-interactions remain $\tau$-independent, once expressed in terms of canonically normalized fields (see \cite{LowESugra} for more details). As explained in more detail in \pref{sssec:DilatonBD}, it also ensures that $\tau$ couples to ordinary matter as does a Brans-Dicke scalar.

\subsection{Dark sector properties}
\label{ssec:SUSYsecProps}

We now compute the main physical properties of the new dark particles in the dilaton and gravity supermultiplets: the complex scalar $\cT$, its dilatino partner, $\xi$ (and how this mixes with the gravitino). This section shows why these particles are light and collects predictions for their masses and couplings for later use.

\subsubsection{Axio-dilaton}
\label{sssec:DilatonAxion}

The leading contributions to the kinetic term for the field $\cT$ is given by \pref{TkinL0}
\be \label{Ltaukin}
 - \frac{ \cL_{\rm kin}}{\sqrt{-g}} 
 \simeq  \frac{3 M_p^2}{4\tau^2} \Bigl[\partial^\mu \tau \partial_\mu \tau + \partial^\mu  \mfa \, \partial_\mu \mfa  \Bigr] 
 = \frac12 (\partial \chi)^2 +  \frac{3 M_p^2}{4} \, e^{-2\zeta \chi/M_p} (\partial \mfa)^2  \,,
\ee
where the axion field is defined by $\mfa =2 \,\hbox{Im} \, \cT$ and the canonically normalized dilaton is
\be \label{chizetadef}
  \tau = \tau_0 \, e^{\zeta \chi /M_p} \qquad \hbox{with} \qquad
  \zeta = \sqrt{\frac23} \,.
\ee

\subsubsection*{Dilaton mass}

The potential for the dilaton is given by \pref{VFBasicFormzzmin22}, $V_\ssF = U(\ln\tau) /\tau^4$, and so looks like a modulated exponential potential in terms of $\chi$,
\be
  V_\ssF \simeq \frac{U(\ln\tau)}{\tau^4} =  U(\chi) \, e^{-\lambda \chi/M_p} \quad \hbox{with} \quad \lambda := 4 \sqrt{\frac23} \,.
\ee
Whether this potential has a minimum or not depends on the function $U$ in and so also on the details of how $\mfK$ depends on $\ln \tau$. \S\ref{ssec:DilationStabilization} describes a simple choice for $\mfK(\ln\tau)$ that ensures there is a minimum for $\tau = \tau_{0}$ (with $V(\tau_0) > 0$) and because $U$ comes as a simple rational function of $\ln \tau$ the minimum can be at exponentially large values like $\tau_0 \sim 10^{30}$ assuming only that ratios amongst the parameters are order $\ln \tau_0 \simeq 70$.

Because the potential is dominantly exponential, its derivatives near the minimum are generically of the same order as the potential itself,\footnote{This estimate is not changed by any dependence on $\ln\tau$ of the form entertained in later sections.}
\be \label{V0derivsexp}
  V'_0 := V'(\chi_0) \sim - \frac{\lambda V_0}{M_p} \quad \hbox{and} \quad
  V''_0 := V''(\chi_0) \sim \frac{\lambda^2 V_0}{M_p^2} \,. 
\ee
If $V_0$ is to describe the present-day Dark Eneryg density (more about which in \S\ref{sssec:DilatonCosmology}) then its potential energy must dominate the present-day Hubble scale, $H_0^2 \simeq V_0/(3M_p^2)$, and so $V_0 \simeq (10^{-2} \, \hbox{eV})^4$. But this means the dilaton mass is also of order $m_\chi^2 \sim V''(\chi_0) \sim H_0^2$ where $H_0 \sim 10^{-32}$ eV. 

It is usually difficult to get scalars to be naturally light enough to be cosmologically active, but this is actually {\it generic} for gravitationally coupled scalars whose mass comes from a potential that naturally describes Dark Energy. This makes them phenomenologically important because scalars  this light are generically visible in tests of gravity. Later sections investigate some of these constraints, and how they might be avoided using the mechanism described in \cite{ADScreening}.

\subsubsection*{Axion interactions}

Consider next the axionic part of $\cT$. Eq.~\pref{Ltaukin} shows that the target-space dilaton-axion interactions can be interpreted as a dilaton-dependent axion decay constant of order\footnote{To pin down the numerical factors requires knowing if $\hat a$ is periodic, since then the requirement that the period be $2\pi$ removes the freedom to rescale the fields.}
\be \label{faprediction}
     f_a \sim \frac{M_p}{\tau}  
\ee
which is of order $10^{-3}$ eV for the numerical benchmarks of \pref{baseline1}. This is an unusually small decay constant, and for traditional axion models one might expect this to cause a breakdown of the low-energy derivative expansion governs the EFT at energies of order $f_a$. 

In simple models this breakdown of EFT methods can indicate the approach of a critical point, with another field, $\varphi$, becoming light enough to combine with $\mfa$ into a complex field $\varphi \,e^{i\mfa}$ that linearly realizes the axion shift symmetry in the unbroken phase. This need not be how it works in supergravity, however, since the axion lagrangian $f_a^2 (\partial \mfa)^2$ there often instead arises in dual form, $f_a^{-2} H^{\mu\nu\lambda} H_{\mu\nu\lambda}$ where $H = \exd b$ is the field strength for a 2-form gauge potential $b_{\mu\nu}$ with $f_a \partial_\mu \mfa \propto \epsilon_{\mu\nu\lambda \rho} \partial^\nu b^{\lambda \rho}$. The axion shift symmetry need not be linearly realized at scales above $f_a$; in the dual formulation it is instead the gauge symmetry $b \to b + \exd \Lambda$ that is important.

But even for supergravity the scale $f_a$ usually indicates a breakdown of effective methods. 
In string theory (for example, taking the Type IIB axio-dilaton as representative) the decay constant is similarly suppressed, with $f_a / M_p \propto 1/\tau \propto g_s$ where $g_s$ is the string coupling. In this example $f_a$ indeed indicates the breakdown of EFT methods that occurs at the string scale (which is systematically below the Planck scale when $g_s \ll 1$). 

Having such a small decay constant for $\mfa$ could well indicate the advent of new physics in the gravity sector at eV scales. Supersymmetric large-extra-dimension models \cite{SLED} provide concrete examples of what this physics might be, with $\tau$ playing the role of the radion for the large dimensions and $\mfa$ being its axionic partner. In this case the new physics that kicks in at eV scales is extra-dimensional: unitarity becomes restored by the tower of Kaluza-Klein modes for the dual 2-form field living in the extra dimensions. This would be consistent with the gravitino mass \pref{gravitinomassF} also being at this scale, as appropriate if it were a specific KK mode of an extra-dimensional field. 

Notice that the existence of such physics does not matter for applications (such as to cosmology or in the solar system) whose energies are much smaller than $f_a$. It also need not affect the discussion of higher-energy couplings of the axio-dilaton to Standard Model particles -- to the extent that extra-dimensional models are our guide -- since these would be restricted to a brane within the extra dimensions and so remain largely four-dimensional. 

Even if present, having extra dimensions at this scale need not undermine use of the above 4D nilpotent EFT, at least for the two most frequent applications. First, it does not affect applications (such as to cosmology or in the solar system as considered in later sections) whose energies are much smaller than $f_a$. Second, it also need not affect the discussion of higher-energy interactions amongst Standard Model particles -- at least to the extent that extra-dimensional models are our guide -- since these would be restricted to a brane within the extra dimensions and so remain four-dimensional. 

\subsubsection*{Axion mass}

The size of the axion mass is more model-dependent, even though its shift symmetry cannot be broken without undermining the no-scale structure of the scalar potential for $\tau$. Unbroken shift symmetry makes the axion massless unless the symmetry is gauged, in which case the axion is eaten by a gauge boson to acquire nonzero mass through the Higgs mechanism. 

Gauging the axion shift symmetry forces the replacement $\partial_\mu \mfa \to \partial_\mu \mfa -  A_\mu$ in the action, where $A_\mu$ is the relevant gauge field. Working in the gauge $\mfa = 0$ then turns the axion kinetic term of \pref{Ltaukin} into a gauge-boson mass term 
\be
  - \frac{\cL_{\rm mass}}{\sqrt{-g}} = \frac{3  M_p^2}{4\tau^2}  \, A_\mu A^\mu \,.
\ee
The size of the resulting mass depends on whether or not the gauge-boson kinetic term is $\tau$-dependent, and this depends on whether the holomorphic gauge kinetic function $\mff$ is a constant or proportional\footnote{Having $\mff \propto T$ implies the existence of an anomalous coupling $\mfa\, \epsilon^{\mu\nu\lambda\rho} F_{\mu\nu} F_{\lambda \rho}$, and so can only be present if there are also other charged fields to provide cancelling gauge anomalies.} to $T$: $\mff = \mff_0 = 1/e^2$ or $\mff \propto T$. The corresponding physical gauge-boson (and so also axion) mass then is of order $M_\ssA \sim M_p/(\mff \, \tau)$ and so is given by one of
\be
   M_\ssA \sim \begin{cases} {e M_p}/{\tau}  \sim 10^{-3}\, \hbox{eV} & \hbox{if $\mff_0 \sim e^{-2} \sim \cO(1)$} \\
   {M_p}/{\tau^{3/2}} \sim 10^{-18} \, \hbox{eV} & \hbox{if $\mff \propto T$} \end{cases} 
\ee

Gauging the axionic symmetry also introduces new complications to the scalar potential however, because in a supersymmetric theory it implies that $K$ depends on $\tau$ only through the combination $K = K(T + \ol T - \cA)$ where $\cA$ is the scalar superfield that contains the gauge potential $A_\mu$. This implies a contribution to the scalar potential involving the gauge-field auxiliary field D, of the form of a field-dependent Fayet-Iliopoulos term: $\delta V \sim  K_\ssT \hbox{D}$ and so after D is eliminated naively contributes to the D-term scalar potential an amount $V_\ssD \sim K_\ssT K_{\ol\ssT}/($Re $\mff)$. This would be $\cO(1/\tau^2)$ if $\mff \sim \mff_0$ and $\cO(1/\tau^3)$ if $\mff \propto T$, which in either case would be large enough to overwhelm the $\cO(1/\tau^4)$ term found above. This need not be a problem if other supermultiplets containing fields charged under the gauge symmetry exist (as anomaly cancellation typically requires in any case when $\mff \propto T$) since then these other fields adjust\footnote{Indeed, there is a sense that the standard adjustment of charged fields to find the potential's D-flat directions -- such as occurs automatically in the MSSM for example -- is a special case of the relaxon mechanism described above for $\phi$.} to ensure that D$=0$.

We see from this discussion that the axion could simply be massless, or it could be eaten by the Higgs mechanism in a meal that necessarily involves the presence of other light fields (the gauge multiplet itself at the very least) in the supersymmetric sector. Depending on the $\tau$-dependence of the gauge couplings such mixings would give the axion a mass that could be as large as $M_p/\tau \sim 10$ eV or as low as $M_p/\tau^{3/2}  \sim 10^{-18}$ eV. 

\subsubsection{Dark fermions}
\label{sssec:DarkFermions}

The supersymmetric sector necessarily also involves very light fermions: both the gravitino -- {\it c.f.}~eq.~\pref{gravitinomass} -- and the dilaton's light partner $\xi$ (the dilatino), in addition to the usual Standard Model neutrinos. For later use we here summarize the leading features of this fermionic sector.  

For simplicity we assume the gravity-dilaton sector not to break lepton number, with lepton number broken only by the neutrino masses themselves. In practice this allows us to ignore mixing between the gravitino/dilatino sector and the neutrinos, though there is clearly much interest in exploring the phenomenology allowed by more general assumptions. 

Under these circumstances the couplings in the dilatino/gravitino sector are the usual ones predicted by supergravity, and so have the form
\bea 
   \frac{\cL_{\rm fermion}}{\sqrt{-g}} &=&  - \frac{i}2 \, \epsilon^{\mu\nu\lambda\rho} \ol \psi_\mu \gamma_5 \gamma_\nu D_\lambda \psi_\rho - \frac{1}{2} \,m_{3/2} \,\ol \psi_\mu \gamma^{\mu\nu} \psi_\nu \\
   && \qquad - \left[ \frac12 K_{\ssT\ol\ssT} \ol \xi \gamma_\ssR\Dsl \xi - \frac12 \mfm_\xi \, \ol \xi \gamma_\ssL \xi + \frac{\mfm_{\xi g}}{M_p} \, \ol \psi_\mu \gamma_\ssR \gamma^\mu \xi + \hbox{h.c.} \right] + \cdots \nn
\eea
where $m_{3/2}$ is given in \pref{gravitinomass} and the ellipses represent other terms (like 4-fermi interactions) whose form is not required in what follows. The supergravity expression for $\mfm_{\xi g}$ (evaluating at $X = 0$) is given by \cite{FreedmanVanProeyen}
\be \label{mfmxig}
   \mfm_{\xi g} = \frac{1}{\sqrt2\; M_p^2} \, e^{K/(2M_p^2)} D_\ssT W \simeq \frac{1}{\sqrt2 \; \cP^{3/2} M_p^2} \left(- \frac{3 w_0}{\cP} \right) 
   \sim - \frac{w_0}{ \tau^{5/2} M_p^2}    \,,
\ee
while that for $\mfm_\xi$ (using $W_\ssT = 0$) is 
\bea \label{mfmxi}
   \mfm_\xi =
   m_{\ssT\ssT} &=& \frac{1}{M_p^2} e^{K/(2M_p^2)} D_\ssT D_\ssT W \simeq \frac{1}{M_p^2} e^{K/(2M_p^2)}\left[  \frac{K_\ssT}{M_p^2} \, D_\ssT W -\Gamma^\ssA_{\ssT\ssT} D_\ssA W   \right] \\ 
     &\simeq& \frac{1}{M_p^2} e^{K/(2M_p^2)} \left[ \left( \frac{K_\ssT}{M_p^2} -K^{\ssT\ol\ssA} K_{\ssT\ssT\ol\ssA} \right) D_\ssT W  -K^{\ssX\ol\ssA} K_{\ssT\ssT\ol\ssA} \,  D_\ssX W \right] 
   \simeq \frac{3w_0}{\tau^{7/2} M_p^2} \,, \nn
\eea
which uses $K_\ssT \simeq -3M_p^2/\cP$ while $K^{\ssT \ol\ssA} K_{\ssT\ssT\ol\ssA} \simeq -2/\cP$ and $K^{\ssX \ol\ssA} K_{\ssT\ssT \ol\ssA} \simeq \cO(1/\cP^3)$. 

These expressions show that $\xi$ acquires a mass partially by mixing with the gravitino (because the supersymmetry breaking contribution $D_\ssT W \neq 0$ causes it to contribute to the Goldstino) and partially through a direct mass term. Keeping in mind that the $\xi$ kinetic term is proportional to $K_{\ssT\ol\ssT} \sim M_p^2/\cP^2$ shows that both these contributions are of the same order as the gravitino mass.

\section{Naturalness issues}
\label{ssec:LoopsAndTN}

Knowing the forms for the couplings between ordinary particles and the dilaton multiplet allows a more explicit treatment of technical naturalness. Our exploration of this comes in three parts, each of which is considered in turn. First \S\ref{ssec:Loops} estimates whether loops of heavy particles preserve the choices that have been made to achieve a small scalar potential. Then, \S\ref{ssec:DilationStabilization} provides an explicit stabilization mechanism that produces exponentially large value for $\tau$ without introducing unusually large parameters into the scalar potential. An explicit understanding of stabilization is necessary because it is the vacuum value of $\tau$ that controls both electroweak and cosmological-constant hierarchies. Finally \S\ref{ssec:UVcomplete} explores the extent to which our choices are compatible with the properties of known stringy UV completions.

\subsection{Loops}
\label{ssec:Loops}

We start by asking whether loops of Standard Model particles preserve the main choices made to this point: ($i$) use of the nilpotent supergravity form of the lagrangian; ($ii$) stability of the small vacuum energy within this supergravity formulation; ($iii$) stability of the small scalar masses for the $\phi$ and $\cT$ fields; and ($iv$) possible origins of the assumed $\ln \tau$ dependence in $k$ (that play a role once we ask how $\tau$ becomes stabilized at the large values).

\subsubsection{Supergravity form}

The entire discussion presupposes that the lagrangian has a supergravity form specified at the two-derivative level by a K\"ahler function $K$, a superpotential $W$ and a gauge kinetic function $\mff_{\alpha\beta}$. In particular this is what ensures terms like $|w_\ssX|^2/\tau^2$ arise in the scalar potential as a perfect square; an important precondition for this term becoming small as the relaxon field seeks its minimum. This structure is ultimately dictated by supersymmetry and its requirements for how auxiliary fields appear in the lagrangian, since it is the elimination of these that give the scalar potential its special form. The key role of auxiliary fields underlines the potential importance of non-propagating auxiliary fields for EFTs in general and for naturalness arguments in particular \cite{LuisForm, MyForm, Burgess:2020qsc}. 

Are these features robust to quantum loops? Ref.~\cite{LowESugra} argues that they are, and does so by investigating how the effective couplings in the functions $K$ and $W$ evolve as heavy nonsupersymmetric particles are integrated out. The remainder of this section argues why this robustness also can be seen on more general symmetry grounds. 

The physical assumption that justifies coupling to supergravity is that the mass splitting $\Delta m_g$ within the graviton multiplet (and the dilaton multiplet $T$) is much smaller than the corresonding splittings $\Delta m_\SM$ within multiplets containing Standard Model particles. This hierarchy seems likely to be natural given that supermultiplet splittings are of order $\Delta m^2 \sim g F$ where $F$ is the supersymmetry-breaking {\it vev}~and $g$ is a measure of the multiplet's coupling to it. Maintaining $\Delta m_g \ll \Delta m_\SM$ should only require Standard Model fields to couple more strongly to the supersymmetry breaking sector than does the weakest force of all: gravity.

It is the because $\Delta m_g \ll \Delta m_\SM$ that we can make our Wilsonian UV/IR split somewhere in between: $\Delta m_g \ll \Lambda \ll \Delta m_\SM$. Since we are only interested in Standard Model scales at energies below $\Lambda$, we are free to integrate out the SM superpartners to obtain a nonsupersymmetric matter sector coupled to supergravity. 

Consider first how this theory looks in the global-supersymmetry limit $M_p \to \infty$. In this limit the low-energy sector contains the Standard model coupled to the Goldstone fermion $G$ \cite{Volkov:1973ix} -- and like for any Goldstone particle these couplings are dictated by the supersymmetry algebra itself. An {\it arbitrary} non-supersymmetric theory can be made globally supersymmetric (for free) by appropriately coupling a Goldstone fermion to it. As is always true when global symmetries break spontaneously, the only symmetry information that survives well below the symmetry breaking scale is encoded in the couplings of the appropriate Goldstone fields \cite{Weinberg:1968de, Coleman:1969sm, Callan:1969sn} (for a textbook description see \cite{EFTBook}). 

The basic claim of ref.~\cite{Komargodski:2009rz} is that there is no loss of generality in describing these low-energy goldstino couplings in terms of the supersymmetric interactions of constrained superfields coupled to a nilpotent goldstino multiplet $X$, and this ultimately is what guarantees that the Wilsonian EFT (for global supersymmetry) at any scale can be captured for some choice of the functions $K$, $W$ and $\mff_{\alpha\beta}$. There is no loss of generality because an arbitrary nonsupersymmetric theory can be made supersymmetric `for free' in this way. Because this framework is so general, it in particular must remain valid for the Wilsonian action as successive nonsupersymmetric Standard Model particles are integrated out. 
 
Once the Standard Model sector is coupled to the goldstino in this way its couplings to the graviton multiplet are dictated by symmetry through the usual rules for gauging supersymmetry \cite{WB, FreedmanVanProeyen}. 

\subsubsection{Vacuum energy} 

Consider next the size of the vacuum energy within this supergravity framework, since this is the quantity that is normally never naturally small. The key assumptions in \S\ref{ssec:QNSRelaxation} are $(a)$ that all functions like $\mfK$ and $w_0$ depend on the generic UV scale $M$ simply as they should on dimensional grounds; and $(b)$ that the lagrangian can be organized into a series of powers of $1/\tau$, with the scalar potential starting off at order $1/\tau^2$. Neither of these properties are changed by contributions to the lagrangian due to loops of Standard Model particles. 

To this end consider, for instance, a loop contribution to $V$ obtained by integrating out a Standard Model particle of mass $m$, given that formulae like \pref{SMmassvstau} show that this mass is given by $m \simeq \mfm/\sqrt \tau$ where $\mfm \sim yM$ for some Yukawa coupling $y$. The dangerous part of this loop is generically given by
\be
   \delta V \sim \frac{m^4}{16\pi^2} \sim \frac{\mfm^4}{16\pi^2 \tau^2}  \sim \frac{y^4 M^4}{16\pi^2 \tau^2} \,.
\ee
This has precisely the $1/\tau^2$ dependence required to be interpreted as a contribution to $V$ coming from a correction to $w_{\ssX}$. Furthermore, the size of this correction is of order $\delta w_\ssX \sim \mfm^2/4\pi \sim y^2 M^2/4\pi$, consistent with assuming $w_\ssX \sim M^2$ (and with the results of \cite{LowESugra}).

Contributions such as these to $w_\ssX$ are irrelevant to the value of $V_{\rm min}$ to the extent that they do not remove the property that a zero of $w_{\ssX}$ exists for some choice of $\phi$. They only change the precise value of the field, $\phi_0$, for which this minimum exists. It is for this reason that $V_{\rm min}$ can be stable against integrating out Standard Model particles. Central to this stability is the scale-invariant form of the $\tau$-dependence of Standard Model masses. 

The one exception to the general assignment of $M$ as a UV scale is the even larger value chosen  for $\mu_\ssW$. Such large values for $\mu_\ssW$ are known to be consistent with string compactifications, \footnote{The precise bound is $W_0\leq \cV^{1/3}$ with $\cV$ the overall volume. Although generically for flux superpotentials $W_0/M_p^3$ tends to be of order 1-100, but higher and smaller values are also allowed. Notice that the bound is saturated when the gravitino mass $m_{3/2}=M_p W_0/\cV$ is of the same order as the overall Kaluza-Klein scale $M_{\KK}\sim M_p/\cV^{2/3}$, at which point the 4D EFT ceases to be valid.} and generically arise whenever the gravitino mass is close to the Kaluza Klein scale \cite{Cicoli:2013swa}. Once chosen, the value of $W_0$ remains unchanged as particles are integrated out. It is not affected by heavy supersymmetry-breaking effects because these always involve $X$ in the low-energy theory. Standard non-renormalization arguments \cite{NRTheorems} protect $W_0$ when integrating out heavy supersymmetric physics.

\subsubsection{Scalar masses}

Besides the small value of $V_{\rm min}$ there are two types of light scalar fields, whose masses must also be protected from loops if the model is to be technically natural. 

\medskip\noindent{\it Dilaton mass}

\medskip\noindent
The lightest scalar in the problem is the dilaton $\tau$ itself, whose mass is shown in \S\ref{sssec:DilatonAxion} to be of order the present-day Hubble scale. Can such a small scalar mass be stable against integrating out UV physics?

We argue here that it is, through the mechanism identified some time ago in \cite{AndyCostasnMe}. The main point can be seen from eq.~\pref{V0derivsexp} which shows that the derivatives of the scalar potential are proportional to the value of the potential itself when evaluated at the minimum. This is an automatic consequence for the exponential potential that is dictated at leading order by the nonlinearly realized accidental scale invariance that underlies the model's construction. 

Approximate scale invariance links the small dilaton mass to the small value of $V_{\rm min}$ and so the dilaton mass is guaranteed to be naturally small once the cosmological constant itself is. The generality of this argument makes it likely that any gravitationally coupled scalar appearing in $V$ should acquire a similar mass, pointing to a world where multiple scalars might be equally light.

\medskip\noindent{\it Relaxon mass}

\medskip\noindent
A second relatively light scalar is the relaxon field $\phi$, which must be light enough to remain in the low-energy EFT defined below the electron mass. If $\phi$ were not this light it would not be present to remove the dangerous $|w_\ssX|^2$ contribution from the scalar potential generated once the electron is integrated out. \S\ref{ssec:QNSRelaxation} arranges $\phi$ to be this light through two choices. First it is assumed that $W$ does not contain a term like $M X \Phi \in W$ that contributes a UV-sensitive linear term in $\phi$ to $w_\ssX$. It is only because of this that the $\phi$ mass is controlled by the dimensionless coupling $g$. This choice is natural since it can be enforced through a symmetry like $\phi \to - \phi$ (or more generally by a continuous rephasing symmetry, provided that the corresponding massless Goldstone boson causes no trouble once $\langle \phi \rangle \neq 0$). 

The second choice required to keep $m_\phi$ small asks for the hierarchy $gv \ll M$ amongst the model parameters. We argue that this is also technically natural, and it is key for this argument that the $\phi$ mass again comes from the leading supersymmetry-breaking piece of the superpotential: $w_\ssX$. If $w_\ssX$ has a zero for some $\phi = v$ then because it is a function only of $\phi^2$ its dependence near this zero very generically has the form $w_\ssX = g (\phi^2 - v^2) + \cdots$ where $g$ is dimensionless and we have seen that large $\phi$-independent Standard Model contributions $\delta w_\ssX \sim M^2$ imply $gv^2 \sim M^2$. This corresponds to the superpotential used in earlier sections having the form
\be \label{WphiXform}
   W = w_0 + X \Bigl[ g  (\Phi^2 - v^2) + \cdots \Bigr]  \,.
\ee
Having $gv^2 \sim M^2$ is completely consistent with $m_\phi^2 \sim (gv)^2 \sim gM^2 \ll M^2$ provided that $g \ll 1$, showing that the hierarchy $m_\phi \sim gv \ll M$ relies solely on the dimensionless coupling $g$ being small. But small dimensionless couplings can be completely natural given that loop corrections to a marginal term like $g X  \Phi^2 \in W$ depend only logarithmically on the mass of the particle in the loop.

How does one see in components that a quadratic term in $\phi$ is has a marginal (dimensionless) coupling rather than a relevant (positive power of mass) one? This occurs because the component part of $X \Phi^2 \in W$ is $F^\ssX \phi^2 \in V$ and so is dimension-four (rather than dimension-two) due to the presence of the auxiliary field $F^\ssX$. 

\subsubsection{Logarithmic $\tau$-dependence}

Although logarithmic mass dependence arising from quantum loops do not threaten naturalness arguments, they do play an important role in the next section's stabilization mechanism and do so because heavy-particle masses in this model are generically $\tau$-dependent. For instance the one-loop contribution to the scalar potential obtained when integrating out a particle of mass $m$ is strictly speaking not simply $\delta V \sim m^4$, but more accurately is given by
\be
   \delta V \sim \pm  \mfc \,\frac{ m^4}{16\pi^2} \ln \left( \frac{m^2}{\mu^2} \right) = \pm \frac{\mfc \, \mfm^4}{16\pi^2 \tau^2} \ln \left( \frac{\mfm^2}{\tau \mu^2} \right) \,,
\ee
rather than simply being proportional to $m^4$. Here $\mfc$ is a calculable number,  $\mu$ is an arbitrary renormalization scale,  the upper (lower) sign applies for bosons (fermions) and the last equality assumes $m^2 = \mfm^2/\tau$, as found above for Standard Model particles.

Similar logarithms appear quite generally for other effective operators once loop corrections are included, such as corrections that change expressions like \pref{psikinL} to
\bea \label{psikinLlog}
   \cL_{{\rm kin}\,\psi} &\simeq&   - \frac{3}{\tau} \left\{ 1 + \mfc_1 \, \frac{\alpha_g}{4\pi} \ln\left( \frac{m^2}{\mu^2} \right) + \mfc_2 \left[ \frac{\alpha_g}{4\pi} \ln\left( \frac{m^2}{\mu^2} \right) \right]^2 + \cdots \right\} \ol\psi \Dsl \psi \\
  &\simeq&  - \frac{3}{\tau} \left\{ 1 + \mfc_1 \, \frac{\alpha_g}{4\pi} \ln\left( \frac{\mfm^2}{\tau\mu^2} \right) + \mfc_2 \left[ \frac{\alpha_g}{4\pi} \ln\left( \frac{\mfm^2}{\tau\mu^2 } \right) \right]^2 + \cdots \right\} \ol\psi \Dsl \psi \,,\nn
\eea
and so on for other effective operators. Here $\alpha_g$ is a dimensionless coupling constant and  the $\mfc_i$ are again calculable numbers. Because the logarithmic dependence on $\tau$ is related to logarithms of mass scales, standard renormalization-group arguments allow them to be resummed to all orders in $\alpha_g \ln \tau$, such as is done when writing the running of dimensionless couplings in the form
\be \label{gaugerunning}
  \frac{4\pi}{\alpha(\mu)} = \frac{4\pi}{\alpha(\mu_0)} + \mfb \, \ln \left( \frac{\mu^2}{\mu_0^2} \right) \,.
\ee
$\tau$-dependence enters into expressions like these once these couplings are evaluated at physical masses $\mu^2_0 = m^2 = \mfm^2/\tau$, such as when matching across particle thresholds. 

The logarithmic $\tau$-dependence hidden within the running of gauge couplings shown in \pref{gaugerunning} actually plays a crucial role in later phenomenology because it also predicts the $\tau$-dependence expected for dynamical scales like the QCD scale
\be
    \Lambda_\QCD = \mu_0 \, \exp \left[ - \frac{1}{2\mfb \, \alpha_s(\mu_0)} \right]= \frac{\mfm_\ssZ}{\sqrt\tau} \, \exp \left[ - \frac{1}{2\mfb \, \alpha_s(M_\ssZ)} \right] \,.
\ee
Phenomenological success (such as tests of the equivalence princple) relies on this sharing (at leading order in $1/\tau$) the same $\tau$-dependence as do other Standard Model masses, since at a microscopic level it is the QCD scale that determines most of the mass of macroscopic objects through its contribution to the nucleon mass. 

Although the above considerations make it sound like {\it all} effective interactions must depend on $\ln \tau$, this need not actually be the case. One way to see why is to notice that $\tau$ appears inside the logarithm in expressions like \pref{psikinLlog} multiplied by $\mu$, but $\mu$ always cancels out of physical observables. This happens in detail because effective couplings are matched across different thresholds, leaving physical results depending only on the logarithm of physical mass ratios 
\be
   \ln \left( \frac{m^2_1}{m^2_2} \right) = \ln \left( \frac{\mfm_1^2}{\mfm^2_2} \right) \,,
\ee
from which all powers of $\tau$ cancel provided that $m_1$ and $m_2$ both share the same $\tau$-dependence (as is in particular true for all Standard Model particles to leading order in $1/\tau$).  

What {\it can} introduce $\ln\tau$ dependence into the effective lagrangian is the presence within loops of particles with masses that depend differently on $\tau$. The gravitino and the axion and/or dilatino fields in the $T$ multiplet are examples of particle whose masses depend on $\tau$ differently than for Standard Model particles, and there is no reason why the same should not occur for particles in the UV sector as well. If two species of particles have masses $m_i^2 \propto \tau^{-p_i}$ with $p_1 \neq p_2$, and if these particles can appear together in loops then $\ln \tau$ dependence in couplings can arise through factors like 
\be
   \ln\left( \frac{m_1^2}{m_2^2} \right) = A_0 - (p_1 - p_2) A_1 \ln \tau \,,
\ee
for appropriate constants $A_0$ and $A_1$. As is argued in \S\ref{ssec:UVcomplete} below, the existence of UV particles with different $\tau$-dependence in their masses is in fact very plausible in candidate UV completions. It is this observation that motivates our including $\ln \tau$ dependence in the effective coupling functions considered in \S\ref{ssec:QNSRelaxation}.

\subsection{Dilaton stabilization}
\label{ssec:DilationStabilization}

The story to this point describes a natural hierarchy, but only does so {\it if} the field $\tau$ takes acceptably large values. We now argue that the potential for $\tau$ actually does have minima at such large values, and that this can be achieved without losing control of the underlying approximations. We describe two mechanisms for doing so, starting first with the mechanism assumed when making the estimates in \S\ref{sssec:SUSYScales} and then sketching an alternative that (at face value) shows promise for providing additional suppression of the vacuum energy (but which we have so far been unable to exploit).

\subsubsection{Logarithmic stabilization}
\label{sssec:LogStab}

We start with an example of dilaton stabilization that exploits the dependence $U= U(\ln \tau)$ in the potential \pref{VFBasicFormzzmin22}, following the ideas in \cite{AndyCostasnMe}.  The main attraction of this mechanism is its ability to produce exponentially large values of $\tau$ using only a mild hierarchy among the input parameters, whose ratios need not be smaller than $\sim 1/60$. 
 
To make things concrete suppose the function $k$ appearing in \pref{VFBasicFormzzmin22} is given by
\be
    k = \cK + (X + \ol X) \kappa + X \ol X \cZ
\ee
where $\kappa$ and $\cZ$ are $\tau$-independent\footnote{This assumption is just to simplify expressions, the general case with both $\kappa$ and $\cZ$ depending on $\ln\tau$ works in the same way as below but with more cumbersome expressions.} but $\cK$ acquires a dependence on $\ln\tau$ through the running of some UV-sector dimensionless coupling $\alpha_g$. In this case the potential \pref{VFBasicFormzzmin22} becomes
\be \label{VFBasicFormzzminzs1NAPPa}
  V_\ssF \simeq  +\frac{3|w_0|^2}{\tau^4} \left[  \frac{\cK'-\cK''}{ 1+ 2 \kappa^2 /\cZ}   
 \right] =: \frac{\cC}{\tau^4} \Bigl(\cK' - \cK'' \Bigr)   \,,
\ee
where $\cC := 3|w_0|^2/(1+2\kappa^2/\cZ)$ and primes denote differentiation with respect to $\ln\tau$. 

To evaluate these derivatives write the perturbative expansion of $\cK$ in the form
\be \label{cKvsalpha}
   \cK \simeq \cK_0 + \cK_1 \left( \frac{\alpha_g}{4\pi} \right) + \frac{\cK_2}{2} \, \left( \frac{\alpha_g}{4\pi} \right)^2 + \cdots 
\ee
with
\be \label{betafunctionNAPP}
\tau \frac{\exd}{\exd \tau} \left( \frac{ \alpha_g}{4\pi}\right) =: \beta(\alpha_g) =  b_1 \left( \frac{\alpha_g}{4\pi} \right)^2 + b_2 \left( \frac{\alpha_g}{4\pi} \right)^3 + \cdots \,.
\ee
The solution for the $\tau$-dependence of $\alpha_g$ to leading order in $\alpha_g$ becomes
\be \label{alphavstauRGNAPP}
   \frac{4\pi}{\alpha_g} = b_0 - b_1 \ln \tau \,,
\ee
for some integration constant $b_0$. This solution neglects $\alpha_g \ll 1$ while working to all orders in $\alpha_g \ln\tau$, which is valuable if minimization occurs in the regime $\ln \tau \sim 1/\alpha_g$ (as it will). For example, with the couplings normalized as above the constant $b_1$ appropriate to $N$ charged fermions would be
\be\label{b1vsN}
   b_1 = \frac{4N}{3} \,.
\ee

Using \pref{cKvsalpha} and \pref{betafunctionNAPP} to evaluate the derivatives in \pref{VFBasicFormzzminzs1NAPPa} gives  
\bea \label{VFBasicFormzzminzs1NAPP}
  V_\ssF &\simeq&  \frac{\cC}{\tau^4} \Bigl[\cK_1 b_1 \left( \frac{\alpha_g}{4\pi} \right)^2 + \Bigl(\cK_1 b_2 +\cK_2 b_1- 2 \cK_1 b_1^2 \Bigr) \left( \frac{\alpha_g}{4\pi} \right)^3 \\
  &&\qquad\qquad\qquad +\Bigl(\cK_1 b_3 + \cK_2 b_2 + \cK_3 b_1- 5\cK_1 b_1 b_2 -3 \cK_2 b_1^2  \Bigr) \left( \frac{\alpha_g}{4\pi} \right)^4 + \cdots \Bigr]   \,. \nn
\eea
This potential can have a minimum at $\tau = \tau_0$ for $\alpha_0 = \alpha_g(\tau_0)$ consistent with using perturbative methods provided there is a mild hierarchy amongst the coefficients $\cK_i$. In particular, if $|\cK_2/\cK_3| \sim \cO(\epsilon)$ and $|\cK_1/\cK_3| \sim \cO(\epsilon^2)$ for some smallish $\epsilon \sim 1/60 \ll 1$, then 
\be
   \frac{\partial V_\ssF}{\partial \tau} =  \frac{\cC}{\tau^5} \Bigl(-4\cK' +5\cK''-\cK''' \Bigr)
   \simeq  -4b_1 \left[ \cK_1 \left( \frac{\alpha_0}{4\pi} \right)^2  + \cK_2 \left( \frac{\alpha_0}{4\pi} \right)^3   +  \cK_3  \left( \frac{\alpha_0}{4\pi} \right)^4 \right] \,,
\ee
where the last equality drops the coefficients of $\alpha_0^n$ that are subleading in $\epsilon$. 

\begin{figure}[t]
\begin{center}
\includegraphics[width=120mm,height=60mm]{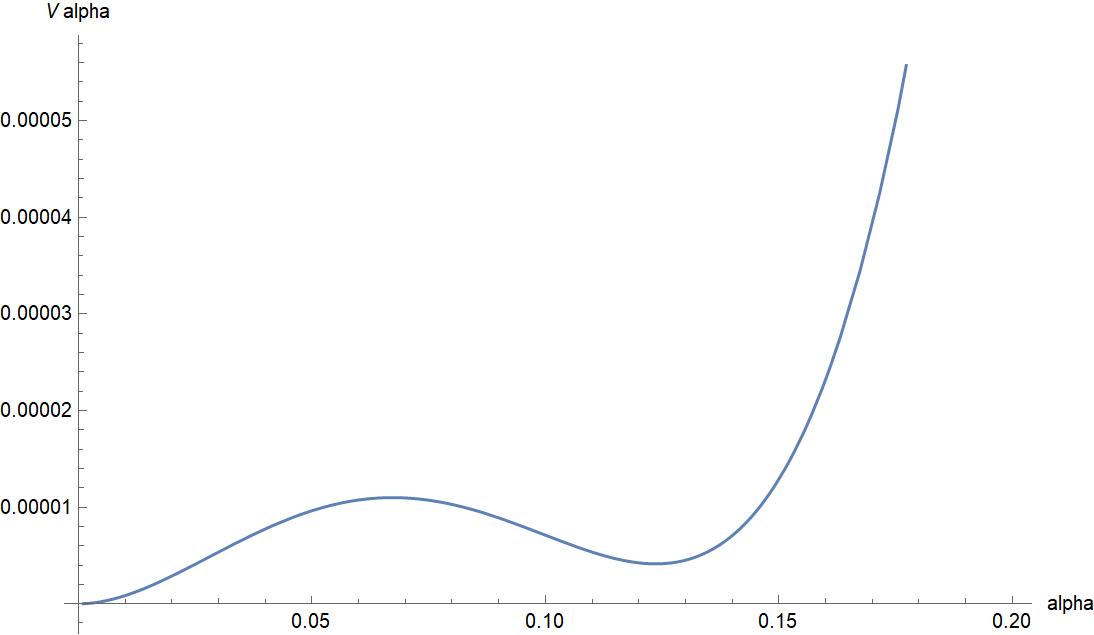} 
\caption{A sketch of the potential $U(\alpha)$ {\it vs} $\alpha$ where $V_\ssF = U(\alpha)/\tau^4$. The plots are obtained from \pref{VFBasicFormzzminzs1NAPP} using the representative values $\cK_1/ \cK_3 = 0.01$ and $ \cK_2/\cK_3 = - 0.133$ (arbitrary scale).} \label{Fig:Valpha} 
\end{center}
\end{figure}

The solutions to $V_\ssF' = 0$ at leading order in $\epsilon$ therefore are
\be \label{valofalphaatmin}
   \alpha_{0\pm} \simeq  \frac{ 1}{2} \left[ - \frac{ \cK_2}{\cK_3} \pm \sqrt{ \left(\frac{ \cK_2}{\cK_3} \right)^2 -  \frac{4 \cK_1}{ \cK_3}} \right] \sim \cO(\epsilon) \,.
\ee
For $\cK_1$ and $\cK_3$ positive and $\cK_2$ negative with $\cK_2^2 > \cK_1 \cK_3$ there are two real roots for which both $\alpha_0$ and $V_\ssF$ can be positive, with the minimum (maximum) being the root $\alpha_{0+}$ (or $\alpha_{0-}$). Because $\alpha_0 \simeq \cO(\epsilon)$ eq.~\pref{alphavstauRGNAPP} implies $\ln \tau_0 \sim 1/\alpha_0 \sim \cO(1/\epsilon)$ at the minimum, provided the constants $b_0$ and $b_1$ are order unity. In principle the values of $\alpha_0$ and $\tau_0$ can be adjusted independently by choosing $b_1$ appropriately (such as by choosing $N$ in \pref{b1vsN}).

The values of the potential and its second derivative 
\be \label{VFBasicFormzzminzs1NAPPx}
 \frac{\partial^2 V_\ssF}{\partial \tau^2} \simeq \frac{\cC}{\tau^6} \Bigl(20\cK' -29\cK'' +10\cK''' -\cK''''\Bigr) \,.
\ee
at these stationary points also turn out to be proportional to $\Bigl[ \cK_1  (\alpha_0/4\pi)^2 + \cK_2 (\alpha_0/4\pi)^3 + \cK_3 (\alpha_0/4\pi)^4  \Bigr]$ at leading order in $\epsilon$, showing that both $V_\ssF$ and $\partial^2 V_\ssF/\partial \tau^2$ are of order $\epsilon^5/\tau_0^4$ when evaluated at the minimum, rather than the naive $\epsilon^4/\tau_0^4$. All of these features are visible in the illustrative plot shown in Fig.~\ref{Fig:Valpha}, which uses a parameter set for which $\epsilon \sim 0.1$ to plot $U(\alpha)$ against $\alpha$, where $V_\ssF = U(\alpha)/\tau^4$.  This is the origin of the factors of $\epsilon$ seen in \pref{mvacest}, which in turn lead to the numerical estimates of \pref{baseline1} in \S\ref{sssec:SUSYScales}. 

\subsubsection{An alternative stabilization scenario}
\label{sssec:AltStab}

The dependence of $k$ on $\ln\tau$ described in the previous section leads to a potential whose minimum can easily occur at the extremely large values of $\tau$ that appear in the benchmark values \pref{baseline1}. For these choices the potential can be very small at its minimum but even so it is only as small as the observed Dark Energy density for extremely tiny values of the parameter $\epsilon \sim 10^{-5}$. But if $\tau \sim 10^{30}$ and $\epsilon \sim 10^{-5}$ then the leading $\epsilon^5 w_0^2/\tau^4$ contribution to the potential is so small that the next-to-leading $w_0^2/\tau^5$ contribution becomes competitive. 

This observation suggests exploring situations where this naively subdominant $1/\tau^5$ contribution might actually dominate. This is actually what happens if $k$ does not depend on $\ln \tau$, since in this case $K = - 3M_p^2 \ln(\tau - k)$ is a no-scale model whose scalar potential vanishes for all $\tau$. Although we have argued that a $\ln\tau$-dependence to $k$ is plausible, it is also not compulsory and in its absence\footnote{A natural way to have $\ln\tau$-dependence first arise in $h$ rather than in $k$ would be if the relevant coupling $\alpha_g$ were itself proportional to $1/\tau$ (as happens if it is the coupling of a gauge field whose kinetic function is $\mff_{ab} \propto T$). In this case $h$ might be expected to be only linear in $\ln\tau$, for which RG resummation is not needed.} it is the $h/\tau$ contribution of \pref{RelaxKW} that dominates the potential and generate $V_{\rm min} \propto w_0^2/\tau^5$. We therefore recompute the action without neglecting $h$, which -- recalling \pref{kexpX} -- we take to have the form
\be \label{hexpX}
  h = \frac{1}{M_p^2} \Bigl\{ \mfH(\Phi,\ol{\Phi},\ln \tau) + \Bigl[ X  \mfH_\ssX(\Phi,\ol{\Phi},\ln\tau) + \hbox{h.c.} \Bigr] +  \ol{X} X \mfH_{\ssX\ol\ssX}(\Phi,\ol{\Phi},\ln\tau) \Bigr\}\,.
\ee

The calculation is relatively easy because $h$ contributes to formulae like \pref{KInv0} or \pref{TkinL0} only at through negligible terms that are subdominant in $1/\tau$. The {\it only} place where nonzero $h$ actually matters is in the potential $V_\ssF$ because when $k_\ssT = 0$ the no-scale cancellation makes the $h$-dependent term the dominant piece. The result for $V_\ssF$ is also simple to read off because it has the same form as \pref{VFtauexp} but with the substitutions
\be
  \mfK_{\ssT\ol\ssX} \to \frac{\mfH_{\ol\ssX}}{\tau^2} - \frac{\mfH_{\ssT\ol\ssX}}{\tau}  \quad \hbox{and} \quad
  \mfK_{\ssT \ol\ssT} \to - \frac{2 \mfH }{\tau^3} + \frac{\mfH_\ssT + \mfH_{\ol\ssT}}{\tau^2} - \frac{\mfH_{\ssT\ol\ssT}}{\tau} \,,
\ee 
while (to leading order) $\mfK_\ssX$ and $\mfK_{\ssX\ol\ssX}$ remain unchanged. Since this substitution makes the $\mfK^{\ol\ssX \ssX} \mfK_{\ssT\ol\ssX} \mfK_{\ssX\ol\ssT}$ terms subdominant to the $\mfK_{\ssT\ol\ssT}$ term, we are led to the following form for the potential
\bea \label{VFtauexph}
  V_\ssF  &\simeq&  \frac{1}{\cP^2} \left\{ \frac13 \, \mfK^{\ol\ssX \ssX} \ol{w_{\ssX}} w_{\ssX}   + \left[ \frac{\mfK^{\ol\ssX \ssX} w_{\ssX} \ol{w_0} }{M_p^2}  \left(\frac{\mfH_{\ol\ssX}}{\tau^2} - \frac{\mfH_{\ssT\ol\ssX}}{\tau}  \right) + \hbox{h.c.} \right] \right. \\
  && \qquad\qquad  \left. + \frac{3 {|w_0|^2}/{M_p^4}}{  1  + 2 \mfK^{\ssX\ol\ssX}  \mfK_\ssX \mfK_{\ol\ssX}/M_p^2 } \left( \frac{2 \mfH }{\tau^3} - \frac{\mfH_\ssT + \mfH_{\ol\ssT}}{\tau^2} + \frac{\mfH_{\ssT\ol\ssT}}{\tau} \right) \right\} \,. \nn
\eea

If the relaxon is now eliminated using its equation of motion it drives $w_\ssX$ to become order $w_0/\tau^2$, and this  means all of the $w_\ssX$-dependent terms now only contribute to $V_\ssF$ at order $w_0^2/\tau^6$. This makes them subdominant to the last line of \pref{VFtauexph}, implying the leading potential for $\tau$ has the advertised $w_0^2/\tau^5$ form:
\be \label{VFtauexphtau}
  V_\ssF  \simeq  \frac{U(\ln\tau)}{\tau^5} \quad \hbox{with} \quad
   U  \simeq \frac{3 |w_0|^2}{M_p^4 } \left[ \frac{ 2 \mfH  - \tau \Bigl( \mfH_\ssT + \mfH_{\ol\ssT} \Bigr) + \tau^2  \mfH_{\ssT\ol\ssT} }{  1  + 2 \mfK^{\ssX\ol\ssX}  \mfK_\ssX \mfK_{\ol\ssX}/M_p^2 } \right] \,.
\ee
This emphasizes that $\mfH$ could well (but need not) depend on $\ln\tau$ through radiative corrections, in the same way as was true for $\mfK$.

Stabilization of $\tau$ can now proceed precisely as above, if $\mfH$ depends on $\ln\tau$ in the appropriate way. An important difference between this case and the one considered in \S\ref{sssec:LogStab} is how suppressed the potential's minimum value, $V_{\rm min}$, is by powers of $\alpha_0/4\pi \simeq \epsilon$. Because $U$ had to vanish if $\mfK$ were $T$-independent it is proportional to derivatives of $\mfK$, making its expansion in powers of $\alpha_g$ start at order $\alpha_g^2$. The same is not true for \pref{VFtauexphtau}, which does not vanish even if $\mfH$ is $T$-indepenent. Consequently although $U = \cO(\epsilon^5)$ when it is constructed from $\mfK(\ln\tau)$ and evaluated at the minimum, the potential \pref{VFtauexphtau} need only be\footnote{If $\alpha_g \propto \epsilon/\tau$ then instead $U \sim \cO(\epsilon)$.} $\cO(\epsilon^3)$. 

Although the extra factor of $1/\tau$ in the potential seems promising, it is compensated by the fact that having $w_\ssX \propto 1/\tau^2$ also implies $F^\ssX \propto K^{\ol\ssX\ssX} w_\ssX$ is smaller by a factor of $\tau$ and so is now $\propto \tau^{-5/2}$. This means that the ratio $V_{\rm min}/(F^\ssX)^2$ is now $\tau$-independent. The upshot is the additional powers of $\tau$ suppress $F^\ssX$ relative to $M_p^2$ but do not suppress $V_{\rm min}$ relative to the supersymmetry breaking scale.

\subsection{Scales and UV constraints}
\label{ssec:UVcomplete}

The point of view taken so far in this paper is to work within an EFT treatment of supergravity coupled to ordinary particles in four dimensions at and below electroweak energies. Our goal was to ensure that the prediction of small vacuum energies can remain stable as ordinary particles are integrated out. There are nonetheless at least five practical reasons to ask how this picture might arise from a UV completion (which we in practice take to be string theory, since this is sufficiently well-developed that questions can be sharply posed):
\begin{itemize}
\item Even if extremely large values like $\tau \sim 10^{26}$ are self-consistent within the low-energy EFT  UV physics relates $\tau$ to other observables in a way that can bring new constraints on its size. (We argue that the most obvious stringy provenance for $\tau$ relates it to the extra-dimensional volume, $\Omega_6$, in string units $\cV := \Omega_6 M_s^6$ through the relation $\cV = \tau^{3/2}$. If so, then extra-dimensosional constraints on $\cV$ impose a new limit not obtainable purely within the low-energy EFT: $\tau \lsim 10^{18}$.) 
\item Even if UV completions should preclude $\tau$ being as large $10^{26}$, they also provide new suppression parameters that would be hard to guess purely from within the EFT. (We explore whether extra-dimensional warping might provide an example of such a suppression, with only mixed results.)
\item Although \S\ref{ssec:Loops} argues that choices for parameters -- {\it e.g.} values for $w_0$ very different than $M_p^3$ -- remain stable as particles are integrated out, one can still ask whether and why the original values chosen in the UV make sense once embedded into a broader framework.  (We argue that they do.)
\item The required EFT couplings ({\it e.g.}~the axion decay constant) can -- but need not (see below) -- require the EFT to break down below TeV energies. If so, UV completions are important at experimentally accessible energies, and are needed to understand why this new physics does not undermine inferences obtained thinking about SM particles within the EFT. (We argue supersymmetric extra dimensions \cite{SLED} plausibly unitarize the cases where new physics intervenes at sub-TeV energies, in which case SM physics is localized to a brane and remains 4-dimensional.)
\item Specific UV completions make predictions with varying reliability. Robust predictions usually rely only on symmetry properties, while more fragile ones depend more sensitively on UV details. Knowing which is which is useful but often beyond the reach of the low-energy EFT. (We argue our central three mechanisms are robust in this way, while other predictions -- such as the value of $w_0$ -- are likely more model-specific.)
\end{itemize}
For these reasons we make preliminary UV connections here, while leaving a fuller study of UV completions for the future.

\subsubsection{A stringy pedigree for $\tau$}

The possibility there might be a connection to extra-dimensional models (including strings) is clear from the form for the dilaton-metric equations whose low-energy implications we discuss in \S\ref{sssec:DilatonCosmology}, since these coincide precisely with the 4D effective description used in ref.~\cite{AndyCostasnMe} motivated by models like \cite{SLED}. Such a convergence of dynamics is not a fluke: it reflects the underlying accidental scale invariance in higher-dimensional supergravity \cite{Salam:1989fm, Burgess:2011rv, SUGRAscaleinv, BMvNNQ, GJZ} that in turn follows from the automatic scale invariances of low-energy string vacua \cite{Burgess:2020qsc, Cicoli:2021rub}.

The extended no-scale property relied on here also suggests a string connection, since it was first identified using explicit loop calculations in string compactifications in \cite{Berg:2005ja} and has been used in concrete mechanisms for moduli stabilisation and inflation in \cite{Cicoli:2008va, Cicoli:2008gp}. It also fits with the general treatments of no-scale supergravity discussed in \cite{Barbieri:1985wq, Burgess:2020qsc}.

\subsubsection*{UV constraints on $\tau$}

The connection to string vacua is the most concrete for Type IIB flux compactifications on Calabi-Yau orientifolds since for these issues of modulus stabilization have been thought through in some detail \cite{Giddings:2001yu}, leading to a rich class of phenomenological constructions \cite{Kachru:2003aw, Balasubramanian:2005zx, Conlon:2005ki}. Moduli are central to these applications because they are naturally light (and so naturally appear in the low-energy 4D EFT), and modulus stabilization is what allows a quantitative calculation of the form of their low-energy scalar potential. 

The number of moduli appearing in any UV completion depends on details of the extra dimensions, but in IIB models there is always at least one modulus associated with the extra-dimensional volume, $\Omega_6$, evaluated in string units: $\cV = M_s^6 \, \Omega_6$. $\cV$ is a natural UV completion for $\tau$ because it is both universal and large: the consistency of describing low-energy physics using a field theory requires $\cV \gg 1$, making expansions in inverse powers of $\cV$ useful and ubiquitous. 

Among other things, the volume modulus determines the relative size of the string and 4D Planck scales through
\be
    M_s \sim \frac{M_p}{\cV^{1/2}}  \,.
\ee
In some geometries all extra-dimensional length scales are similar in size, allowing the volume to be written $\Omega_6 \sim L^6$. For these the volume also indicates the energy threshold above which 4D effective descriptions fail,
\be
    M_\KK = \frac{1}{L} \sim \frac{M_s}{\cV^{1/6}} \sim \frac{M_p}{\cV^{2/3}}  \,.
\ee

When the EFT below $M_\KK$ is a 4D supergravity then the dynamics of a potentially large class of 4D fields -- the K\"ahler moduli -- is described by a supergravity K\"ahler potential that at leading order at large $\cV$ has the form (in Planck units)
\be 
   K \simeq - 2 \ln \cV \,,
\ee
where $\cV$ is regarded as being an implicit function of the complex scalars that represent the K\"ahler moduli and transform in the standard way under 4D supersymmetry. Comparing this to $K \simeq - 3 \ln \tau$ suggests the identification
\be \label{cVvstau}
   \cV \sim \tau^{3/2} \,,
\ee
in which case the string and KK scales are related to $\tau$ by
\be
    M_s = \frac{M_p}{\tau^{3/4}} \quad \hbox{and} \quad
    M_\KK = \frac{M_p}{\tau} \,.
\ee

Both $M_s$ and $M_\KK$ are scales associated with physics not present in the low-energy 4D EFT, but knowing their $\tau$ dependence allows new constraints to be derived for $\tau$, since both should be UV scales from the point of view of the 4D theory. For instance, the constraint that $M_s$ be a UV scale implies $M_s  \gsim 10^4$ GeV and so 
\be \label{voltaubound}
   \cV \lsim 10^{28} \quad \hbox{or} \quad \tau \sim \cV^{2/3} \lsim 4 \times 10^{18} \,.
\ee
The stronger constraint $M_\KK \gsim 10^4$ GeV would instead require $\cV \lsim  10^{21}$ and so $\tau \lsim 10^{14}$ (though, as argued below, is less robust). At face value both of these constraints appear to rule out having $\tau$ as large as $10^{26}$ whenever \pref{cVvstau} holds.

\subsubsection*{Asymmetric compactifications}

The constraint from $M_\KK$ can be evaded if all extra dimensions are not similar in size, but \pref{voltaubound} holds more robustly. This is because the lower bound on $M_\KK$ can be much smaller than $10^4$ GeV \cite{Arkani-Hamed:1998jmv} if these extra dimensions can only be probed only using gravitational strength interactions. When this is so two extra dimensions could be large enough to put $M_\KK$ at sub-eV scales. Having extra dimensions only probed by gravitational interactions requires all SM fields to be trapped on a lower-dimensional surface (such as happens if they are open-string modes localized on a D-brane). Because not all dimensions can be this large any such a compactification must be `asymmetric' in that it involves dimensions with dramatically different sizes. 

The constraint \pref{voltaubound} is insensitive to this type of asymmetry. Taking two dimensions much larger than the other four leads to a total volume $\cV = \cV_4 \cV_2$ where $\cV_4 = M_s^4 \ell^4 $ and $\cV_2 = M_s^2L^2$ with $\ell \ll L$. In terms of these the two kinds of KK scale are 
\be
    M_\KL = \frac{1}{\ell} = \frac{M_s}{\cV_4^{1/4}} = \frac{M_p}{\cV_4^{3/4}\cV_2^{1/2}}
    \quad\hbox{and} \quad
    M_\KS = \frac{1}{L} = \frac{M_s}{\cV_2^{1/2}} = \frac{M_p}{\cV_4^{1/2}\cV_2}\,.
\ee
If (for example) $M_\KS  \sim 10$ eV and $M_\KL \sim 10$ TeV then these imply $\cV_2 \sim 10^{25}$ and $\cV_4 \sim 100$ and so $\cV = \cV_2 \cV_4 \sim 10^{27}$. The string scale remains $M_s \sim M_p/\cV^{1/2} \sim 3 \times 10^{-14} M_p \sim 60$ TeV, which is acceptably above the weak scale (and a bit larger than $M_\KL$). But because the low-energy K\"ahler potential remains $K = - 2 \ln \cV$ it again suggests the identification $\cV = \tau^{3/2}$, and so $\tau \sim \cV^{2/3} \sim 10^{18}$ which does not evade the bound \pref{voltaubound}.

In the absence of other suppression $\tau$ is too small on its own to make $V_{\rm min}$ in \S\ref{sssec:SUSYScales} tiny enough to describe Dark Energy. For instance, with $\tau \sim 10^{18}$ the estimate \pref{mvacest} becomes
\be \label{unwarpedVscales}
   V_{\rm vac} \sim \frac{\epsilon^5 w_0^2}{\tau^4 M_p^2} \sim \frac{\epsilon^5 M_p^4}{\tau^3} \sim \epsilon^5 10^{-54} M_p^4 \sim  \begin{cases} (10^{4} \; \hbox{GeV})^4 & \hbox{if} \quad \epsilon \sim \cO(1) \cr (1 \;\hbox{GeV})^4 & \hbox{if} \quad \epsilon \sim \cO(10^{-3}) \cr (10^{-4} \; \hbox{GeV})^4 & \hbox{if} \quad \epsilon \sim \cO(10^{-6}) \,, \end{cases} 
\ee
and so on. At face value, asymmetric compactifications in themselves do not avoid the constraint \pref{voltaubound} -- provided $\tau$ is identified with the volume modulus -- and so seem unable to achieve the observed dark-energy density (although the vacuum energy is substantially reduced relative to standard constructions).

\subsubsection*{Warping}
\label{sssec:Warping}

UV completions also suggest other sources of hierarchical suppression, and we now ask whether these can help reduce the value of $V_{\rm min}$. The most promising of these involves warping: the phenomenon where the 4D metric has a normalization that depends on an observer's position within the extra dimensions: $g_{\mu\nu}(x,y) = e^{2A(y)} g_{\mu\nu}(x)$. String compactifications are known to produce such warping \cite{Giddings:2001yu}, which (depending on the kinds of fluxes present) can generate strongly warped throats within which $e^A$ can be very small \cite{Klebanov:2000hb}. 

Schematically we regard a strongly warped geometry to be one whose warp-factor satisfies 
\be
    e^A \ll \cV^{-1/6} \ll 1 \,, 
\ee
and within strongly warped geometries physical scales are suppressed by warping in their immediate vicinity. For instance the string scale as measured by the tension of a space-filling 3-brane sitting at a position $y_b$ in the extra dimensions is given by
\be \label{Mswdef}
  M^w_s = M_s\, \Xi(\varpi, \cV) =  \frac{M_p}{\cV^{1/2}} \; \Xi(\varpi, \cV) \simeq \begin{cases} 
  M_p/\cV^{1/2} = M_p/\tau^{3/4} & \hbox{(unwarped)} \cr 
  \varpi M_p/\cV^{1/3} = {\varpi M_p}/{\tau^{1/2}} & \hbox{(strongly warped)} \,, \end{cases}
\ee
where $\varpi := e^{A(y_b)}$ and the calculable function $\Xi \sim \varpi \cV^{1/6} \ll 1$ when $\varpi \ll \cV^{-1/6}$ but is $\cO(1)$ otherwise. This means that a 3-brane tension in a strongly warped throat, $T_3 \sim (M_s^w)^4$, has the volume dependence
\be \label{antibraneT3}
   T_3 \sim \frac{\varpi^4 M_p^4}{\cV^{4/3}} \sim \frac{\varpi^4 M_p^4}{\tau^{2}} \,.
\ee

In particular, UV sources of supersymmetry breaking (such as anti-D3 branes) are energetically attracted to such regions, when they exist \cite{Kachru:2003aw, Kachru:2003sx}. Indeed, the observation that the energy density \pref{antibraneT3} shares the same $\tau$-dependence as does the leading $|w_\ssX|^2/\tau^2$ term of the potential \pref{VFtauexp} shows how the nilpotent field formulation provides a natural supergravity description of antibranes situated in warped throats \cite{Ferrara:2014kva, Kallosh:2014wsa, Antoniadis:2014oya, Aparicio:2015psl, GarciadelMoral:2017vnz}. 

Comparing \pref{Mswdef} to the KK scale also gives a lower bound that $\varpi$ can take in a strongly warped region, since consistency of an extra-dimensional field-theory description requires $M^w_s \gsim M_\KK$ and so $\varpi \gsim \cV^{-1/3}$. Field theoretic descriptions (usually the only ones available) for strongly warped throats therefore only allow
\be\label{warpingboxedin}
  \frac{1}{\cV^{1/6}} \gsim \varpi \gsim \frac{1}{\cV^{1/3}} \quad\hbox{and so}\quad
  \frac{1}{\tau^{1/4}} \gsim \varpi \gsim \frac{1}{\tau^{1/2}}  \,.
\ee

An obvious candidate UV completion for the Yoga models therefore has the SM localized on supersymmetry-breaking antibranes deep within a warped throat. In such a picture Standard Model fields are naturally described in the 4D EFT by constrained superfields coupled to the nilpotent goldstino field $X$ that captures the low-energy limit of a very unsupersymmetric place; the position of the anti-D3 brane attracted to the tip of a warped throat.\footnote{For a recent discussion on how to adapt these tools to a concrete statistical proposal to explain the small cosmological constant see \cite{Li:2020rzo}.} Concrete string scenarios with the Standard Model at an anti-brane are studied in \cite{Cascales:2003wn}. Ref.~\cite{Kallosh:2015nia, Garcia-Etxebarria:2015lif} obtains explicit string constructions for which the goldstino superfield $X$ is unequivocally identified. 

Because the most general effective field theory including warping and a nilpotent superfield has not fully been studied, it is not clear how the different components of the $X$-dependent K\"ahler potential depend on the warp factor. For the purposes of estimates we assume the following form:

\be \label{warpscaling2}
k_\ssX:=\frac{\mfK_\ssX}{M_p} \propto \varpi^a  \,,\quad  
k_{\ssX\ol\ssT} :=\frac{\mfK_{\ssX\ol\ssT}}{M_p} \propto \varpi^a  \,,\quad  
 k_{\ssX\ol \ssX}:=\frac{\mfK_{\ssX\ol \ssX}}{M_p^2}\propto\varpi^b
  \quad \hbox{and} \quad
  w_\ssX \propto \varpi^{(4+b)/2} \,,
\ee
with arbitrary powers $a,b$. The warping dependence of $w_\ssX$ is fixed by obtaining the known $\varpi^4$ result for an anti-brane tension. The choice $a=b$ would correspond to all $X$-dependent terms in $k$ sharing the same warping dependence while $b = 2a$ corresponds to the situation where the warping can be removed from $k$ (but not also from $W$) by rescaling $X$ appropriately.

Eq.~\pref{warpscaling2} predicts the warping dependence of $F^\ssX$ to be
\be \label{FXsoln2}
  \ol F^{\ol\ssX}  \simeq e^{K/2} K^{\ol \ssX \ssX} w_\ssX\sim \frac{ w_\ssX}{\tau^{1/2} \varpi^b}
  \quad\hbox{and} \quad
  w_\ssX\simeq \frac{3 \mfK_{\ssX\ol\ssT}\, w_0}{M_p^2}  \sim \frac{\varpi^a \, w_0}{\tau \, M_p}
\ee
and so  
\be \label{FXsolnno2}
  F^\ssX \simeq \varpi^{a-b} \left(\frac{w_0}{\tau^{3/2}  \, M_p}\right)   \,.
\ee
When $b > a+1$ this leads to an enhancement of $F^\ssX$ and therefore a suppression of $V_{\rm min}$ relative to $(F^\ssX)^2$. To see why, notice that in this case the vacuum energy is:
\be
V_{\rm min}\sim \frac{\epsilon^5 w_0^2}{\tau^4 M_p^2} \sim  \frac{\epsilon^5}{\tau}\;  \varpi^{2(b-a)} \Bigl( F^\ssX \Bigr)^2 \gsim \frac{\epsilon^5 }{\tau} \;  \varpi^{2(b-a)}\Bigl( 10^{-15} \, M_p \Bigr)^4 \gsim  \epsilon^5    \varpi^{2(b-a-1)}\Bigl( 10^{-22} \, M_p \Bigr)^4 \,,
\ee
where the first inequality imposes $F^\ssX \gsim (10^3 \; \hbox{GeV})^2$ and the last inequality uses $\varpi^2/\tau \gsim 10^{-28}$ as required once \pref{Mswdef} is used in $M_s^w \gsim 10^4$ GeV. For instance, choosing $b = a+ 2$ and $\tau \sim 10^{14}$ with $\varpi \sim 10^{-7}$ gives
\be
    V_{\rm min} \gsim \frac{\epsilon^5 \varpi^4 }{\tau} \Bigl( 10^{-15} M_p \Bigr)^4 \simeq  10^{-102} \epsilon^5 M_p^4
\ee
which can be tantalizingly small for reasonable values of $\epsilon$  like $\epsilon \sim 10^{-3}$. Note that these values for the warp factor $\varpi$ and the modulus $\tau$ satisfy the conditions \pref{warpingboxedin} above.

We remark in passing that these choices for warping also change the relation between the gravitino mass and $F^\ssX$, leading to
\be\label{m32warped}
    m_{3/2} = e^{K/2} \frac{|W|}{M_p^2} \sim  \frac{w_0}{\tau^{3/2} M_p^2}  \sim \varpi^{b-a} \frac{ F^\ssX}{M_p} \,,
\ee
a result that can be very small. For instance, choosing as above $b=a+2$ with $\varpi \sim 10^{-7}$ and $F^\ssX \sim (10^3 \; \hbox{GeV})^2$ implies $m_{3/2} \sim 10^{-17}$ eV: at face value well below the experimental bound of $10^{-5}$ eV quoted in \cite{Brignole:1998me, Brignole:1997sk}. We argue in \S\ref{ssec:OtherIssues} why the calculations on which these bounds are based need not be valid for the model described here.

What should we take for the powers $a$ and $b$? Although a full answer to this requires a better understanding of warped compactifications, some insight comes from the more detailed description of how strong warping enters into a 4D EFT  \cite{Bena:2018fqc, Dudas:2019pls,Crino:2020qwk}. It does so through a complex structure modulus, $Y$, whose expectation value determines the shape of the throat, and so warping dependence can be computed given how $Y$ couples to the nilpotent superfield. It can be consistent to incude this particular modulus in the 4D description because warping makes it lighter than the other complex structure moduli. When included it is represented using an ordinary unconstrained superfield (similar to $T$) and its inclusion in the 4D EFT reproduces the known calculations of warping with nilpotent superfields.

Integrating in this new modulus $Y$ in the 4D effective field theory introduces the following new terms in the superpotential:
\be \label{CCmodulusW}
     W=W_0+U(Y)+\sigma Y^2 X \quad \hbox{with} \quad
     U(Y) = Y^3 (\mfa\ln Y+\mfb) \,,
\ee
where $\mfa,\mfb$ and $\sigma$ are constants, of which $\mfa$ and $\mfb$ capture the presence of the three-form fluxes that fix the corresponding three-cycle. Solving $W_\ssY = 0$ for $Y$ gives the warp factor $\varpi^2 := \langle Y \rangle \simeq e^{-\mfb/\mfa} \ll 1$, and so in this picture how this appears in the rest of the action is controlled by the action's $Y$-dependence. For example, choosing the K\"ahler potential
\be
K=-3\ln\cP, \qquad \cP=\tau -k- \alpha Y\ol Y- \beta (X\ol Y+\ol X Y)-\gamma Y\ol Y X \ol X
\ee
with $\alpha,\beta,\gamma$ order-one coefficients gives, after substituting for $\langle Y \rangle$, the warping dependence assumed in \pref{warpscaling2} with $a=4$ and $b=2$. Even though this appears to be a promising source of hierarchies, it remains to be shown that this effective field theory actually emerges from the physics of the anti-brane EFT.\footnote{In particular the potential presence of an term of the form $(Y+\ol Y) X\ol X$ would change the results to $a=b=2$ in which case the predictions for warping would not differ from the unwarped case.}

\subsubsection{Further UV information: fluxes and more moduli}
 
With the above estimates of scales in mind, we briefly summarize some of the other implications that a UV completion using string theory can have for some of the other choices made in the Yoga scenario.

\subsubsection*{Supersymmetry and fluxes}

Several types of three-form fluxes, $F_{\ssM\ssN\ssP}$ and $G_{\ssM\ssN\ssP}$, play a central role in Type IIB vacua, where they help stabilise the complex-structure moduli and the string dilaton.\footnote{Not to be confused with the dilaton field of this paper.} The fluxes do not fix the K\"ahler moduli, whose potential remains flat at tree level because it has a no-scale form \cite{Giddings:2001yu}. This no-scale structure arises along the lines described in \S\ref{sssec:OtherDark} and has its roots in the compactification's underlying accidental scale invariance.

Fluxes also contribute to the way that compactifications break supersymmetry, and because of this it is fluxes that determine the UV value of the parameter $w_0$. The contributions of fluxes to the low energy superpotential turns out to be given by \cite{Gukov:1999ya}
\be \label{GVW}
   W = \int_X \Bigl( F_3 - i S G_3 \Bigr) \wedge \Omega \,,
\ee
where $S$ is the string dilaton and $\Omega$ is the holomorphic harmonic (3,0)-form that exists on any Calabi-Yau space $X$. 

For the present purposes notice that because \pref{GVW} gives $W$ as an integral over the extra dimensions its size need not be set purely by the string scale and so can grow with extra-dimensional size. (Having $W$ be larger than string scale in string frame means it is larger than Planck scale in 4D Einstein frame.) Ref.~\cite{Cicoli:2013swa} examines precisely how big this allows the parameter $w_0$ to be within the 4D EFT and argues it is bounded above by 
\be
    w_0 \lsim M_p^3\, \cV^{1/3} \sim M_p^3\, \tau^{1/2} \,.
\ee
They further argue that the upper limit is saturated, $w_0 \sim M_p^3 \tau^{1/2}$, when the gravitino mass, $m_{3/2} \sim w_0/(\cV M_p^2) \sim w_0/(\tau^{3/2} M_p^2)$ is of order the KK scale $M_\KK \sim M_p/\cV^{2/3} = M_p/\tau$ (as considered by our earlier 4D estimates). 

But this does not mean $w_0$ {\it must} be large; systematic searching \cite{Demirtas:2021nlu} has also found many solutions where $w_0$ can be as small as $10^{-90}$ in Planck units. Small $w_0$ turns out to be of interest when $F^\ssX \sim (M_s^w)^2$, as is often true for warped string compactifications.

\subsubsection*{Logarithmic potentials}

Generating exponentially large field values, as we have found here, embeds well into the `large-volume scenario' (LVS) string vacua \cite{Balasubramanian:2005zx, Conlon:2005ki} (if $\tau$ is the volume modulus), for which similar arguments are used to show that the volume of the extra dimensions can be exponentially large. Logarithmic dependence of moduli within the K\"ahler potential has also been considered within a string context \cite{Conlon:2010ji} where its effects on KKLT \cite{Kachru:2003aw} or LVS moduli stabilisation was investigated. More recently this log-dependence was considered in \cite{Weissenbacher:2019bfb, Antoniadis:2019rkh} to provide a new mechanism for moduli stabilisation. In particular \cite{Antoniadis:2019rkh} uses it to realise large volumes even in the case of one K\"ahler modulus. 

The RG-based stabilisation mechanism used here goes back to the earlier extra-dimensional examples of \cite{AndyCostasnMe} has not been used in a string context, but may be useful there for modulus stabilisation in string theory independent of our current scenario. If so, it may provide a new approach in which the volume is naturally exponentially large with a built-in de Sitter uplift that builds on a controlled approximation in powers of a small coupling $\alpha_g$, for which RG resummation allows the exploration of the regime $\ln \tau \sim 1/\alpha_g$ without losing control of the underlying expansion (as was done here).

\subsubsection*{Stringy provenance for $\phi$}

The relaxon field $\phi$ is the one ingredient to this picture whose UV provenance is the least obvious (although the related general phenomenon of charged scalars adjusting to minimize $D$-term potentials at zero is very common). One way to think about $\phi$ is that it is a field that interpolates between regions that turn off and turn on a large positive contribution in the scalar potential. If the positive energy were to dominate in the early universe then any such a field that can turn it off through slow evolution would naturally play the role of an inflaton. 

From this point of view a UV provenance for $\phi$ is equivalent to finding stringy origins for a nonsupersymmetric inflaton. Among the many fields that present themselves, the most obvious candidate is the modulus describing brane-antibrane separation within the extra dimensions. This naturally arises in a sector that badly breaks supersymmetry (the antibrane) and also naturally interpolates between regions where a large positive energy (the antibrane tension) can be present or not. In this picture the region where the positive energy turns off would be places where brane-antibrane annihilation removes this underlying energy. This suggests a natural embedding of our framework into brane-antibrane inflationary models.\footnote{{\it Note added:} further exploration confirms this connection, with the stabilization mechanism explored here improving them by removing the $\eta$ problem these models usually have \cite{Burgess:2022nbx}.} 

\subsubsection*{Other moduli}

The constraint \pref{voltaubound} rules out values as large as $\tau \sim 10^{26}$. But it could also be that $\tau$ should not be identified with the volume modulus. In general $\tau$ could also correspond to a combination of the many other moduli that are generically present in Calabi-Yau orientifolds, including blow-up or fibre moduli. (The warped geometries described in \S\ref{sssec:Warping} are a special case of additional suppression coming from other -- in that case, complex-structure -- moduli.) The generic accidental scale invariance generally ensures other K\"ahler moduli can also be included into the low-energy EFT while preserving the no-scale condition \cite{Burgess:2020qsc}, along the lines described in \S\ref{sssec:OtherDark} below. 

Although it is generic to have many moduli, it is also true that their masses generically depend on the overall volume $\cV$ and the flux superpotential $W_0$. Even if $\tau \neq \cV^{2/3}$, having $\tau$ so large would imply the volume is also large even if $\tau$ is some other modulus. If so, the cosmological moduli problem \cite{Coughlan:1983ci, Banks:1993en,deCarlos:1993wie} has to be addressed (see also \cite{Conlon:2007gk}). This demands that light and long-lived fields not overclose the universe. Roughly speaking generic $\cO(M_p)$ initial amplitudes for these fields can become dangerous once they begin to oscilate around their minima because these oscillations can dominate the energy density of the universe today ($\rho_\chi\geq \rho_c$) unless their mass satisfies $m_\chi \leq 10^{-26}$ eV.  If this bound is satisfied they may contribute to dark matter. Any heavier unstable modulus that decays into Standard Model daughters must do so quickly in order not to ruin nucleosynthesis and therefore needs to be  heavier than $30$ TeV. This problem does not arise if the potential at its minimum can be suppressed down to Dark Energy scales, because then our earlier bounds imply $m_\chi \sim 10^{-32}$ eV, but must be faced otherwise.

\subsubsection*{$\tau$-dependent masses} 

Another feature playing an important role in the EFT was the $\tau$-dependence of masses, and the possibility that different species of particles could depend differently on $\tau$ (since this would allow logarithmic dependence on mass ratios to turn into a dependence on $\ln\tau$). String compactifications very generically contain a suite of states that have a hierarchy of masses parametrised by differing dependence on the volume $\cal V$:
\be
M_p\geq M_s\propto \frac{M_p}{{\cal V}^{1/2}}\geq M_{kk}\propto \frac{M_p}{{\cal V}^{2/3}}\geq m_{3/2}\propto \frac{M_p|W|}{\cal V}\cdots
\ee
and so can naturally lead to logs of $\tau$ for $\tau\sim {\cal V}^{2/3}$. Moreover, the $\tau$-independence of physical Yukawa couplings (like those involving the Higgs field) argued here also generically happens for the Yukawa couplings of particles localized on a (anti) D3 brane \cite{Conlon:2006tj}. 

It is clear that many ingredients of our scenario have natural counterparts in concrete UV completions from string compactifications, but the general bound \pref{voltaubound} that follows from the simplest identification $\cV = \tau^{3/2}$ implies that some non-trivial engineering is required to fully achieve Dark Energy scales from string constructions. If this should turn out to be possible, the string landscape might then provide a welcome alternative to anthropic arguments to explain the Higgs mass and dark energy scales. The anisotropic compactification scenario presented in \cite{Cicoli:2011yy} already includes some of the required ingredients and may provide a natural place to start. We leave further considerations on how to realize this scenario in string theory to future work.

\section{Phenomenological issues}
\label{sec:Phenomenology}

The next two sections take the point of view that values as large as $\tau \gsim \tau^{26}$ can be reconciled with a UV completion along the lines of \cite{SLED} at eV scales and above (where the axion coupling tells us that the EFT for $\tau$ this large must break down). We do so because the main phenomenological challenges and opportunities provided by the relaxation/supersymmetry mechanism described above arise at energies lower than this, and can be examined independent of any search for a UV embedding. All we assume about SM fields is that their masses are $m \propto \tau^{-1/2}$ as found above; a result we rederive in this section on more robust grounds that are likely to apply to the brane-world framework suggested by string models and \cite{SLED}.

We identify the main arenas of concern and/or opportunity to be astrophysical. In particular, this section describes general constraints, focussing in particular tests of gravity such as in the solar system.  Cosmology is the focus of \S\ref{sssec:DilatonCosmology}. Our aim here is not to be exhaustive, but instead to identify some of the model-building issues that are likely necessary to accommodate our approach to the cosmological constant problem.  The core message is that although the models considered here are potentially constrained by many observations, it is not clear that these constraints need be fatal (and some offer tantalizing opportunities). 

\subsection{The EFT relevant to astrophysics} 
\label{ssec:LEDSEFT}

Because the main arena is astronomy and cosmology, for phenomenological purposes it is useful to strip away all UV details and examine the low-energy effective lagrangian relevant at scales well below the electron and $\phi$-scalar mass, dropping all subleading powers of $1/\tau$. This lagrangian has the form 
\be\label{VLEEFT}
   \cL = - \frac{M_p^2}{2} \,\sqrt{-g}\;  R + \cL_{{\rm dil}} + \cL_{\SM} +   \cL_{{\rm dark}} + \cL_{\rm int}\,,
\ee
where the first term is the usual Einstein-Hilbert action, whose canonical form shows the remainder of the lagrangian is given in Einstein frame. $\cL_{\rm dil}$ describes the leading interactions of the dilaton supermultiplet, both with itself and with the graviton supermultiplet. The next two terms respectively describe the couplings of the graviton and dilaton-axion to Standard Model degrees of freedom ({\it i.e.}~ordinary matter), and to any other non-dilaton dark low-energy degrees of freedom, while $\cL_{\rm int}$ contains all other mutual interactions (such as 4-fermi interactions) of these fields. 

We next collect the expressions for each of these in turn, to leading order in $1/\tau$. 

\subsubsection{Standard Model sector}

The above discussion determines the Standard Model contribution $\cL_{\SM}$ in \pref{VLEEFT} to leading order in $1/\tau$. Keeping only terms out to dimension 4 we have
\be \label{SMVLEEFT}
    \frac{\cL_{\SM}}{\sqrt{-g}} = - \frac14 \, F_{\mu\nu} F^{\mu\nu} - \sum_a \ol \psi_a \Bigl[ \cDsl + m_{ab}(\tau)  \Bigr] \psi_b \,,
\ee
where $F_{\mu\nu} = \partial_\mu A_\nu - \partial_\nu A_\mu$ is the standard electromagnetic field strength, and the sum over $a$ runs over any spin-half particles, $\psi_a$, relevant (which for cosmology or solar-system applications might be neutrinos or heavier but stable everyday particles like electrons and baryons). The derivative $\cD_\mu$ is the spin-half version of the derivative defined in \pref{DilatonCovD} below.

Although $m_{ab} \propto \delta_{ab} \tau^{-1/2}$ for stable everyday particles, there are two natural options to assume for the $\tau$-dependence of neutrino masses. One is to assume that they also have masses proportional to $\tau^{-1/2}$. But the presence of supersymmetry in the dark sector also predicts there to be other sterile fermions present that could mix with Standard Model neutrinos. Some of these must be light (such as the superpartners of the graviton or the dilaton), but others could be very heavy. Those that are heavy would have to have masses unsuppressed by $\tau$, and so once they mix with left-handed neutrinos through a mixing term that scales like other SM masses -- {\it i.e.}~like ${\tau}^{-1/2}$ -- the see-saw mechanism would predict physical light-neutrino masses to be order $m_\nu \propto \tau^{-1}$; providing an attractive explanation of the otherwise coincidental similarity between neutrino masses and the cosmological constant scale: $V_{\rm min} \sim m_\nu^4$.

\subsubsection{Dilaton sector}

The supergravity framework is unusually specific about some parts of the dark sector. 
The most constrained of these new dark particles are the members of the dilaton multiplet itself, which we divide into its bosonic and fermionic parts: $\cL_{\rm dil} = \cL_{{\rm dil}\,b} + \cL_{{\rm dil}\,f}$.

The bosonic part contains the kinetic terms \pref{Ltaukin} for the axion and dilaton together with the scalar potential computed in \S\ref{sec:Relaxation}:
\be \label{LdilbForm}
  - \frac{\cL_{{\rm dil}\,b}}{\sqrt{-g}} =  \frac12 \Bigl( \partial^\mu \chi \, \partial_\mu \chi +M_p^2 \, e^{- 2\zeta \chi/M_p} \, D^\mu \mfa \, D_\mu \mfa \Bigr) + V(\chi) \,,
\ee
with ({\it c.f.}~eq.~\pref{chizetadef}) $\zeta = \sqrt{\frac23}$. Unbroken axion shift symmetry prevents $V$ from depending on $\mfa$. The stabilization mechanism of \S\ref{ssec:DilationStabilization} leads to a potential like \pref{VFBasicFormzzmin22}
\be \label{VvsUExp}
  V(\chi, \mfa) = U \, e^{-4\zeta \chi/M_p} \,,
\ee
where $U = U(\ln \tau) = U(\chi)$ is what remains after $\phi$ has relaxed to suppress $\mfw_{\ssX}$, and might be a rational function of $\ln\tau$. In the examples of \S\ref{ssec:DilationStabilization}, the dependence of $U$ on $\ln \tau$ gives $V$ a minimum at $\tau \sim \tau_0$ where $\ln \tau_0 \sim 1/\alpha_g \sim 1/\epsilon \sim 60$, and when evaluated at its minimum $U$ is of order $\epsilon^5$ and so $V_{\rm min} \sim \epsilon^5 m_\TEV^8/M_p^4$ as in \pref{mvacest}.  

As discussed in \S\ref{sssec:DilatonAxion} the axion shift symmetry might be gauged. If not then $D_\mu \mfa = \partial_\mu \mfa$ in \pref{LdilbForm}, and if so then $D_\mu \mfa = \partial_\mu \mfa - A_\mu$ for some dark-sector gauge boson. This implies the axion could, but need not, be massless, depending on whether or not gauging allows it to be eaten via the Higgs mechanism. If gauged its mass could be either of order $M_p/\tau^{3/2} \gsim M_\TEV^3/M_p^2 \sim 10^{-14}$ eV or $M_p/\tau \gsim M_\TEV^2/M_p \sim 10$ eV, depending on how the dark photon kinetic term depends on $\tau$. 

A potential axion mass strongly affects how the axio-dilaton responds to gravitating objects. If its mass is around 10 eV then its Compton wavelength is comparable to the size of an atom and it can be integrated out and cannot mediate macroscopic forces relevant to tests of gravity. If its mass is $10^{-12}$ eV then its Compton wavelength is of order $10^4$ km, or about a tenth of a light-second, and although its implications for solar system tests are less clear it is also unlikely to be relevant for testing gravity on longer scales than this. In either case we shall see that the dilaton-matter coupling likely poses a problem because it is too large to have escaped detection. If the axion were massless, however, then it can mediate long-range forces and its presence can be exploited as in \cite{ADScreening} to help hide the axio-dilaton from tests of gravity.  

The fermionic contribution to $\cL_{\rm dil}$ consists of the leading parts of the gravitino and the dilatino actions, 
\bea
   \frac{\cL_{{\rm dil}\,f}}{\sqrt{-g}} &=&  - \frac{i}2 \, \epsilon^{\mu\nu\lambda\rho} \ol \psi_\mu \gamma_5 \gamma_\nu D_\lambda \psi_\rho - \frac{1}{2} \,m_{3/2} \,\ol \psi_\mu \gamma^{\mu\nu} \psi_\nu \\
   && \qquad\qquad - \frac12 \, \ol \xi \Dsl \xi - \frac12\left[ \mfm_\xi \, \ol \xi \gamma_\ssL \xi + \frac{\mfm_{\xi g}}{M_p} \, \ol \psi_\mu \gamma_\ssL \gamma^\mu \xi + \hbox{h.c.} \right] \nn
\eea
where the gravitino mass, $m_{3/2}$, is given by \pref{gravitinomass} and so is proportional to $w_0^2\, \tau^{-3/2}$ and of present-day numerical size $m_\TEV^3/M_p^2 \sim 10^{-14}$ eV. The functions $\mfm_\xi$ and $\mfm_{\xi g}$ are similarly given by eqs.~\pref{mfmxig} and \pref{mfmxi}, and generically also give $\xi$ a mass that is of order $m_{3/2}$. Notice that the gravitino-dilatino mixing parameter $\mfm_{\xi g}$ vanishes when $D_\ssT W = 0$, but the direct mass term $\mfm_\xi$ need not also do so because of the target-space curvature terms in \pref{mfmxi}.  

\subsubsection{Other dark ingredients}
\label{sssec:OtherDark}

The remaining term $\cL_{\rm dark}$ of \pref{VLEEFT} describes any other of the more model-dependent light supersymmetric degrees of freedom that might be present, of which there are two types of natural candidates. 

\subsubsection*{Dark gauge sector}

The first type of additional dark sector could consist of a dark-sector gauge boson, which could (but need not) include one that gauges the axion shift symmetry and so eats the axion to get a mass. As discussed in \S\ref{sssec:DilatonAxion} the presence of such a gauge boson often also implies the existence of other dark degrees of freedom, such as the corresponding gaugino as well as possible charged supermultiplets whose presence can be required to cancel anomalies and/or to allow the corresponding gauge auxiliary fields, D, to be minimized at zero (see {\it e.g.}~\cite{Williams:2011qb}). 

\subsubsection*{Dark moduli}

A second natural sector to have in the supersymmetric world are new superfields $S^i$ corresponding to extensions of the no-scale sector. These kinds of extensions are well-motivated by specific UV examples -- for which they can arise as compactification moduli in addition to the scaling field $T$ (see {\it e.g.}~the discussion in \S\ref{ssec:UVcomplete}). When present, they generically acquire a Hubble-scale mass for the same reasons that the dilaton $\tau$ does.

The single field $T$ can be extended to a more general sector $Z^\ssA = \{ T, S^i \}$ without ruining the no-scale cancellations that lead to an acceptable Dark Energy density because all that is needed is for the $T$ sector to be scale invariant at leading order and to remain a no-scale model at subleading order in $1/\tau$. Both of these properties also hold for more complicated sectors. For example imagine if our starting K\"ahler potential were to replace \pref{RelaxKW} by 
\be \label{RelaxKModuli}
   K = -3 M_p^2 \ln \cP \quad \hbox{with} \quad \cP = \cF(\tau, \sigma^i) - k + \frac{h}{\tau} + \cdots \,,
\ee
where $\sigma^i = S^i + \ol S^i$ and scale invariance is built in by requiring $\cF$ to be a homogeneous degree-one function:
\be \label{HomoDegOne}
   \cF(\lambda \tau, \lambda \sigma^i) \equiv \lambda \cF(\tau, \sigma^i) 
\ee
for all $\tau$ and $\sigma^i$. This property ensures $\cF$ can always be written with an overall factor of $\tau$ scaled out: $\cF(\tau,\sigma^i) = \tau \, F(x^i)$ where $x^i := \sigma^i/\tau$. 

As is easy to check, eq.~\pref{HomoDegOne} implies that $K$ defined by \pref{RelaxKModuli} satisfies the no-scale identity $K^{\ol\ssA \ssB} K_{\ol\ssA} K_\ssB \equiv 3 M_p^2$ provided we drop all terms involving $h$ (and higher orders in $1/\tau$) and that $k$ is independent of $\tau$ and $\sigma^i$. This means that the scalar potential $V_\ssF$ vanishes identically for all $T$ and $S^i$ so long as these fields do not appear in $k$, and when they do appear the potential must be proportional to derivatives of $k$, similar to \pref{VFtauexpagain} in the single-modulus case. These properties ensure that the arguments for the low-energy potential being order $w_0^2/\tau^4$ remain true in this more complicated setting. 

For phenomenological purposes, what is important is that the additional scalars $\sigma^i$ generically have masses at the present-day Hubble scale, as argued above. This makes these fields also relevant for tests of gravity, and allows them to be active during cosmology. Their axion counterparts can be massive or massless depending on the details, much like for the axion partner of $\tau$ discussed above.

\subsection{Axio-dilaton couplings and tests of gravity}
\label{ssec:DilatonMatterPhenomenology}

At first sight, the extremely small dilaton mass and its nominally gravitational-strength couplings to matter makes the dilaton's competition with gravity likely the most constraining part of our phenomenological story. Because of its importance we start by rederiving the dilaton-matter couplings in a more transparent way than in \S\ref{ssec:MatterCouplings}, showing these couplings follow very generally from the accidental scale invariance. 

The good news is that scale invariance forces the dilaton-matter couplings to have a `quasi-Brans Dicke' form \cite{ScalarTensorTests}, and this is good because it means that it automatically satisfies all of the very stringent tests of the Equivalence Principle \cite{EPTests}. We then show that this same scale invariance actually predicts the dilaton behaves as an honest-to-God Brans-Dicke scalar \cite{BransDicke}, whose coupling is predicted quite generally to be\footnote{The traditional Brans-Dicke parameter $\omega$ is related to the coupling as defined here by $2\mfg^2 =1/(3+2\omega)$.} $\mfg = - {1}/{\sqrt 6} \simeq - 0.408$, even in the general multi-modulus case. 

The bad news is that couplings this large are naively ruled out by solar system tests of GR. But all is not lost: we summarize how having the axion couple to matter in addition to the dilaton provides a way to evade this constraint, even for extremely small axion couplings, using the mechanism proposed in \cite{ADScreening}.

\subsubsection{Quasi-Brans Dicke scalar}

For the present purposes we define a quasi-Brans Dicke scalar to be a scalar-tensor theory for which the scalar field $\chi$ couples through a lagrangian density of the form
\be \label{QBDLagr}
    \cL= - \sqrt{- g}\left[ \frac{M_p^2}{2} \, \cR + \frac12 \, (\partial \chi)^2 \right] + \cL_m(\tilde g, \tilde \psi, \tilde h) \,.
\ee
Here $\cL_m$ is the lagrangian density for representative spin-half and spin-zero matter fields ($\tilde \psi$ and $\tilde h$) whose defining feature is that the Einstein-frame metric and the scalar $\chi$ only enter into $\cL_m$ through the Jordan-frame metric 
\be \label{JFtoEF}
   \tilde g_{\mu\nu} = A^2(\chi) \; g_{\mu\nu} \,,
\ee
for some function $A(\chi)$. The Brans Dicke special case \cite{BransDicke} is the choice
\be \label{BransDickeA}
   A(\chi) = e^{\mfg \chi/M_p}
\ee
where $\mfg$ is one of the ways to write the dimensionless Brans-Dicke coupling parameter. 

Because particle kinetic and mass terms are by assumption $\chi$-independent in Jordan frame, their $\chi$-dependence in Einstein frame is predicted to be universal ({\it i.e.}~depend only on the matter-particle spin); for instance for spinless and spin-half fields
\be
   \sqrt{-\tilde g}\;\Bigl[  \tilde g^{\mu\nu} \partial_\mu \tilde h^* \, \partial_\nu \tilde h + {\tilde{e}_a}{}^\mu \, \ol{\tilde\psi} \gamma^a D_\mu \tilde\psi \Bigr] = \sqrt{-g}\Bigl[ g^{\mu\nu} A^2(\chi)\, \partial_\mu \tilde h^* \, \partial_\nu \tilde h +  e_a{}^\mu A^3(\chi)\, \ol{\tilde\psi} \gamma^a D_\mu \tilde\psi \Bigr] \,.
\ee
The kinetic terms are put into canonical form by redefining $\tilde h = A^{-1} h$ and $\tilde\psi = A^{-3/2} \psi$, at the expense of modifying the covariant derivatives: $\partial_\mu \tilde h = A^{-1} \cD_\mu h$ and $D_\mu \tilde \psi = A^{-3/2}\cD_\mu \psi$, with
\be \label{DilatonCovD}
 \cD_\mu h = \partial_\mu h -  \left( \frac{\partial_\mu  A}{A}  \right)  h \quad \hbox{and} \quad
 \cD_\mu \psi = D_\mu \psi  - \frac{3}{2} \, \left( \frac{\partial_\mu  A}{A}\right)  \psi \,.
\ee

In both cases Einstein-frame particle masses depend universally on $\chi$ with
\be \label{EFparticlemass}
   m_{\EF} = m \, A(\chi) \,,
\ee
and so all mass ratios are $\chi$-independent, although their ratio with the Planck mass is not.\footnote{Since these are physical conclusions they are frame-independent and so can be seen equally well in either Jordan or Einstein frame.} The Brans-Dicke coupling function $\eta(\chi)$ is defined by the matter contribution to the $\chi$ field equation,
\be \label{DilatonFE}
  \Box \chi(x) +\frac{ \eta(\chi)}{M_p} \, g_{\mu\nu}  T^{\mu\nu} = 0 \quad\hbox{where}\quad
    \frac{\eta(\chi)}{M_p} :=  \frac{\partial}{\partial \chi} \ln A(\chi) \,,
\ee
and $T^{\mu\nu}$ is the usual matter stress energy tensor as defined by varying the Einstein-frame metric. Notice $\eta = \mfg$ is a constant in the pure Brans Dicke special case defined by \pref{BransDickeA}.

\subsubsection{The dilaton as a Brans-Dicke scalar}
\label{sssec:DilatonBD}

The above definitions precisely capture the matter couplings of the dilaton $\tau$, provided we omit the scalar potential\footnote{For cosmological purposes the scalar potential must be included, but its contributions are negligible for this section's focus: scalar-tensor constraints in non-cosmological settings.} and restrict to leading order in $1/\tau$. To see why, recall that the supergravity action can be derived in superspace starting from an expression 
\be \label{XSugraKW}
   \cL = \int \exd^2\Theta \, 2\cE \left[ \frac38 \Bigl( \ol \cD^2 - 8 \cR \Bigr) \, e^{-K/3} + W \right] + \hbox{h.c.} \,,
\ee
where $W$ and $K$ are the superpotential and K\"ahler potential. The gravitational lagrangian in this frame\footnote{For aficianados: we imagine fixing superconformal invariance here using a $K$-independent compensator, and so the Einstein-Hilbert action is not in canonical form.} turns out to be proportional to $M_p^2\, e^{-K/(3M_p^2)}$, and so the Weyl rescaling \pref{JFtoEF} required to reach Einstein frame is given by 
\be \label{AchiLeading}
   A(\chi) = e^{K/(6M_p^2)} = \frac{1}{\cP^{1/2}}  
   \,,
\ee
which uses $K = -3 M_p^2 \ln\cP$ with $\cP \simeq \tau - k + \cO(1/\tau)$. 

Furthermore when $W$ is independent of $T$ then $\tilde g_{\mu\nu}$ also agrees with the Jordan-frame as defined in \pref{QBDLagr} because it is the frame for which the particle masses found in \S\ref{ssec:MatterCouplings} for ordinary matter fields like $Y_\pm$ are $\tau$-independent.\footnote{$\tau$-independent in that they depend at most logarithmically on $\tau$ through $k$.} Using \pref{AchiLeading} in \pref{EFparticlemass} shows more directly why the masses of all ordinary particles turned out proportional to $\cP^{-1/2}$ at leading order (as was restated in \pref{SMVLEEFT}).

The dilaton therefore couples like a quasi-Brans Dicke scalar, and using eq.~\pref{AchiLeading} in \pref{DilatonFE} gives the Brans-Dicke coupling function
\be \label{EtaVsKT}
  \eta(\chi)   
  = \zeta \,\frac{\partial  \ln A}{\partial \ln\tau}  \simeq \frac{\zeta\,\tau  K_\ssT}{6M_p^2}  \simeq - \frac{ \zeta}{2} = - \frac{1}{\sqrt 6} \simeq -0.408 \,,
\ee
showing that $\tau$ is a pure Brans Dicke scalar (for which $\eta$ is $\chi$-independent) to leading order in $1/\tau$, with coupling  $\mfg = - \zeta/2 \simeq -0.408$.  

\subsubsection*{Multiple moduli}

We now pause to show that the same conclusion also holds very generally when the no-scale sector has multiple moduli, as in \pref{RelaxKModuli}, subject to the scaling condition \pref{HomoDegOne}. 

To this end suppose we have several scaling moduli, $Z^\ssA = \{ T, S^i \}$, with kinetic term
\be
   \cL = - \sqrt{-g} \Bigl[ K_{\ssA \ol\ssB}(z) \, \partial_\mu Z^\ssA \, \partial^\mu \ol Z^{\ol\ssB} + V \Bigr] + \cL_m\,,
\ee
where $K_{\ssA\ol\ssB} = \partial_\ssA \partial_{\ol\ssB} K$ is the K\"ahler metric built using \pref{RelaxKModuli}, and $z^\ssA := Z^\ssA + \ol Z^{\ol\ssA}$. In this case the scalar field equation \pref{DilatonFE} generalizes to
\be \label{DilatonFEcosMF}
   \Box Z^\ssA + \Gamma^\ssA_{\ssB\ssC} \partial_\mu Z^\ssB \, \partial^\mu Z^\ssC  -   K^{\ol\ssB \ssA}\, \partial_{\ol\ssB} V +  \eta^{\ssA} \; g_{\mu\nu}  T^{\mu\nu} = 0  
\ee
where (as usual) $K^{\ol\ssB\ssA}$ and $\Gamma^\ssA_{\ssB\ssC} := K^{\ol\ssD\ssA} K_{\ssB\ssC\ol\ssD}$ are respectively the inverse metric and the Christoffel symbol built from the target-space K\"ahler metric $K_{\ssA\ol\ssB}$. The matter-coupling function appearing in \pref{DilatonFEcosMF} is defined just as for the single-field case, with
\be \label{DilCoupMF}
   \eta_{\ol\ssB} = \frac{ \partial_{\ol\ssB} A }{ A} = \frac{K_{\ol\ssB}}{6M_p^2}  \quad
   \hbox{and so} \quad
   \eta^\ssA = K^{\ol\ssB \ssA} \eta_{\ol\ssB} = \frac{K^{\ol\ssB\ssA} K_{\ol\ssB}}{6M_p^2}  \,.
\ee

The same arguments as given above also show that multiple-modulus models are QBD scalars, though in this case the Weyl rescaling factor, $A$, appearing in \pref{JFtoEF} becomes 
\be \label{AforModuli}
   A = e^{K/(6M_p^2)} = \frac{1}{\cF^{1/2}}  \,,
\ee
once \pref{RelaxKModuli} is used with $\cP$ truncated to leading order: $\cP = \cF(z+\bar z)$. The scale-invariance condition \pref{HomoDegOne} satisfied by $\cF$ turns out to make the coupling $\eta^\ssA$ surprisingly simple, since it implies the identity
\be \label{simpleDilatonCoupling}
    K^{\ol\ssB\ssA} K_{\ol\ssB} = - z^\ssA \,,
\ee
that follows by multiply differentiating the scaling identity $K(\lambda z) \equiv K(z) - 3 M_p^2 \ln \lambda$. This last result shows that the dilaton coupling function $\eta^\ssA = -  z^\ssA/(6M_p^2)$ is completely insensitive to the details of the choice of the K\"ahler potential for the moduli fields, assuming only that it is scale invariant (in the sense of \pref{HomoDegOne}).

In the absence of an axion mass or gauge-boson/axion mixing the axio-dilaton corresponds to the special case of a single field $Z^\ssA \to T$ with $K = - 3M_p^2 \ln \tau$. For this choice eq.~\pref{DilatonFEcosMF} reduces to
\be \label{DilEqMGeoTNR}
      \Box T  - \frac{2}{\tau} \,\partial_\mu T \, \partial^\mu T - \frac{\tau^2}{3M_p^2} \; V_\ssT - \frac{\tau}{6 M_p^2}\; g_{\mu\nu}  T^{\mu\nu}  = 0 \,,
\ee
whose imaginary part gives the axion evolution equation\footnote{This neglects any axion mass or gauge mixing, should this exist.}
\be \label{DilEqMGeoTNRi}
      \Box \mfa  - \frac{2}{\tau} \,\partial_\mu \tau \, \partial^\mu \mfa = 0 
\ee
and whose real part agrees with \pref{DilatonFE} once expressed in terms of the canonical field $\chi$.

\subsubsection{Solar system tests of gravity}
\label{ssed:DilatonPheno}

Modern precision tests of gravity -- be it within the solar system, using binary pulsars or from cosmology -- pose serious challenges for any theory with very light scalars (for a review see {\it e.g.}~\cite{TestsofGR}). Many of the strongest constraints, such as tests of the equivalence principle \cite{EPTests} or on variations of the fine-structure constant \cite{alphaTests}, are automatically evaded by theories in the QBD class,\footnote{Having all Standard Model masses and the QCD scale depend on the same power of $\tau$ is important for this statement, which is another feature generically satisfied by our model.} so the above demonstration that our model falls into this category (to leading order in $1/\tau$) makes these not worrisome (at least at leading orders in the PPN expansion).

With these limits evaded, the most restrictive bounds on light dilaton-like fields come from current-epoch tests of General Relativity within the Solar System \cite{Cassini} and binary pulsars\footnote{Gravitational wave emission by compact objects might eventually also provide competitive bounds, but do not do so yet \cite{GWBounds}.} \cite{Pulsars}.  (We return to the challenges and opportunities such a picture raises for concordance cosmology in \S\ref{sssec:DilatonCosmology} below.) Such constraints arise because eq.~\pref{DilatonFE} -- or, for multiple-field models, eqs.~\pref{DilatonFEcosMF} and \pref{simpleDilatonCoupling} -- ensures that matter acts as a source to the dilaton field and so $g_{\mu\nu}$ and $\tilde g_{\mu\nu}$ differ from one another in a predictable way, even in the vacuum away from the gravitating masses themselves. Since matter particles move (in the eikonal limit) along geodesics of $\tilde g_{\mu\nu}$ rather than $g_{\mu\nu}$, their motion reveals both the discrepancies between these two metrics and the change in $g_{\mu\nu}$ due to the dilaton stress energy (see \cite{ADScreening} for more details on how this works in this particular model). 

Current measurements -- {\it e.g.}~using the Cassini probe\footnote{Although solar system tests are currently the most restrictive, pulsar measurements are likely to become competitive in the near future \cite{DoublePulsar}.} \cite{Cassini}  -- agree with the predictions of general relativity, and this leads to the following constraint on the parameterized post-Newtonian (PPN) parameter $\gamma_\PPN$ \cite{PPNDefs, TestsofGR}:
\be \label{CassiniBound}
   \gamma_\PPN - 1 < 1.5 \times 10^{-5}  \,.
\ee
PPN effects also cause gravitational and inertial masses to differ from one another, leading to a slightly weaker lunar-laser-ranging constraint on the Nordvedt parameter
\be \label{Nordtvedt}
 |\eta_\ssN | \lsim 5 \times 10^{-5} \quad \hbox{with} \quad \eta_\ssN = 4 \beta_\PPN - \gamma_\PPN - 3 \,.
\ee
Comparing these with the Brans-Dicke result
\be \label{PPNBD}
 \gamma_\PPN = \frac{1-2\mfg^2}{1+2\mfg^2} = \frac{\omega + 1}{\omega + 2}
   \quad \hbox{and} \quad
   \beta_\PPN = 1
\ee
shows that \pref{CassiniBound} is inconsistent (by several orders of magnitude) with the prediction \pref{EtaVsKT}. 

How might these bounds be avoided? Evidently not by changing the size of the predicted coupling, \pref{EtaVsKT}, since the above arguments show this only relies on the no-scale structure that underpins the suppression of the cosmological constant itself. A different approach to evading solar-system bounds does not try to suppress all of the scalar couplings to matter, but instead uses the nonlinearities of the scalar self-interactions -- that is, the $\Gamma^\ssA_{\ssB\ssC}$ term in the equations of motion \pref{DilatonFEcosMF}  (along the lines used in other contexts in \cite{TargetSpace})  -- to channel any scalar fields that are produced into directions that do not contribute much to the Weyl factor $A$ and so are suppressed in their influence on the motion of test particles. One seeks to evade the constraint \pref{CassiniBound} by changing the prediction \pref{PPNBD} instead of the prediction \pref{EtaVsKT}.

\subsubsection*{A new screening mechanism}

The couplings of the axion to the dilaton turn out to provide (for free) a simple and novel example of this type \cite{ADScreening}. The K\"ahler metric $K_{T\ol T} = - 3M_p^2/\tau^2$ implies a target-space curvature whose effects can channel field development away from the dilaton and towards the axion despite the dilaton having the dominant coupling to matter. This changes the prediction \pref{PPNBD} because test-particle motion can be insensitive to the axion because it does not appear at all in $A$. 

The details of how this works can be found in \cite{ADScreening}, which solves the coupled axion-dilaton equations \pref{DilEqMGeoTNR} supplemented by a small direct axion-matter coupling 
\be\label{cJdef}
  \sqrt{-g} \; \cJ = 2 \left( \frac{\delta S_m}{\delta \mfa} \right) \neq 0 \,. 
\ee
It is useful when making estimates to assume the axion source profile to be proportional to the energy density, $\cJ(x) = \varepsilon_a \, \rho(x)$, with $\varepsilon_a$ constant and small. There are several reasons to expect $\varepsilon_a \ll 1$ to be small. The microscopic axion-matter coupling might itself be small. Or it might be small because the parity properties of the axion cause it to couple to parity-odd quantities -- such as local spin density -- that do not add up coherently for macroscopic objects (unlike for energy). Or both these reasons could apply at the same time. 

In weak gravitational fields\footnote{For solutions in strong fields see \cite{Burgess:1994kq}.} the equations exterior to a spherically symmetric source turn out to have the following simple general solutions:
\be \label{avsrexact}
    \mfa(r) =  
    \alpha - \beta \tanh X \quad \hbox{with} \quad X(r) :=  \frac{\beta\gamma}{r} + \delta 
\ee
and
\be \label{tauvsrexact}
   \tau(r) = \frac{\beta}{\cosh X(r)} \,,
\ee
with four integration constants $\alpha$, $\beta$, $\gamma$ and $\delta$. These solutions descibe motion along geodesics of the target-space metric that turn out to be semicircles in the $\tau$--$\mfa$ plane since they imply
\be \label{Semicircle2}
   \tau^2  +  (\mfa - \alpha)^2  = \beta^2\,.
\ee

The four integration constants are determined by the values of the two fields at spatial infinity,
\be \label{Asymptoticatau}
   \mfa_\infty = \alpha - \beta \tanh \delta \quad\hbox{and}\quad
   \tau_\infty = \frac{\beta}{\cosh \delta} \,,
\ee
and two first integrals of the field equations that relate the derivatives of the solutions to integrals of the source energy density $\rho$ and axion source density $\cJ$ (defined in \pref{cJdef}):
\be \label{Aconsradint2}
   \gamma = R^2 \left( \frac{\mfa'}{\tau^2} \right)_{r=R} = - \frac{1}{3M_p^2}\int_0^R \exd r \; r^2 \cJ(r) \simeq -\frac{2}{3}\varepsilon_a GM \,,
\ee
and
\be \label{Sconsradint2}
   \gamma \alpha = R^2 \left( \frac{\tau'}{\tau} + \frac{\mfa \, \mfa'}{\tau^2}\right)_{r=R} = - \frac{1}{3M_p^2} \int_0^R \exd r\; r^2 \Bigl[\rho(r) + \mfa(r) \,\cJ(r) \Bigr] \simeq \frac23\, GM\,,
\ee
where $R$ is the radius of the gravitating source and we identify its mass as $M = 4\pi \int_0^R \exd r\, r^2 \rho$. 

These boundary conditions show that $\beta$ and $\delta$ are dictated by the values of the fields at infinity, while $\gamma$ and $\alpha$ are dominantly controlled by the properties of the source. Clearly $\gamma$ in particular can be made as small as desired by making the axion-matter couplings smaller, but the product $\gamma \alpha$ is held fixed if $\rho$ remains unchanged. This limit $\gamma \to 0$ corresonds to the semicircle degenerating into a vertical line in the $\mfa$--$\tau$ plane. The screening mechanism is established by showing that {\it e.g.}~solar-system observables are all suppressed by $\gamma$ (and sometimes by $\delta$) without compensating factors of $\alpha$.   

\subsubsection*{Post-Newtonian analysis}

To compute the sizes of the PPN parameters the key observation is that test particles built from ordinary matter move along geodesics of the Jordan-frame metric (at least in the absence of other -- {\it e.g.}~electromagnetic -- forces). They do so because it is this metric that appears in their kinetic terms and so is the one to which standard arguments apply for quantum motion in the eikonal limit. 

Consequently PPN parameters measure how the Jordan-frame metric, $\tilde g_{\mu\nu} = A^2 g_{\mu\nu}$, deviates from the metric predicted by GR. There are two sources for this deviation: ($i$) the position dependence within the Weyl factor $A$, and ($ii$) deviations of the Einstein-frame metric $g_{\mu\nu}$ from the predictions of GR (that arise becaues of the stress energy of the light scalar fields). Both of these quantities are suppressed by the axion coupling. 

To see why consider first the Weyl factor, 
\be
   A^2 = e^{K/(3M_p^2)} = \frac{1}{\tau} \simeq \frac{1}{\beta} \cosh \left(-  \frac{2\beta\varepsilon_a GM}{3r} + \delta \right) \simeq \frac{1}{\tau_\infty} -     \frac{2\varepsilon_a \sinh \delta}{3} \left(\frac{GM}{r} \right)+ \cdots \,,
\ee
which shows that its position dependence is suppressed by $\epsilon_a$ (and by $\delta$ when this is small). The deviation of $g_{\mu\nu}$ from the predictions of GR are controlled by the scalar stress energy, which for the above solutions contribute only to the $(rr)$-component of the Einstein-frame Ricci curvature, evaluating to \cite{ADScreening}
\be \label{TTEinstrrMAD}
   \cR_{rr} =  - \frac{3}{4} \left[ \frac{(\tau')^2 + (a')^2}{\tau^2} \right] 
   = - \frac{3\gamma^2 \beta^2}{4r^4} = - \frac{\varepsilon_a^2 \beta^2}{3} \left( \frac{GM}{r^2} \right)^2  \,.
\ee
Notice that this expression uses \pref{Semicircle2} but does {\it not} expand in powers of $GM/r$. Eq.~\pref{TTEinstrrMAD} shows that the scalar field stress energy is also suppressed by the axion coupling $\varepsilon_a$.

Because both kinds of deviations from GR come suppressed by axion couplings there is no surprise that the PPN parameters that result are also suppressed. For instance the values for the PPN parameters $\gamma_\PPN$ and $\beta_\PPN$ found in \cite{ADScreening} are
\be
  \gamma_\PPN - 1 = - \frac{2\varepsilon_a \beta \tanh\delta}{3+\varepsilon_a \beta \tanh\delta}  \quad\hbox{and}\quad
  \beta_\PPN - 1 =  \frac{\varepsilon_a^2 \beta^2}{9(\cosh\delta+\frac13 \varepsilon_a \beta \sinh\delta)^2} \,.
\ee
The main point is that $\varepsilon_a$ (and so $\gamma$) is controlled by axion-source strength $\cJ$ through \pref{Aconsradint2} and so can be made arbitrarily small simply by coupling the axion more and more weakly to matter (see \cite{ADScreening} for details). Any direct non-metric effects of the axion on particle motion also vanish in this limit.

One might worry about other, non PPN, effects that could constrain these models. These can be explored by integrating out the structure of the source (be it the Earth, Moon or the Sun) and examining the EFT for its centre-of-mass motion. At lowest order in derivatives this becomes:
\be
   S_{\rm pp} = - \int \exd s \sqrt{\tilde g_{\mu\nu} \dot x^\mu \dot x^\nu} \Bigl( \tilde m + \cdots \Bigr) = - \int \exd s \sqrt{g_{\mu\nu} \dot x^\mu \dot x^\nu} \Bigl( A \tilde m + \cdots \Bigr) \,,
\ee 
where $\tilde m$ is the Jordan frame inertial mass, in terms of which the Einstein-frame mass is $m = \tilde m A \simeq \tilde m/\sqrt\tau$. To very good approximation the Jordan-frame mass, $\tilde m$, is independent of the scalar fields $\tau$ and $\mfa$,  and using this when varying $S_{\rm pp}$ with respect to the source position $x^\mu$ is one way to see why the source moves along geodesics of $\tilde g_{\mu\nu}$.

But $\tilde m$ {\it does} acquire scalar-field dependence once gravitational self-energy is included, because this contributes $\delta \tilde m/\tilde m \propto GM/R$. Although the ratio of different particle masses, $\tilde m_1/\tilde m_2 = m_1/m_2$ remains field-independent (and this is why {\it e.g.}~electromagnetic binding energies do not introduce scalar-field dependence), the ratio $M/M_p$ does depend on scalar fields (in both Jordan frame and Einstein frame). This might be expected to cause objects to deviate from Jordan-frame geodesic motion, and one might worry that this effect applied to the Earth and the Moon could cause effects in the very precise lunar-laser-ranging experiments at the $GM_\oplus/R_\oplus \sim 10^{-9}$ level (which would be ruled out). 

The effects of having field-dependence in $\tilde m$ is to alter the geodesic equation to include a new term $\delta (\ddot x^\mu) \propto \partial_\alpha \tilde m \, \tilde g^{\mu\nu} \partial_\nu \phi^\alpha$, which is therefore proportional to local gradients of the scalar fields. However eqs.~\pref{avsrexact} (applied to the field of the Sun) show that all scalar gradients are proportional to 
\be
   \frac{\partial X}{\partial r} = \frac{\beta \gamma}{r^2} = - \frac{ 2\beta \varepsilon_a}3 \left( \frac{GM}{r^2} \right) 
\ee
and so is again suppressed by the strength of the axion-matter coupling. 

We regard evading constraints on Brans-Dicke scalars as one of the main model-building challenges for our proposal, and the fact that an evasion mechanism comes pre-paid is a welcome surprise, particularly given that the other known screening mechanisms -- such as those in \cite{Khoury:2003aq} (for a review see also \cite{Burrage:2017qrf}) -- do not seem to work for our set-up.\footnote{They do not do so because they typically seek matter-dependent and vacuum contributions to the scalar dependence that push the light field in opposite directions, thereby creating a new matter-dependent minimum about which long-range forces are suppressed. We have been unable to get these to work for Yoga models -- and string models more generally (see however \cite{Hinterbichler:2010wu, Nastase:2013ik}) -- because the underlying scale invariance makes both matter and vacuum contributions always consistent with the runaway to the scale-invariant limit $\tau \to \infty$.}   Moving forward we seek to know how robust and widespread are such mechanisms, but their presence makes constraints coming from cosmology all the more interesting. See also \cite{Cicoli:2012tz,Acharya:2019pas,Cicoli:2020noz} for related discussions.

\subsection{Other issues}
\label{ssec:OtherIssues}

This section closes with a brief summary of some of the other phenomenological issues for this model.

\subsubsection*{Axion constraints}

The phenomenology of the axion is largely shaped by the prediction \pref{faprediction} of an unusually small decay constant $f_a \sim {M_p}/{\tau} \sim 10 \, \hbox{eV}$. Depending on the model it may be one of: massless; essentially massless (if axion shift-symmetry anomalies allow $W \propto e^{-bT}$ so that $m_a\sim e^{-b\tau}$); very light (with $m_a \sim M_p/\tau^{3/2} \leq 10^{-12}$ eV); or of order $m_a\sim M_p/\tau \sim 10$ eV.  

We take the point of view that the new physics that intervenes at eV scales is extra-dimensional, in which case the physics above these scales is the physics of an extra-dimensional Kalb-Ramond field, $B_{\ssM\ssN}$, of which the dual two-form gauge potential $b_{\mu\nu}$ given by $\partial_\mu \mfa \propto \epsilon_{\mu\nu\lambda\rho} \partial^\nu b^{\lambda \rho}$ is a single KK mode. In this case the phenomenology above eV energies is the phenomenology of supersymmetric large extra dimensions, and is dominated by missing energy constraints from astrophysical and accelerator observations \cite{Atwood:2000au, MSLED}.

At energies below eV scales the phenomenology is more along the lines for a traditional axion.
In any case, the observational limits are very mild for the axion-mass options considered here (for recent reviews on axions, including a comprehensive discussion of bounds, see for instance \cite{Marsh:2015xka, Hook:2018dlk, Choi:2020rgn}). A massless (or essentially massless) axion can contribute to the Dark Energy story in an interesting way by complicating how the dilaton evolves at very late times in cosmology (see \S\ref{sssec:DilatonCosmology} for a preliminary discussion). In the very light case it might fit into the black hole superradiance regime for some types of black holes. Axion-matter couplings seem important to help the dilaton escape solar-system tests \cite{ADScreening}, and play a similar role for the viability of dilaton cosmology. But if axion-matter couplings lie in a particular range (not to strong -- to avoid thermalization -- and not too weak) they can also cause problems by providing too efficient an energy-loss channel for red-giant stars and supernovae. But none of these constraints seem particularly dangerous for this model. 

\subsubsection*{Gravitino and dilatino constraints}

Similar remarks apply to the gravitino/dilatino system. Eq.~\pref{gravitinomassF} shows that the gravitino in this scenario is very light: $m_{3/2}\sim m_\TEV^2/M_p \sim 10^{-5}$ eV. \S\ref{sssec:DarkFermions} argues that the dilatino has a similar mass.  These are too light for these particles to be proper dark matter candidates, even though they can be considered essentially as stable (since  gravitational-strength couplings imply a decay time of order $M_p^2/m_{3/2}^3$). But, precisely because they are so light (with masses $\ll 1 $ eV) they are also cosmologically harmless since there is no danger for them to over-close the universe (see for instance \cite{Feng:2010gw, Kawasaki:2008qe} and references therein).

Collider bounds provide a general constraint on gravitino/goldstino production \cite{Brignole:1998me, Brignole:1997sk}, since these are constrained by searches for producing invisible final states in coincidence with a photon (say). At face value the non-observation of these gravitino-production reactions at colliders preclude a gravitino with mass smaller than $10^{-5}$ eV (see also \cite{Hebecker:2019csg}). It is the equivalence theorem that makes these such strong bounds, because although the gravitino is only gravitationally coupled, at energies high compared with $m_{3/2}$ longitudinal polarizations dominate gravitino production and the production of these is well-approximated by the production of goldstinos as predicted in the $M_p \to \infty$ limit. We sketch this argument in Appendix \ref{AppC}.

The unwarped models described here evade these bounds, precisely because $F^\ssX$ as defined in \pref{FXEF} has been designed to be large -- see for example eq.~\pref{muXest}. However, these bounds are at first sight a concern for the warped models described in \S\ref{sssec:Warping}, for which the gravitino mass prediction \pref{m32warped} can be much smaller (if $b > a)$. Because this bound is really a  constraint on the goldstino emission using the 4-dimensional goldstino emission rate as described in Appendix \ref{AppC}, it is really a constraint on the scale $\mfF := m_{3/2} M_p$ that controls goldstino emission at low energies. But the prediction for the emission rate as a function of $\mfF$ breaks down for energies $E \gsim \sqrt\mfF$, and $\mfF$ can be as low as $(10 \; \hbox{eV})^2$ in the scenarios described in \S\ref{sssec:Warping}. In the warped models the 4D goldstino EFT description breaks down at these low energies, for much the same reasons as for the axion-production rate at the same scales. The production rate must be computed within the UV completion that kicks in at these scales  (such as is done in \cite{Atwood:2000au, MSLED} if the UV completion consists of supersymmetric extra dimensions) in order to be compared with collider experiments whose energies are at TeV scales.

It has recently been argued that models with a light gravitino can undergo catastrophic gravitino production \cite{Kolb:2021xfn}, although subsequent studies argue that when supersymmetry is linearly realized such production does not arise once goldstino alignment within the light fermion spectrum is carefully tracked. But this leaves open the possibility that it can arise\footnote{Even if it were to occur it would signal the breakdown of the nonlinearly realized EFT rather than physical gravitino emission.} in nonlinearly realized models (like the framework used here) for which constrained fields like $\phi$ exist and evolve nontrivially with time \cite{Dudas:2021njv, Antoniadis:2021jtg}. Although this can be an issue for cosmological applications it is not an issue here because the field $\phi$ simply sits at its local minimum while other fields evolve. It is also typically not expected to be an issue for inflationary applications for which fields like $\phi$ only roll very slowly.

\subsubsection*{Relaxon constraints}

The relaxon also raises some issues, though we have found nothing fatal. A key property of the relaxon is that its nonzero {\it vev} shifts to cancel out part of the vacuum energy. But it was also true that $\phi$ had to carry a conserved charge (either for the discrete symmetry $\phi \to - \phi$ for the real $\phi$ considered in the main text, or a rephasing symmetry $\phi \to e^{i\theta} \phi$ if $\phi$ were complex) in order to forbid the appearance of a term $M \Phi X \in W$ where $M \sim M_p$ is the UV scale. Although such a term also gives a mass to $\phi$ that is of order $m_\phi \sim M/\sqrt\tau$, in this case radiative corrections to the mass are not controlled by corrections to a dimensionless coupling $g$ and so might destabilize the choices leading to $m_\phi < m_e$. Naturalness of the small $\phi$ mass seems to require such a symmetry.

Either type of symmetry can cause phenomenological trouble once  spontaneously broken, with $\phi \to -\phi$ potentially leading to the formation of domain walls in the early universe, and $\phi  \to e^{i\theta} \phi$ leading to a massless Goldstone boson in the continuous case. Neither of these need be deadly, since (for example) domain walls can be inflated away and Goldstone bosons need not couple significantly to observable matter. The side-effects of radiative stability of $\phi$ properties seem to be among the model's less attractive features.

Since the $\phi$ mass is not too far below the electron mass it is high enough that the effective range of any $\phi$-mediated force acts only over submicron scales, making macroscopic forces due to $\phi$-exchange not a worry for tests of gravity. It could have particle physics implications if coupled to Standard Model particles, though nothing requires these to be a level that would have been detected. 

The most model-independent coupling between $\phi$ and Standard Model particles is to the $\phi^2 \, \cH^\dagger \cH$ coupling to the Higgs in the $|w_\ssX|^2/\tau^2$ part of the scalar potential. This would be generic if $w_\ssX$ were a linear combination of both $w_\ssX \ni g(\Phi^2 - v^2)$ and $w_\ssX \ni g_\ssH (H^\dagger H - v_h^2)$ terms, and if present would contribute to the decay width of the Higgs boson into unseen decay daughters (for which the branching ratio cannot be larger than 24\% without conflicting with observations \cite{PDG}). 

Besides being generic, a $\phi^2 H^\dagger H$ coupling arising in $w_\ssX$ is also likely to be unsuppressed by powers of $1/\tau$ given that both the $\phi$ and Higgs fluctuation fields have kinetic terms that are themselves suppressed by $1/\tau$. This is precisely the argument used in \S\ref{ssec:MatterCouplings} (see also \cite{LowESugra}) for the quartic Higgs self-interactions themselves. The $\phi^2 H^\dagger H$ coupling is, however, suppressed by $g g_\ssH$, where $g$ is the small $\phi$ self-coupling that is argued above -- see the discussion below eq.~\pref{gvvalues} -- to be of size $g \sim y_e^2$, where $y_e$ is the small electron Yukawa coupling.

\section{Axio-dilaton cosmology}
\label{sssec:DilatonCosmology}

We turn now to a preliminary discussion of cosmological issues, and work through the good exercise of comparing the gross features of yoga models with $\Lambda$CDM cosmology.  (A full discussion of cosmology, including fluctuations, goes beyond the scope of this paper.) An important difference between this and previous sections is that  the scalar potential cannot be neglected (as opposed, say, to solar-system tests of gravity). Indeed, the framework we propose predicts that dilaton evolution in the presence of this potential provides a specific type of quintessence model for Dark Energy, but (unusually) does so with the small masses and potential energies involved seeming to arise in a natural way. 

The field equations describing cosmology obtained by varying the action built from $\cL_{{\rm dil}\,b} + \cL_\SM$ as given in \pref{LdilbForm} are  
\be \label{TTEinstcos}
  \cR_{\mu\nu} + e^{-2\zeta \hat\chi} \,\partial_\mu \mfa\, \partial_\nu\mfa+  \partial_\mu \hat\chi \, \partial_\nu \hat\chi+ \frac{1}{M_p^2}\left[   V(\hat\chi)\,  g_{\mu\nu} +  T_{\mu\nu} - \frac12 \, g^{\lambda\rho} T_{\lambda\rho} \, g_{\mu\nu} \right] = 0 \,,
\ee
\be \label{DilatonFEcos}
   \Box \hat\chi + \zeta \, e^{-2\zeta \hat\chi}   \, \partial_\mu \mfa \, \partial^\mu \mfa + \frac{1}{M_p^2} \left( - \frac{\partial V}{\partial \hat\chi} + \mfg\, g_{\mu\nu}  T^{\mu\nu}\right) = 0  
\ee
and
\be \label{AxionFEcos}
  \partial_\mu \Bigl[ \sqrt{-g} \; e^{-2\zeta \hat\chi} \, \partial^\mu \mfa \Bigr]   = - \frac{\sqrt{-g} \, \cJ}{3M_p^2}  
\ee
where $\hat\chi := \chi/M_p$ and $\mfg = - \frac12 \, \zeta \simeq -0.41$ and we assume $\mfa$ is massless.

For cosmological applications we assume the axion source $\cJ$ --- defined in \pref{cJdef} --- is proportional to the number density of baryonic matter,\footnote{More precisely, we assume for simplicity here that $a^3 \cJ$ is independent of both $\chi$ and cosmic time, where $a(t)$ is the cosmic scale factor. } and so satisfies $\cJ = \mfg_a n_\ssB$ with a constant effective axion-matter coupling $\mfg_a$ that is small in the sense that $|\cJ| \ll \rho_m$. Having $\cJ$ smaller than $\rho_m$ seems natural for two reasons: the density of baryons can be smaller than the dark matter density, and the parity-odd nature of the axion usually keeps microscopic contributions to $\cJ$ from summing coherently within macroscopic bodies (unlike the energy density).
 
For homogeneous solutions in a spatially flat FRW spacetime, $g_{\mu\nu} \exd x^\mu \exd x^\nu = - \exd t^2 + a^2(t) \,\exd \bfx^2$, eq.~\pref{AxionFEcos} becomes
\be \label{axioncosmo}
    \left( \ddot \mfa + 3H \dot \mfa - \frac{2\zeta}{M_p^2} \dot{\hat\chi} \dot \mfa \right)   e^{-2\zeta \hat\chi}   - \frac{\mfg_a  n_\ssB}{3M_p^2}=0\,.
\ee
 Eq.~\pref{TTEinstcos} similarly boils down to the Friedmann equation
\be \label{Friedmann}
   H^2 = \frac{\rho}{3M_p^2} = \frac{1}{3M_p^2} \left\{ \rho_f + \rho_a + \frac{ \dot\chi^2}{2}  + V\right\} \,,
\ee
Here the subscript `$f$' denotes the contribution of the cosmological fluid, whose energy density and pressure
\be\label{matterprho}
  \rho_{f} = \rho_m(\chi) + \rho_{\rm rad} \quad \hbox{and} \quad
  p_{f} = p_{\rm rad} = \frac13 \, \rho_{\rm rad} \,,
\ee
contain the contributions from of ordinary radiation and nonrelativistic (ordinary and dark) matter. Finally, the dilaton equation \pref{DilatonFEcos} reduces to
\be \label{CosmologyEqs}
   \ddot{\hat\chi} + 3H \dot{\hat\chi} + \zeta \, e^{-2\zeta \chi/M_p}   \, \dot\mfa^2  +\frac{1}{M_p^2} \Bigl[ V'(\hat\chi) + \mfg \, \rho_{m}(\hat\chi) \Bigr]  = 0 
\ee
with prime denoting differentiation with respect to $\hat\chi$ and the scalar potential given by \pref{VvsUExp}
\be \label{VUlambda}
   V = U \, e^{-\lambda \hat\chi} \,, 
\ee
where $\lambda = 4\zeta = 4\sqrt{\frac{2}{3}} \simeq 3.3$ is a known dimensionless parameter and $U$ is the polynomial or rational function of $\hat\chi$ used to stabilize $\tau$ in \S\ref{ssec:DilationStabilization}.

Eq.~\pref{matterprho} deliberately emphasizes how the Einstein-frame matter density implicitly depends on $\chi$, since this is important when computing how it evolves. To identify how $\rho_m$ depends on $\chi$, start from the observation that the total number of nonrelativistic particles is given in Einstein frame by $\sqrt{-g} \; n$, implying that the Einstein frame particle density $n \propto a^{-3}$ falls inversely proportional to the Einstein-frame volume as the universe expands. Writing $\rho_m = m(\hat\chi) \, n$ with particle mass $m(\hat\chi) = \mfm \, A(\hat\chi) \propto \tau^{-1/2} = e^{-\frac12\,\zeta\, \hat\chi}$, then implies that the Einstein-frame energy density for nonrelativistic matter evolves as $\rho_m \propto A(\hat\chi)/a^3$ as the universe expands. The factor of $A(\hat\chi)$ can be important during epochs where $\hat\chi$ evolves.

\subsection{Dilaton dark energy}

Can the above system describe Dark Energy as revealed to us by observations? We here first explore `dilaton-only' cosmologies for which the axion-matter coupling vanishes: $\cJ = 0$, in which case the trivial solution with constant $\mfa$ satisfies the axion equation \pref{axioncosmo}. We return to more complicated axionic evolution below. 

Viable dilaton cosmologies require the present-day scalar energy density to be positive and to make up around 75\% of the total, with an equation of state $w = p_\chi/\rho_\chi \lsim - 0.9$ \cite{PlanckDE}. It  also must not ruin the other successes of $\Lambda$CDM, and so (for example) $\rho_\chi$ cannot be more than a few percent of the total energy density at nucleosynthesis \cite{BBNRad}. Fluctuations about such configurations must also be consistent with CMB measurements and what is known about its consistency with large-scale structure (LSS).

There are also constraints on how far the field $\chi$ itself can roll because changes to $A(\chi)$ correspond to changes in particle masses (relative to the Planck mass, which in Einstein frame is fixed). Particle masses cannot vary by more than a few percent between nucleosynthesis and now, without ruining otherwise successful standard predictions \cite{Alvey:2019ctk}. They also cannot vary too much in the relatively recent universe (redshifts $z \lsim 7$) without unacceptably changing the properties of spectral lines observed once stars and quasars turn on \cite{PlanckDE, Mould:2014iga}. In principle particle masses are also constrained not to be too different at recombination, though (intriguingly) a few percent change in particles masses at that time changes the precise epoch of recombination in a way that can help resolve \cite{Sekiguchi:2020teg} current tensions \cite{HubbleTension} between CMB and local measurements of the present-day Hubble scale.

We are able to satisfy these constraints, but only if the factor $U$ appearing in the scalar potential allows the existence of a local minimum, at a position we denote $\chi = \chi_0$. (\S\ref{ssec:DilationStabilization} shows that this is relatively easy to arrange and can be done for acceptably large $\tau_0 = e^{\zeta  \hat\chi_0}$ using only parameters that are no larger than around 60.) Successful cosmology requires some structure in $U$ because there are several things that go wrong when $U = U_0$ is constant. First, if the scalar energy density were ever to dominate (as Dark Energy now does) then it must be slowing rolling, with $w+1 \simeq \dot \chi^2/V \lsim 0.1$ in order to be sufficiently close to $w = -1$. But slow-roll solutions with $V = U_0 e^{-\lambda \hat\chi}$ satisfy $3H \dot{\hat\chi} \simeq -V'/M_p^2 \simeq \lambda V/M_p^2$ and $3M_p^2 H^2 \simeq V$ and so would predict 
\be \label{quintessencelambda}
  \frac{\dot \chi^2}{V} = \frac{\dot{\hat\chi}^2 M_p^2}{V} \simeq \frac{(V'/3H)^2}{M_p^2V} \simeq \frac{\lambda^2}3 \simeq 3.6 \,,
\ee 
which is clearly too large. 

Alternatively, if the scalar field does not initially dominate (as must have been true in the past, such as during the recombination or nucleosynthesis epochs) then with $V = U_0 e^{-\lambda \hat\chi}$ it turns out never to dominate at later times. It does not do so (at least when $\lambda \simeq 3.3$) because it robustly gets drawn into a late-time attractor scaling solution of the form 
\be
  \chi = \chi_0 + \frac{n}{\lambda} \ln \left( \frac{a}{a_0} \right) \,,
\ee
where the dominant energy density is assumed to scale with universal expansion like $\rho \propto a^{-n}$ (such as $n=4$ for radiation or $n=3$ for matter domination). In this solution the scalar energy density scales as a fixed fraction of the dominant energy density \cite{Tracker}. More precisely, if
$\lambda^2 > n$ the tracker solution very robustly predicts the late-time scalar energy fraction to take a time-independent value: $\rho_\chi/\rho = n/\lambda^2$ (as we have also verified numerically). Since for the case of interest $\lambda^2 \simeq 10.9$ this applies in both radiation and matter dominated universes, implying in these cases $\rho_\chi/\rho_{\rm rad} \simeq 0.37$ or $\rho_\chi/\rho_m \simeq 0.28$. Both of these are much too small to describe Dark Energy.

\subsubsection*{Dilaton cosmologies with more potential}

\begin{figure}[t]
\begin{center}
\includegraphics[width=100mm,height=50mm]{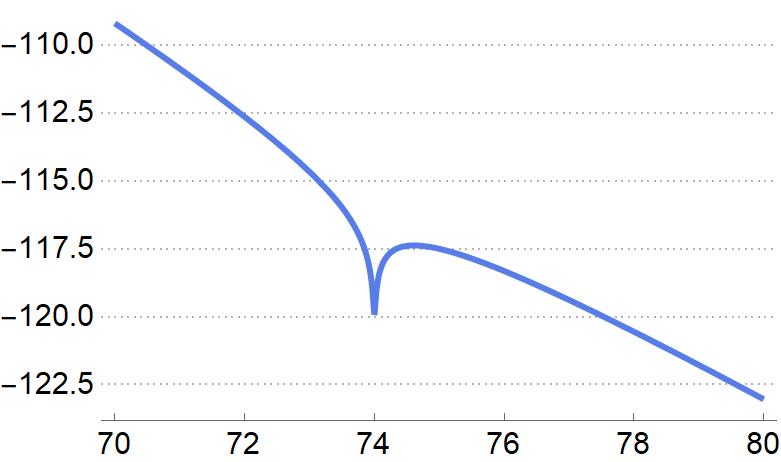} 
\caption{Log (base 10) of the scalar potential {\it vs} the canonical field $\chi$ in Planck units. This plot uses $u_1 = 0.027027$, $u_2 = 0.00036523$ and $U_0 = \epsilon^5$ with $\epsilon = 1/30$. \label{Fig:Vplot} }
\end{center}
\end{figure}

Things work much better when $U(\chi)$ allows a minimum at $\chi = \chi_0$ provided the late-time evolution becomes dominated by the fixed vacuum energy $V(\hat\chi_0)$, because in this case the scalar sector can behave at late times like a cosmological constant. For concreteness' sake we use the potential $V = U \, e^{-\lambda \hat\chi}$ with
\be\label{etaUchoices}
  U \simeq U_0 \left[1 - u_1 \hat\chi + \frac{u_2}2 \, \hat\chi^2 \right] \,, 
\ee
where earlier sections tell us $\lambda = 4 \zeta \simeq 3.3$ and $U_0 \sim \epsilon^5 M_p^4$ for $\epsilon \simeq 1/400$ implying present-day cosmology should occur for $\hat\chi \sim 75$ since then $\tau = e^{\zeta \hat\chi} \sim e^{60} \sim 10^{26}$ (as required in earlier sections) and $\epsilon^5 e^{-\lambda \hat \chi} \sim \epsilon^5 e^{-240} \sim 6 \times 10^{-105} \epsilon^5$. These choices are the minimal ones that allow $V$ to be positive for all $\hat\chi$ (which is true provided  $2u_2 \ge u_1^2$) and have a local minimum: $V'(\hat\chi_0) = 0$ with $V''(\hat\chi_0)$ positive (see Fig.~\ref{Fig:Vplot}), with
\be\label{chi0def}
    \hat \chi_0 = \frac{1}{\lambda} \left[ 1 + \frac{\lambda u_1}{u_2} - \sqrt{1 + \frac{\lambda^2 u_1^2}{u_2^2} - \frac{2\lambda^2}{u_2}} \right]
\ee
showing that $\hat\chi_0$ can be order 75 simply by requiring a mild hierarchy amongst the positive parameters $u_1$ and $u_2$. The root not shown in \pref{chi0def} is a local maximum, $\hat\chi_{\rm max} > \hat\chi_0$, that separates the minimum from the runaway as $\hat\chi \to \infty$.

In the absence of the axion and for sufficiently small $\mfg$ these choices give evolution that resembles the cosmic histories described in \cite{AndyCostasnMe}. Small $\mfg$ is required because when $\mfg <0$ (as in our case) the matter coupling in \pref{CosmologyEqs} relentlessly pushes the dilaton to larger values of $\chi$. When $\dot\mfa = 0$ the only force (besides Hubble friction) pushing in the opposite direction is the scalar potential $V'$ in the range $\hat\chi_0 < \hat\chi < \hat\chi_{\rm max}$ for which $V'(\hat\chi) > 0$. Trapping in the local minimum only happens when $V'$ successfully competes with the matter force, and so requires $|\mfg| \rho_m \lsim |V'|$, but because $V' \sim V(\hat\chi_0)$ in the range where $V' > 0$ this can only occur once $|\mfg|\rho_m$ falls below the present-day Dark Energy density, which only happens at very late times for $\mfg$ order unity.

\begin{figure}[t]
\begin{center}
\includegraphics[width=140mm,height=70mm]{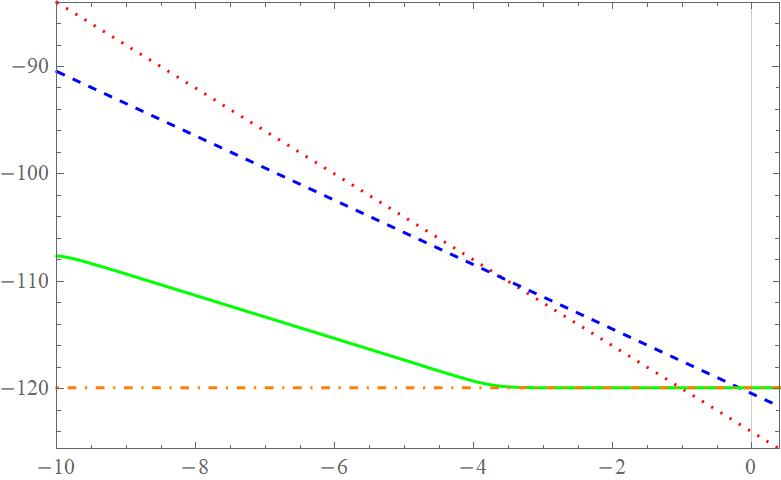} 
\caption{A log-log plot (base 10) of the energy density in matter (dashed), radiation (dotted) and the dilaton's total (solid) and potential (dot-dashed) energy densities as a function of universal scale factor, with $a = 1$ representing the present, with dilaton initially chosen moving near the potential's minimum. For illustrative purposes we choose no axion evolution and a small matter-dilaton coupling $|\mfg| \sim 10^{-5}$ (colour online).} \label{Fig:Eplot1} 
\end{center}
\end{figure}

\begin{figure}[t]
\begin{center}
\includegraphics[width=140mm,height=70mm]{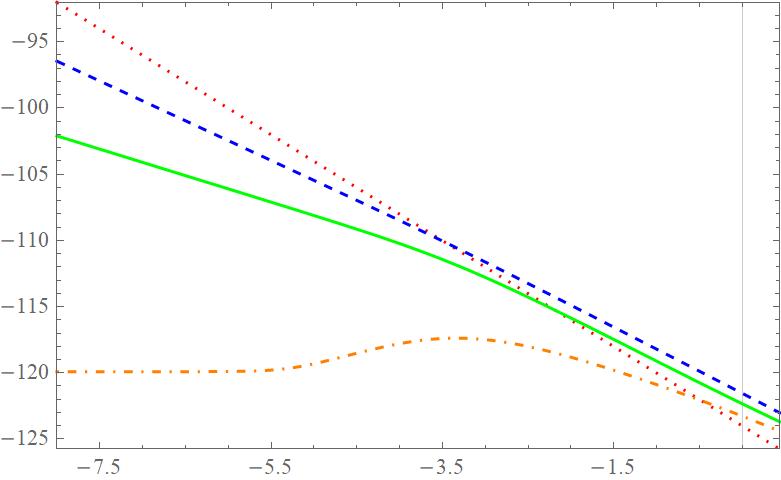} 
\caption{A log-log plot (base 10) of the energy density in matter (dashed), radiation (dotted) and the dilaton's total (solid) and potential (dash-dotted) energies as a function of universal scale factor, with $a = 1$ representing the present, with initial kinetic energy near BBN not small. We choose no axion evolution but use $\mfg = -0.408$ as predicted by our model. In this case the large dilaton-matter coupling sweeps the dilaton out to a runaway past the potential's minimum.} \label{Fig:Eplot2} 
\end{center}
\end{figure}

Fig.~\ref{Fig:Eplot1} plots the energies in radiation, matter and the scalar field for a sample evolution in which $|\mfg| \sim 10^{-5}$, showing how in these circumstances the dilaton becomes trapped at late times despite having initially fairly large kinetic energy. Fig.~\ref{Fig:Eplot2} shows how the evolution instead proceeds if $\mfg = - \frac12\zeta \simeq -0.408$ as predicted for this model in earlier sections. In this case the stronger matter coupling draws the dilaton evolution into an attractor scaling solution early in the matter-dominated epoch (though driven by the matter coupling rather than the potential, as described in \cite{AndyCostasnMe}) and the matter-dilaton coupling sweeps the dilaton over the barrier and out to the runaway at $\chi \to \infty$. This is a serious problem when constructing potentially viable cosmologies, but (similar to \cite{ADScreening}) a small axion-matter coupling can save the day in an interesting way. 

\subsection{Including axion evolution}
\label{sssec:AxioEvolution}

Keeping in mind the above experience with the dilaton, we now revisit the possibility that $\cJ \neq 0$, and so assume nonzero coupling $\mfg_a$. In this case the axion-matter interaction prevents the axion from remaining constant cosmologically, and (similar to what happened in \S\ref{ssed:DilatonPheno}) this can ameliorate the implications of large dilaton coupling. 

When $\mfg_a = 0$ the axion evolution equation \pref{axioncosmo} describes a conservation law stating
\be \label{Ldef}
   L := a^3 \; e^{-2\zeta \hat\chi} \, \dot \mfa 
\ee
is time independent. In the absence of dramatic $\chi$ evolution constant $L$  describes the draining of any initial axion kinetic energy
\be \label{rhoavsL}
   \rho_a = \frac{M_p^2}{2} \, e^{-2\zeta \hat\chi} \dot \mfa^2 = \frac{L^2M_p^2}{2a^6} \, e^{+2\zeta \hat\chi}
\ee
due to Hubble friction. Initially rolling axion fields stop and thereafter remain constant.

By contrast, when $\mfg_a \neq 0$ baryonic matter acts as a source for axion `charge' and \pref{axioncosmo} describes how this winds the axion up by increasing $L$ with constant rate 
\be
  \dot L = \frac{\mfg_a}{3M_p^2} \; a^3(t) n_\ssB(t) = \frac{\mfg_a n_{\ssB 0}}{3M_p^2}  \,,
\ee
where the combination $a^3 n_\ssB$ is time-independent (and is conveniently given by $n_{\ssB 0} = n_\ssB(t_{\rm now})$ where $t_{\rm now}$ is the present-day time, for which $a(t_{\rm now}) = 1$). It is natural to assume $L = 0$ when $t = 0$ and if so this integrates to give:
\be \label{Lvsa}
  L(a) =  \frac{\mfg_a  n_{\ssB 0}}{3M_p^2}\, t(a) \,,
\ee
where $t(a)$ is the cosmic time as a function of scale factor. 

For a universe sequentially dominated by radiation and matter the Hubble scale is given by $H^2(a) = \frac12 H_{\rm eq}^2[(a_{\rm eq}/a)^3 + (a_{\rm eq}/a)^4]$ where the radiation and matter densities are equal when $a = a_{\rm eq}$ and $H(a_{\rm eq}) = H_{\rm eq}$. This implies the expression for $t(a)$ to be used in \pref{Lvsa} is  
\be \label{tvsa}
  t(a) = \frac{2\sqrt2}{3 H_{\rm eq}} \left[ 2 + \left( \frac{a}{a_{\rm eq}}  + 1\right)^{1/2} \left( \frac{a}{a_{\rm eq}} - 2\right) \right] \,.
\ee
Notice that for $a \ll a_{\rm eq}$ this predicts that $L \propto t(a) \propto a^2$ during radiation domination but when $a \gg a_{\rm eq}$ it instead predicts $L  \propto t(a) \propto  a^{3/2}$ during matter domination. 

Using these scaling rules in eq.~\pref{rhoavsL} predicts (if $\chi$ does not roll dramatically) the axion kinetic energy scales like
\be
   \rho_a \propto \begin{cases} a^{-2} & \hbox{(radiation dominated)} \\ a^{-3} & \hbox{(matter dominated)} \end{cases}
\ee
showing how the baryon-driven wind-up of the axion makes the axion energy density grow relative to other components during radiation domination, though it remains a fixed fraction of the total density once the universe is matter dominated.

Cosmological evolution is governed by the Friedmann equation \pref{Friedmann} together with the dilaton evolution equation \pref{CosmologyEqs}, which after using \pref{Ldef} becomes
\be \label{CosmologyEqs2}
   \ddot {\hat\chi} + 3H \dot {\hat\chi} + \frac{1}{M_p^2} \left[ V'(\hat\chi) + \zeta \left( \frac{L^2M_p^2}{a^6} \,  e^{+2\zeta \hat\chi}  - \frac{\rho_{m0}}{2a^3} \, e^{-\frac12 \zeta( \hat\chi - \hat\chi_0)} \right)\right] = 0 
\ee
where (as before) primes denote differentiation with respect to $\hat \chi = \chi/M_p$ and we use the prediction $\mfg = - \frac12 \zeta \simeq -0.408$ with $\zeta = \sqrt{\frac23} \simeq 0.8165$. Notice that the $\chi$-dependence of the axion and matter couplings make the $\chi$ evolution appear as if it were moving in the presence of a time-dependent potential
\be\label{VeffDilaton}
  V_{\rm eff}(\hat\chi, a) := V + \frac{L^2(a)M_p^2}{2a^6} \,  e^{+2\zeta \hat\chi}  + \frac{\rho_{m0}}{a^3} \, e^{-\frac12 \zeta (\hat\chi-\hat\chi_0)} \,.
\ee
At late times this always approaches $V$ but the other two terms can dominate at early times when the matter and axion energy densities are larger than $V$. 

\subsubsection*{Viable axio-dilaton cosmologies}

For viable cosmologies we again seek solutions whose dilaton becomes trapped at late times at the minimum of the scalar potential, with $\chi = \chi_0$ and $\dot \chi_0$ small enough to preclude escaping over the local maximum and running off to infinity. Such solutions ensure that Dark Energy at late times resembles a cosmological constant. 

The possibilities for finding such solutions that can trap the dilaton into the potential's local minimum are much greater once the axion is included, because the axion term in \pref{VeffDilaton} provides a force that always pushes $\hat\chi$ towards smaller values. Intuition for why it does so can be found by solving the axio-dilaton evolution in the limit where the dilaton-matter and axion-matter couplings ($\mfg$ and $\mfg_a$) are negligible. In this case $L$ is to a good approximation conserved and the coupled axio-dilaton equations can be explicitly solved under the assumption that the Hubble scale is dominated by another energy source for which $\rho(a) \propto a^{-n}$ (such as radiation, say, for which $n = 4$). The solutions are derived very much along the lines pursued in \cite{ADScreening}, leading to
\be\label{semisolutions}
    \mfa =A - B \tanh X(t) \quad \hbox{and} \quad \tau = \frac{B}{\cosh X(t)} 
\ee
where $A$ and $B$ are integration constants and (as before) $\tau = e^{\zeta\hat\chi}$ while
\be
   \dot X(t) = - B L \left( \frac{a_0}{a} \right)^3 \,.
\ee
This last equation integrates to give
\be
  X(t) = X_0 - B L a^3_0 \int_{t_0}^t \frac{\exd u}{a^3(u)} = X_0+\frac{B L t_0}{3s-1} \left[1- \left( \frac{t_0}{t} \right)^{3s-1}  \right]
\ee
where $a(t) = a_0 (t/t_0)^s$ with $s = 2/n$. In particular, eqs.~\pref{semisolutions} imply these solutions trace out a semicircle in the upper-half $\tau$--$\mfa$ plane, given by
\be\label{circle}
    \tau^2 + \left(\mfa - A \right)^2 = B^2 \,
\ee 
and it is the $L$-dependent force pushing $\chi$ to smaller values that causes $\chi$ to accelerate as needed to remain on this semicircle.

Returning to viable cosmologies with nonzero $\mfg$ and $\mfg_a$: as an existence proof for solutions with $\chi = \chi_0$ and $\dot{\chi} \simeq 0$ at late times we ask when $\chi = \chi_0$ can be a time-independent solution to \pref{CosmologyEqs2}. Since $V'(\chi_0)  = 0$ such solutions exist when the other terms in $V'_{\rm eff}$ obtained from \pref{VeffDilaton} cancel:
\be \label{CosmologyEqsBal}
  \frac{L^2(a)}{a^6} \,  e^{+2\zeta \hat\chi_0}  - \frac{\rho_{m0}}{2a^3M_p^2}   = 0 \,.
\ee
Remarkably, this is a time-independent condition during a matter-dominated universe because in this case \pref{Lvsa} and \pref{tvsa} imply $L(a) \propto a^{3/2}$. Eq.~\pref{CosmologyEqsBal} requires that the present-day value $L(t_0) = L_0$ should satisfy 
\be
   L_0^2 = \frac{\rho_{m0}}{2M_p^2} \, e^{-2 \zeta \hat \chi_0} = \frac{\rho_{m0}}{2M_p^2 \tau_0^{2}} \,,
\ee
and so $L_0^2 \lll \rho_{m0}/M_p^2$ and $\rho_a(t_0) \lll \rho_{m0}$. 

With these choices eqs.~\pref{Lvsa} and \pref{tvsa} imply $L_{\rm eq} = L(a_{\rm eq})$ at radiation-matter equality is of order 
\be
  L_{\rm eq} \simeq L_0 \left( \frac{a_{\rm eq}}{a_0} \right)^{3/2} \quad\hbox{and so} \quad
  \frac{\rho_a(a_{\rm eq})}{\rho_m(a_{\rm eq})} \simeq  \frac{\rho_a(a_{0})}{\rho_{m0}} \lll 1
\ee
because the evolution of $\chi$ plays no role in the evolution of $\rho_m$ when $\chi$ is fixed at $\chi_0$. Pushing back further, into the radiation dominated regime, the field $\chi$ can no longer remain anchored at $\chi_0$ because $L^2/a^6 \propto 1/a^2$ implies \pref{CosmologyEqsBal} can no longer be satisfied for all times. Furthermore, assuming $\chi$ does not evolve dramatically, the evolution of $L(a)$ during radiation domination also implies that the axion term becomes less and less important relative to the matter term (so far as dilaton evolution is concerned) the further back in time one goes, given that they were equal at $a = a_{\rm eq}$. 

In this regime the dilaton evolution during radiation domination proceeds as if the axion were not present, and so closely resembles the evolution discussed earlier (and found in \cite{AndyCostasnMe}) for dilaton-matter evolution. In particular, Hubble damping indeed suppresses $\chi$ evolution even when the dilaton kinetic energy is much greater than its potential energy. 
 
\begin{figure}[t]
\begin{center}
\includegraphics[width=140mm,height=70mm]{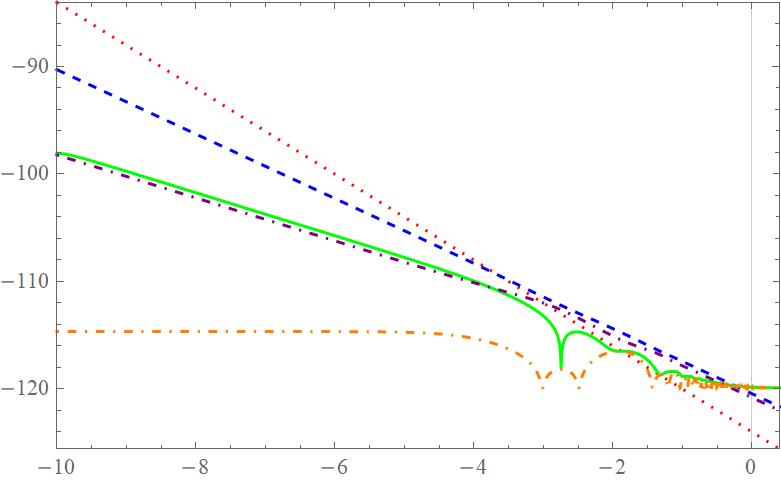} 
\caption{A log-log plot (base 10) of the energy density in matter (dashed), radiation (dotted) and the dilaton's total (solid) and potential (orange dash-dotted) energies and axion energy (purple dash-dotted) as a function of universal scale factor,  with $a = 1$ representing the present, with initial axion and dilaton kinetic energies nonzero and $\mfg = -0.408$ as predicted by our model. The axion energy executes a crossover from $\rho_a \propto a^{-2}$ to $\rho_a \propto a^{-3}$ when passing from radiation to matter domination, as described in the text, and the competition between axion-dilaton and matter-dilaton couplings keeps the dilaton near its present value long enough for the scalar potential to take over.} \label{Fig:Chiplot} 
\end{center}
\end{figure}

We have explored how sensitive successful cosmology is to initial conditions, and find that those stated above can be relaxed (see for example the plot of Fig.~\ref{Fig:Chiplot}) provided that there is sufficient axion energy density present. This implies a lower bound on the axion-matter coupling, but this lower bound turns out to be very small (the plot in Fig.~\ref{Fig:Chiplot} corresponds to $\cJ/\rho_{m} \simeq 1.05 \times 10^{-25}$). Couplings this small are also large enough to evade solar-system bounds for the dilaton through the `homeopathy' mechanism of \cite{ADScreening}, yet are easily small enough to keep the axion itself undetected in solar-system tests.

\begin{figure}[t]
\begin{center}
\includegraphics[width=140mm,height=70mm]{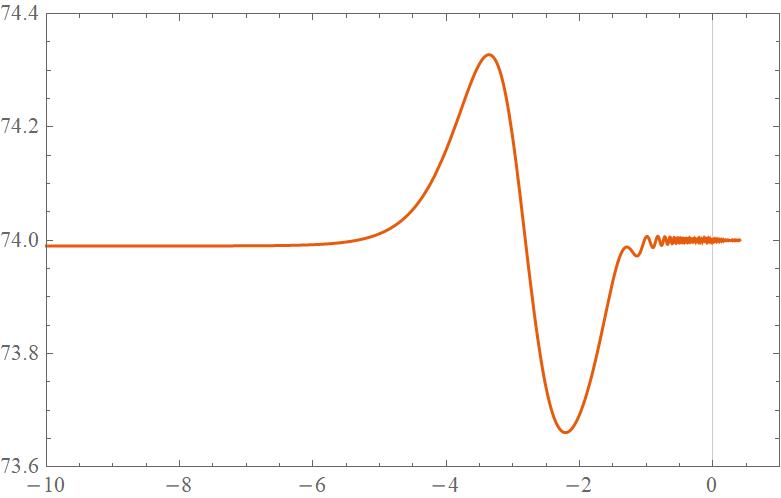} 
\caption{The evolution of the scalar field, $\hat\chi = \chi/M_p$, as a function of $\mfz = \log_{10} a$ where $a$ is the universal scale factor and $a = 1$ represents the present, using the same parameters used in Fig.~\ref{Fig:Chiplot}. This shows that the field $\chi$ does not evolve unacceptably between nucleosynthesis ($a \simeq 10^{-10}$) and now, though the transition in axion behaviour at radiation-matter equality ($a \simeq 3 \times 10^{-4}$) naturally produces variations of in $\chi$ near recombination ($a \simeq 10^{-3}$) at the several percent level (and somewhat larger just before and after). } \label{Fig:Eplot3} 
\end{center}
\end{figure}

\begin{table}
\begin{center}
\begin{tabular}{c|c|c|c|c|c}
$\omega_{r}$ & $\omega_m$ & $\omega_\chi$ & $\omega_a$ & $\omega_c = \omega_m + \omega_a$ & $w$ \\
\hline
$5 \times 10^{-5}$ & $0.18$ & $0.76$ & $0.080$ & $0.26$ & $-0.99$ 
\end{tabular}
\end{center}
\caption{The present-day fraction of the total energy density in the various cosmic fluids, with $\omega_i = \rho_i/\rho_{\rm tot}$ for radiation ($r$), dark matter ($m$), the dilaton ($\chi$) and axion ($a$), for the cosmology shown in Figs.~\ref{Fig:Chiplot} and \ref{Fig:Eplot3}. Notice that the effective total dark matter density is $\omega_c = \omega_m + \omega_a$ in this model. The last column gives the equation of state parameter, $w = p_\chi/\rho_\chi$, for the dilaton fluid.} \label{Table1}
\end{table}

The cosmology shown in Fig.~\ref{Fig:Chiplot} satisfies the main phenomenological smell-checks one would require before launching more detailed studies. For instance, the scalar fields' energy densities are less than a few percent of the total at nucleosynthesis (indeed, in the scenario shown they are well below this bound). The predictions for the present-day fraction of the total energy budget in the various ingredients are given in Table \ref{Table1}, as is the dilaton's equation of state parameter $w = p_\chi/\rho_\chi$. These are also broadly consistent with standard $\Lambda$CDM cosmology, though with a few interesting wrinkles.

The wrinkles are revealed once the time-dependence of the energy densities is included. In the scenario shown, for example, 44\% of the total apparent Dark Matter at present consists of axions, though the precise mix changes with time as the universe evolves. Similarly, the oscillations of the dilaton around its minimum also imply possibly detectable oscillations in the Dark Energy equation of state parameter in the relatively recent past. 

More stringent constraints come from the change in particle masses (relative to $M_p$) between now and past epochs, that occur because of their dependence on the dilaton: $m(\chi) = \mfm \,e^{-\frac12 \, \zeta \hat\chi}$. These can be seen from the evolution of $\hat\chi(t)$ shown in Fig.~\ref{Fig:Eplot3} (for the same parameters as used in Fig.~\ref{Fig:Chiplot}). For instance, particle masses cannot differ from their current values by more than a few percent at nucleosynthesis without altering otherwise-successful $\Lambda$CDM predictions \cite{Alvey:2019ctk}. Similar bounds coming from CMB observations preclude significant mass changes near recombination (more about which below). Measured spectral lines also strongly constrain mass changes after stars form at redshifts around 7. 

These conditions require the dilaton profile not to differ too much from its current value in these windows where these masses are well-measured, and the profile shown in Fig.~\ref{Fig:Eplot3} is chosen to try to minimize these. As the figure shows, the constraint at BBN is fairly easy to satisfy. Although one might think that not allowing the dilaton to move far over cosmological times should also strongly constrain the amount of kinetic energy it can initially have, the ruthless efficiency of Hubble friction turns out to make this not so; cosmic evolution usually drains the scalar kinetic energy before it rolls very far. 

We find it to be fairly generic to have excursions in $\tau$ shortly after radiation-matter crossover because this changes the nature of the underlying scaling solutions towards which evolution is attracted. Late-time oscillations of the dilaton as it settles into the potential's local minimum are also common features of most scenarios we have so far found, though the details of their amplitude and timing depend somewhat on the precise initial conditions. Such oscillations in particle masses are likely to be strongly constrained by observations of spectral lines unless they are damped out before the end of the Dark Ages. 

A more detailed exploration of these and other cosmological consequences goes beyond the scope of this paper, but is clearly of great interest to further explore. 

\subsubsection*{Axion as Dark Matter}

As mentioned above, because the axion energy density falls like $1/a^3$ during matter domination it contributes to the total observed Dark Matter abundance at late times. Indeed in the specific scenario described above it contributed just shy of half of the dark matter at the present epoch.  

This suggests it is logically possible that the axion itself could provide all of the dark matter. This possibility might seem surprising at first sight, because normally an axion with the couplings and masses entertained here would not be considered a dark-matter candidate. It would normally be excluded to the extent that these properties are inconsistent with standard production mechanisms, such as when axions arise from coherent oscillations about the minimum of a symmetry-breaking potential. 

Inapplicability of such production mechanisms need not be a problem in the present instance, however, because the winding up of the axion due to a small axion-matter coupling -- as predicted by \pref{Lvsa} and \pref{tvsa} -- is a production mechanism in itself. We have explored this numerically and have verified that we can generate a cosmologically significant axion energy density that when dominant falls as $1/a^3$ (as would be required of dark matter). This can be done despite such a cosmology being unable to profit from the balancing of dilaton-axion and dilaton-matter forces, however, and we have verified that trapping of the dilaton can be done in this case under certain circumstances. 

The possibility of uniting Dark Matter and Dark Energy into a unified axio-dilaton picture is a beautiful direction that is well worth more detailed exploration, though we leave this for future work.

\subsection*{Potential relevance to the $H_0$ puzzle}

A suggestive consequence of the scenario proposed here is the dilaton evolution that is predicted at relatively late epochs, particularly immediately after radiation-matter equality. This is a fairly generic phenomenon: particle masses can evolve coherently relative to the Planck scale as the dilaton evolves (both in space and in time) although mass {\it ratios} do not change. Such changes in particle masses are detectable, and we have tried to minimize this motion when seeking cosmological solutions to avoid many of the bounds that exclude gross changes to particle masses.

It is tantalizing in this context that having particle masses change by a few percent around the epoch of recombination --- which conveniently occurs shortly after radiation-matter equality ---  has been identified \cite{Sekiguchi:2020teg} as being one of the few viable options for resolving the current Hubble tension \cite{HubbleTension} through changes to fundamental physics (see also \cite{Cyr-Racine:2021alc, Zahn:2002rr} for related ideas involving scale invariance and the Hubble tension). 

In particular ref.~\cite{Sekiguchi:2020teg} studies how changes to cosmological parameters can avoid ruining the great success of CMB observations. They find that a correlated change at recombination to the electron mass, $m_e$, baryon density, $\omega_b$ and dark-matter density, $\omega_c$, can drop out of the gross features of the CMB (though not from those more detailed features that depend on non-equilibrium effects). Defining for any variable $X$ the quantity $\Delta_\ssX = \ln(X/X_0)$ where $X_0$ is the `baseline' local value, ref.~\cite{Sekiguchi:2020teg} finds CMB observations remain largely unchanged if
\be \label{NeededDelta}
  \Delta_{m_e} = \Delta_{\omega_b} = \Delta_{\omega_c} \,.
\ee

Remarkably, nonzero changes of this type can actually improve the description of CMB physics because they alter the epoch of recombination by an amount $\Delta_{a_\star} = - \Delta_{m_e}$. This in turn alters the value of $H_0$ inferred from CMB measurements in a way that can bring it into line with non-CMB measurements of $H_0$. Although the change in $H_0$ implied by \pref{NeededDelta} would also alter BAO observations, ref.~\cite{Sekiguchi:2020teg} argues these changes can be compensated for by a small nonzero spatial curvature of order $\omega_k \simeq - 0.125 \Delta_{m_e}$, leading to a change to the Hubble scale of $\Delta_h = 1.5 \Delta_{m_e}$. In these circumstances a 5\% increase in the electron mass at recombination would account for the 8\% discrepancy between measured values for $H_0$.

Since our model predicts the electron mass could indeed be different at recombination, it is natural to ask whether the conditions \pref{NeededDelta} would also follow if $\tau$ were to differ at recombination from its present-day value. We believe that they do, because this would change all masses by a common factor. For instance, to the extent that the number density of baryons is tied to the CMB photon density, changes to the baryon density at recombination dominantly come from changes to the nucleon mass: $\Delta_{\omega_b} = \Delta_{m_\ssN} = \Delta_{m_e} = -\frac12 \Delta_\tau = - \mfg (\hat \chi - \hat\chi_0)$. A similar argument also plausibly applies to dark matter, provided its mass scales like $\tau^{-1/2}$. 

The prediction $\mfg \simeq - 0.41$ implies a 12\% decrease in $\hat \chi$ ({\it i.e.}~a 10\% decrease in $\tau$) at recombination produces a 5\% increase in particle masses, and as plots like Fig.~\ref{Fig:Eplot3} show, 10\% excursions in $\delta \tau$ are easily obtainable, and naturally happen shortly after radiation-matter equality. Fig.~\ref{Fig:Eplot3} also shows that in principle this can be done without also requiring unacceptably large mass changes at BBN. 

We regard these preliminary cosmological scenarios to be promising enough to warrant more careful exploration of their phenomenological properties, including in particular a more systematic treatment of their fluctuations near recombination and in the universe nearer by, and short of this cannot yet claim that all constraints can be satisfied. We regard the challenge of cosmological model building around these constraints (and the opportunities for new signals) to be a welcome alternative to the head-banging ordeal of seeking a cosmological constant that is technically natural.

\section{Concluding Remarks}
\label{sec:Conclusions}

To summarize: we propose here a framework for understanding the Dark Energy density in a technically natural way that concentrates on the cancelation of the contributions to the vacuum energy coming from Standard Model particles. The framework relies on the following core ingredients: 
\begin{enumerate}[label=(\roman*)]
\item  A very supersymmetric gravity sector coupled to a Standard Model sector in which supersymmetry is badly broken and so non-linearly realised using the formalism of constrained superfields. The low-energy presence of supergravity and goldstino auxiliary fields plays an important role by imposing a supersymmetric form on the low-energy scalar potential.
\item A relaxation mechanism wherein a relaxon field $\phi$ adjusts to suppress the leading vacuum energy. Here $\phi$ is a new  light scalar field that also realizes supersymmetry nonlinearly (as does the Standard Model sector).
\item Approximate accidental scale invariance under which a dilaton field $\tau$ and the metric scale by constant factors, whose breaking is captured by an expansion of the action in powers of $1/\tau$. The field $\tau$ belongs to the gravitationally coupled supersymmetric sector and so comes with axion and dilatino partners.
\end{enumerate}

In the explicit realisation presented here, a minor tuning of lagrangian parameters of order $1/60$ allows the dilaton field to be stabilised at an exponentially large value $\tau\sim 10^{26}$, and this large value explains the size of the electroweak hierarchy inasmuch as Standard Model particles acquire masses of order $M_\TEV \sim M_p/\sqrt\tau$. This exponentially large value for $\tau$ is obtained naturally because the potential arises as a rational function of $\ln\tau$ due to the generic presence of logarithms of mass ratios amongst the UV particles that are integrated out above the weak scale (along the lines proposed in a different context some time ago \cite{AndyCostasnMe}).

The low-energy scalar potential is calculable as a series in $1/\tau$ of the form 
\be\label{Vtauseries}
   V = \frac{V_2}{\tau^2} + \frac{V_3}{\tau^3} + \frac{V_4}{\tau^4} + \cdots 
\ee
and because Standard Model particle masses are $m \propto \tau^{-1/2}$ their loops contribute $\delta V \sim m^4 \propto 1/\tau^2$ and so contribute to $V_2$. Strictly speaking these contributions actually arise as order $m^2$ corrections to $w_\ssX$ \cite{LowESugra}.

The supersymmetry of the gravity sector implies $V_2 \propto |w_\ssX|^2$ is positive since it must turn off in the hypothetical limit where global supersymmetry is unbroken. The relaxation field appears in $V_2$ and it prefers to minimize $w_\ssX \to 0$ in the large-$\tau$ limit. In global supersymmetry it would do so by seeking $V_2 = w_\ssX = 0$, but in supergravity this combination instead becomes Planck suppressed. This suppression does not mean superpartners cannot remain heavy, however, because supersymmetry-breaking fields like $F^\ssX \propto W_0/M_p$ remain at the weak scale despite being Planck suppressed. 

The interplay between scale invariance and supersymmetry (as manifested in `extended no-scale structure') then leads to $V_3$ also vanishing, leaving in $V_4/\tau^4$ a naturally small, positive, cosmological constant\footnote{Note that the dilaton potential $V_4$ must both dominate its kinetic energy and be positive so that Dark Energy dominates the energy density of the universe at late times with an acceptable equation of state.} that is order $(M_\TEV^2/M_p)^4$ (and so is the right size). The fact that supersymmetry is so mildly broken in the gravity sector allows it to protect the series form of $V$ given in \pref{Vtauseries}, as well as the cancellation of the $V_3$ terms. 

The large value of $\tau$ determines both the small cosmological-constant scale, $m_{\rm vac} \propto 1/\tau$, and the electroweak scale, $m_\TEV \propto M_p/\sqrt\tau$, fixing their ratio to be $m_{\rm vac}/m_\TEV \propto 1/\sqrt{\tau}$. This contains the seeds of the oft-made observation that the weak scale is the geometric mean of the Planck and cosmological-constant scales. 

Weinberg's no-go argument applies (as it must since the underlying mechanism relies on scale invariance \cite{CCWeinberg, Burgess:2013ara, CCReviews}) in the sense that quantum corrections  to the scalar potential (such as subdominant powers of $1/\tau$) are present. But Weinberg's argument does not say how large these corrections must be and supersymmetry -- through the cancellation of $V_3$ -- is what keeps them small.

The presence of auxiliary fields in the scalar potential required by supersymmetry in the gravity sector allows supersymmetry also to have implications for the naturalness of the relaxon $\phi$. This is because its mass term arises from a dimension-three operator $g\, \Phi^2 X \in W$ rather than from a dimension-two operator in $V$; involving $X$ because it is a strong supersymmetry-breaking effect. Having its mass come from $W_\ssX$ both ensures that the $\phi$ mass is proportional to $\tau^{-1/2}$ (and so at most lies at TeV energies) and makes radiative corrections enter through $g$, which is dimensionless. 

Similar arguments may also apply to the Higgs boson itself if its scalar potential also arises as a contribution $\lambda_\ssH (H^\dagger H - v_\EW^2) \in W_\ssX$, although in the Higgs case having a mass $\propto \tau^{-1/2}$ is more important (and more generic). The same physics underlies both the electroweak and cosmological constant scales at a fundamental level (without need for anthropic arguments). 

There are several prices we pay (plus a few opportunities we reap) for this suppression. First, the small size of the vacuum energy in the low-energy theory requires $\tau \gsim 10^{26}$ and then this drives the axion decay constant so low that the low-energy EFT must fail at eV energies. Although plausible UV physics (such as supersymmetric large extra dimensions \cite{SLED}) could plausibly intervene at these scales, it is not yet known how it does so. In particular, the UV pedigree for $\tau$ is not known and so we do not know whether such large values are allowed. This is not an empty worry because if $\tau \sim \cV^{2/3}$, as suggested by the simplest string models, then extra-dimensional constraints preclude it from being larger than of order $10^{20}$.

A second price we pay is the large Brans-Dicke-like coupling of the dilaton to ordinary matter, which flirts with inconsistency with current tests of gravity. Although the model seems to bring its own evasion mechanism -- wherein the $SL(2,R)$-invariant axion-dilaton self-couplings divert dilaton couplings into generating harder-to-detect axionic response \cite{ADScreening} -- there are also numerous potential signals for their presence in tests of gravity and cosmology. 

Some of the low-energy axio-dilaton signals might even solve problems, such as the current puzzle over apparent inconsistencies in measurements of the Hubble scale $H_0$. Whether these eventually kill or verify the model, we regard it as progress to trade the cosmological constant problem for exercises in late-epoch model-building, and leave a more detailed treatment of this mechanism for future research.

The remainder of this section puts our scenario into the context of earlier approaches using similar ingredients and briefly summarizes several open issues.  

\subsection{Relation to other approaches}

All three of our ingredients have been separately used previously in related contexts, so it is useful to clarify what differs from these in our particular framework.

\subsubsection*{Low-energy supergravity}

Approaches to quintessence and Dark Energy that use supergravity equations of motion\footnote{Examples where this happens within an underlying string construction are discussed in \S\ref{ssec:UVcomplete}.} go back more than 20 years \cite{SUGRAQuint}. Even if supersymmetry is valid at high energies, there is a basic question that these constructions do not address: why should it survive down to the extremely low energies required to be relevant to quintessence, given the apparent absence of supersymmetry at the intervening energies containing the observed Standard Model particles? (For recent evaluations of quintessence models within a string framework see \cite{Cicoli:2018kdo, Cicoli:2021skd}.)

This question is special case of a larger problem faced even by nonsupersymmetric quintessence models: why is it legitimate to compute and use a carefully designed quintessence potential entirely within the classical approximation? This is often phrased as the statement that vacuum energies and very small scalar masses are not technically natural  (see eg \cite{Burgess:2013ara}): quantum effects cause scalar masses and potentials to change dramatically once heavier particles are integrated out, making them particularly sensitive to a system's UV sector. This sensitivity can also undermine \cite{LEScalarPC, EFTBook} the usual low-energy arguments that justify using the classical approximation in gravitating systems.

By contrast, addressing these issues is the main motivation of the model presented here. Our tool for doing so is the explicit coupling of low-energy supergravity to a non-supersymmetric Standard Model permitted by the nonlinear realization of ref.~\cite{Komargodski:2009rz} and its coupling to supergravity \cite{Bergshoeff:2015tra, DallAgata:2015zxp, Schillo:2015ssx}, since this allows one to trace the low-energy effects of integrating out non-supersymmetric sectors \cite{LowESugra}. 

\subsubsection*{Relaxation mechanisms}

Relaxation mechanisms that use a field's dynamical evolution to suppress the apparent cosmological constant also have a long history \cite{RelMechs}, and more recently have been applied to the electroweak hierarchy problem \cite{Relaxion}. We feel that none have been entirely convincing as solutions to these problems on their own, and the same would have been true for us if we had not combined relaxation with the other two ingredients. See \cite{Graham:2019bfu} for a recent  interesting proposal in this direction.

\subsubsection*{Scale invariance}

Scale invariance has an equally long history, with both early applications to the cosmological constant problem and elsewhere. Early proposals assumed a scale-invariant action with scaling broken only by quantum anomalies \cite{ScaleAnomalyCC}, leaving open whether such examples exist. They very early ran up against no-go results \cite{CCWeinberg} that identified why even completely unbroken scale invariance cannot prevent the lifting of classically flat dilaton directions. 

Our approach evades some of these issues because scale invariance is only approximate, even at the classical level. The arguments of the no-go theorems broadly apply, inasmuch as the central issue is to quantify the lifting of flat directions by quantum effects. But the no-go arguments do not forbid the use of supersymmetry to suppress (though not completely eliminate) this lifting, as we do here using the low-energy distillation \cite{Burgess:2020qsc} of the extended no-scale structure found in string models \cite{Berg:2005ja, Cicoli:2007xp}.

Approximate scale invariance has also been conjectured to be related to small cosmological constants and the existence of cosmologically light dilatons \cite{Tye:2016jzi} and within an anthropic context \cite{Andriolo:2018dee} within string constructions. These papers (and we) both use the robust link between potentials that predict a small cosmological constant and Hubble-scale dilaton masses described in \cite{AndyCostasnMe}, though in the present paper we provide an explicit mechanism for building a potential with a naturally small vacuum energy that exploits this connection (without the need to resort to anthropic arguments).

Indeed, scale invariance is more broadly suggestive as an ingredient for solving naturalness problems because the essence of these problems is that Nature seems to be closer to scale invariance ({\it i.e.}~some masses or energies are smaller) than we would normally expect. This has led to its exploration for other purposes, such as to higher-derivative theories of gravity (which are more broadly scale invariant in the UV \cite{ConformalGravity}, but bring associated difficulties with ghosts \cite{ButGhosts} -- see however \cite{MaybeNoGhosts}). Scale invariance has also been applied to inflationary models \cite{ConformalInf} or motivating the choices required \cite{Kallosh:2013pby} to obtain inflation using only the Standard Model Higgs as the inflaton \cite{Bezrukov:2007ep}. 

This history teaches that scale invariance, though attractive, seems to come with undesirable extras (such as ghosts in higher-derivative gravity or dangerous dilatons in the models described here). We regard the dilaton to be amongst the more benign of these options, in that its existence need not point towards a fundamental instability, though viability of the model requires threading a minefield of potential observational tests.

\subsection{General implications and future directions}
 
In the picture we paint the small observed size of the dark-energy density points towards a very rich low-energy `dark' sector consisting of supersymmetric gravity, a cosmologically active axio-dilaton multiplet, a somewhat heavier gravitationally coupled relaxon plus possibly other dark ingredients. This picture leaves open a great many interesting directions worth further exploring.

\begin{itemize}
\item{\bf Naturalness:}
The main conceptual issue is to better verify through explicit calculations that the general arguments about nonlinear realization of supersymmetry do indeed preserve the supergravity structure of the scalar potential, along the lines of \cite{LowESugra} but more explicitly in the regime of small $W_\ssX$.
\item{\bf Super-eV completions:}
$\tau \gsim 10^{26}$ plays an important role suppressing the vacuum energy in the low-energy 4D theory and this drives a breakdown of EFT methods at eV energies (due to the small axion decay constant). Can extra dimensions provide the UV completion that unitarizes the theory up to the electroweak scale? If so, how does $\tau$ arise in this completion and can it take the large values that are needed?
\item{\bf Axio-dilaton phenomenology:}
The biggest phenomenological issue is to see whether the required dilaton properties could have evaded contemporary precision tests of gravity, and can be consistent with what we know about cosmology. How robust is the mechanism of \cite{ADScreening}  to dynamical issues and how sensitive is it to the detailed axion couplings? How does it stand up to more detailed studies of cosmology, including issues of structure formation? Can the baryonic wind-up mechanism of \S\ref{sssec:AxioEvolution} allow the axion $\mfa$ to contribute to (or to replace) Dark Matter? Are the predicted variations of masses over cosmological times consistent with observations? Can they help solve the current Hubble tension?
\item{\bf Dark matter}
Our discussion leaves open what plays the role of dark matter, which could be included in the simplest scenarios by supplementing the Standard Model sector by another non-supersymmetric particle whose mass varies as $m \propto \tau^{-1/2}$. More economical options might also be worth exploring, however, including those for which the dark matter mass varies differently with $\tau$ or the option where the axion $\mfa$ itself plays the role of dark matter, as sketched in \S\ref{sssec:AxioEvolution}. Could the relaxon be the dark matter? (Although the $\phi^2 \cH^\dagger \cH$ coupling resembles scalar-portal dark-matter models \cite{ScalarPortal} its small coupling strength is too small to allow the $\phi$ field to be thermally produced, unlike in the minimal case.)
\item{\bf Neutrino physics}
Our picture potentially populates the low-energy world with a rich spectrum of weakly coupled very light fermions, and explains why they are there (some are superpartners to known light bosons, like the graviton). Their presence is a required consequence of the supersymmetry of the gravity sector (and resembles the low-energy sector of \cite{MSLED} in this way). Should these mix with Standard Model neutrinos they would provide natural candidates for light sterile neutrinos (along the lines of \cite{Dienes:1998sb, MSLEDnu}), and could open up new ways to express lepton-number violation at low energies, with possible implications for lepto- and baryogenesis. If super-heavy sterile fermions mix with SM neutrinos (as in the see-saw mechanism) with a Dirac mass $m_\ssD \propto \tau^{-1/2}$ that scales like other SM masses then physical neutrino masses are order $m_\nu \sim 1/\tau$, and so would explain the coincidence between neutrino masses and the cosmological constant scale: $V_{\rm min} \sim m_\nu^4$. What observable features do the resulting neutrino/dark-sector interactions imply?
\item{\bf Pre-BBN cosmology}
The picture also likely modifies pre-nucleosynthesis cosmology in a variety of ways that are worth exploring. Among these are the interplay between $\phi$ and the Higgs $\cH$ in the scalar potential, since if the potential for these both lie within $|w_\ssX|^2$ then only the vev of a linear combination gets fixed by the single condition $w_\ssX = 0$. What would this mean for late-time cosmology and/or the epoch of the electroweak phase transition?

The field $\phi$ also seems designed to be a good inflaton candidate \cite{Burgess:2022nbx}. After all, nonlinearly realized supersymmetry naturally provides large positive (de Sitter-like) vacuum energies (the potential's $|w_\ssX|^2$ term) for any value of $\phi$ away from its minimum and $\phi$ has been designed as a field whose evolution parameterizes changes between large nonzero $w_\ssX$ and vanishing $w_\ssX$. The early-universe evolution of $\phi$ while positive energies dominate therefore provides an attractive picture of inflation in which the inflaton is not completely divorced from Standard Model physics, and changes to $\tau$ become correlated with changes to the size of the observable universe.\footnote{Notice that when $w_\ssX \neq 0$ eq.~(\ref{VFtauexpagain}), which schematically can be written as  $V\sim (A w_\ssX^2\tau^2-Bw_\ssX \tau+C)/\tau^4$, allows a new type of stabilization at smaller values of $\tau \sim w_0/w_\ssX$, showing that inflation can drive nontrivial evolution for the $\tau$ field as well, providing a realization of large scale inflation. Once inflation ends and $w_\ssX\sim 0$, $\tau$ can roll towards its large global minimum $\tau\sim 10^{26}$, correlating the early and late-time hierarchies to the inflationary growth of the volume of the observable universe (as in \cite{Conlon:2008cj, Burgess:2016ygs}).}  The inflationary models found in this way come with scale invariance baked in (much like for the possible UV embeddings described in \S\ref{ssec:UVcomplete}), in a way that is known to help such models agree with observations, along the lines described in \cite{Burgess:2016owb}. We report the details of this scenario, and other `yoga breathing' exercises (relaxed inflation) in a future publication.

\end{itemize}

Yoga Dark Energy ties together many of the scales of physics and so its implications are legion; further investigations are underway into several of these directions.

\section*{Acknowledgements}
We thank Aizhan Akhmetzhanova, Clare Burrage, Michele Cicoli, Ed Copeland, Shanta de Alwis, Emilian Dudas, Nemanja Kaloper, Justin Khoury, Lloyd Knox, Francesco Muia, Jos\'e de Jes\'us Padua Arg\"uelles and Henry Tye for many helpful conversations.  CB's research was partially supported by funds from the Natural Sciences and Engineering Research Council (NSERC) of Canada. Research at the Perimeter Institute is supported in part by the Government of Canada through NSERC and by the Province of Ontario through MRI.  The work of FQ has been partially supported by STFC consolidated grants ST/P000681/1, ST/T000694/1.

\appendix

\section{Useful supergravity formulae}
\label{App:UsefulSG}

This appendix collects some useful formulae encountered in the main text when working with supergravity models.

One of the cumbersome steps is inverting the K\"ahler metric. Consider then a K\"ahler function of the form $K = - 3 \ln \cP$, where $\cP = \cP(Z^\ssA)$. Then
\be \label{KahlerDownP}
  K_\ssA = - \frac{3 \cP_\ssA}{\cP} \,, \quad K_{\ssA \ol\ssB} = - \frac{3 \cP_{\ssA\ol\ssB}}{\cP} + \frac{3 \cP_\ssA \cP_{\ol\ssB}}{\cP^2} 
\ee
and so (as is easy to check) the inverse is
\be \label{KahlerUpP}
  K^{\ol{\ssA}\ssB} = - \frac{\cP}{3} \left[ \cP^{\ol{\ssA}\ssB} + \frac{\cP^{\ol\ssA} \cP^{\ssB}}{\cP - P^2} \right]
\ee
where $P^2 := \cP^{\ol{\ssA}\ssB} \cP_{\ol\ssA} \cP_\ssB$ and $\cP^{\ol{\ssA}\ssB} \cP_{\ssB \ol\ssC} := {\delta^{\ol\ssA}}_{\ol\ssC}$.  With these definitions notice that
\be \label{KKKwrtP}
  K^{\ol\ssA \ssB}   K_\ssB =  \frac{\cP \cP^{\ol\ssA}}{\cP-P^2} \quad \hbox{and} \quad
  K^{\ol\ssA \ssB} K_{\ol\ssA} K_\ssB = \frac{3P^2}{P^2 -\cP}  \,.
\ee

\subsection*{Two sectors coupled through $T$}

Next suppose that the fields divide up into three types: $\{ Z^\ssA \} = \{ T, S^i , z^a\}$ and $\cP = A(T,S) + B(T,z)$. Then the first derivatives become
\be
  \cP_\ssT = A_\ssT + B_\ssT \,, \quad \cP_i = A_i \,, \quad \cP_a = B_a \,,
\ee
and the second derivatives are
\be
 \cP_{\ssT\ol\ssT} = A_{\ssT\ol\ssT} + B_{\ssT\ol\ssT} \,, \quad
 \cP_{\ssT\bar\jmath} = A_{\ssT\bar\jmath} \,,\quad
 \cP_{\ssT\bar c} = B_{\ssT\bar c} \,,\quad
 \cP_{i \bar\jmath} = A_{i \bar\jmath} \,, \quad
 \cP_{a\bar c} = B_{a\bar c} \,, 
 \ee
 and $\cP_{i \bar c} = 0$. The vector of 1st derivatives and the matrix of 2nd derivatives therefore are
 \be
    \cP_\ssA =  \left( \begin{array}{c}
  A_\ssT+B_\ssT  \\ A_i \\ B_a  \end{array} \right)  \quad \hbox{and} \quad
   \cP_{\ssA\bar\ssB} =  \left( \begin{array}{ccccc}
  A_{\ssT\ol\ssT}+B_{\ssT\ol\ssT}  && A_{\ssT\bar\jmath} && B_{\ssT\bar c}  \\  
  A_{i\ol\ssT} && A_{i \bar\jmath} && 0 \\
  B_{a\ol\ssT} && 0 && B_{a\bar c}  \end{array} \right) \,.
\ee
The inverse of the matrix of second derivatives is then
\be
   \cP^{\ol\ssA\ssB} =  \left( \begin{array}{ccccc}
  \alpha  && -\alpha A^j && - \alpha B^{c}  \\  
  -\alpha A^{\bar \imath} && A^{\bar \imath j} + \alpha A^{\bar\imath} A^j && \alpha A^{\bar \imath} B^c \\
  - \alpha B^{\bar a} && \alpha A^j B^{\bar a} && B^{\bar a c} + \alpha B^{\bar a} B^c \end{array} \right) \,,
\ee
where $A^j := A^{\bar\imath j} A_{\ssT \bar\imath}$, $A^i := A^{\bar \jmath i} A_{i \ol\ssT}$, $B^a := B^{\bar c a} B_{\ssT \bar c}$, $B^a := B^{\bar c a} A_{a \ol\ssT}$ and $\alpha$ is given by
\be \label{1alpha}
  \frac{1}{\alpha} := A_{\ssT\ol\ssT}+B_{\ssT\ol\ssT} - A^{\bar \imath j} A_{\ssT \bar \imath} A_{j \ol\ssT} - B^{\bar a c} B_{\ssT \bar a} B_{c \ol\ssT} \,.
\ee
In this case
 \be
    \cP^{\ol\ssA} = \cP^{\ol\ssA\ssB} \cP_\ssB = \alpha \left( \begin{array}{c}
  A_\ssT+B_\ssT - A^i A_i - B^a B_a \\ A^{\bar\imath} (A_{\ssT\ol\ssT} + B_{\ssT\ol\ssT} - A_\ssT - B_\ssT) \\ B^{\bar a} (A_{\ssT\ol\ssT} + B_{\ssT\ol\ssT} - A_\ssT - B_\ssT)  \end{array} \right)  \,,
\ee
which uses \pref{1alpha} when simplifying the result (and in particular to extract an overall factor of $\alpha$). Finally
 \be \label{PsqEq}
   P^2 = \cP^{\ol\ssA} \cP_{\ol\ssA} = \alpha \Bigl[ (A_{\ol\ssT} + B_{\ol\ssT})(
  A_\ssT+B_\ssT - A^i A_i - B^a B_a) +( A^i A_i + B^a B_a) (A_{\ssT\ol\ssT} + B_{\ssT\ol\ssT} - A_\ssT - B_\ssT) \Bigr]  \,,
\ee

\subsection*{One no-scale sector}

With no-scale models in mind, suppose now that $A$ is independent of Im $T$ and Im $S^i$ and that $A$ is a homogeneous degree-one function, so $A(\lambda S^i, \lambda T) = \lambda A(S^i, T)$. When these are true we can write 
\be
  A(S,T) =: \tau \, F(x^i) \quad \hbox{with} \quad x^i := \frac{\sigma^i}{\tau
  } \,,
\ee
with $\tau := T+\ol T$ and $\sigma^i := S^i + {\ol S}^i$ so that
\be
   A_\ssT = F - \frac{\sigma^i F_i}{\tau} \,, \quad
   A_i = F_i  
\ee
where $F_i := \partial F/\partial x^i$ and so on. The second derivatives then become
\be
  A_{\ssT\ol\ssT} =  \frac{\sigma^i \sigma^j F_{ij}}{\tau^3}  \,, \quad
   A_{\ssT \bar \jmath} = - \frac{\sigma^i F_{ij}}{\tau^2}   \,, \quad
   A_{i \bar\jmath} = \frac{F_{ij}}{\tau} \quad \hbox{and so} \quad
   A^{\bar i j} = \tau F^{ij} \,,
\ee
where $F^{ij} F_{jk} = {\delta^i}_k$. It follows that 
\be
  A^j = A^{\bar \imath j}A_{\ssT \bar \imath} =- \frac{\sigma^j}{\tau} \,, \quad
  A^{\bar\imath} =A^{\bar \imath j}A_{j \ol\ssT} =- \frac{\sigma^i}{\tau}   \,, \quad
   A^{\bar \imath j}A_{\ssT \bar \imath}A_{j \ol\ssT} 
   = \frac{F_{ij} \sigma^i \sigma^j}{\tau^3}
\ee
so that 
 \be \label{PdownAForm}
    \cP_\ssA =  \left( \begin{array}{c}
  F - \frac{\sigma^i F_i}{\tau} +B_\ssT  \\ F_i \\ B_a  \end{array} \right)  \quad \hbox{and} \quad
   \cP_{\ssA\bar\ssB} =  \left( \begin{array}{ccccc}
  \frac{F_{ij} \sigma^i \sigma^j}{\tau^3}+B_{\ssT\ol\ssT}  &&  - \frac{\sigma^i F_{ij}}{\tau^2}  && B_{\ssT\bar c}  \\  
  - \frac{\sigma^j F_{ij}}{\tau^2}  && \frac{F_{i j}}{\tau} && 0 \\
  B_{a\ol\ssT} && 0 && B_{a\bar c}  \end{array} \right) \,.
\ee
and so
\be \label{PupABForm}
   \cP^{\ol\ssA\ssC} =  \left( \begin{array}{ccccc}
  \alpha  && -\alpha A^j && - \alpha B^{c}  \\  
  -\alpha A_{\bar \imath} && A^{\bar \imath j} + \alpha A^{\bar\imath} A^j && \alpha A^{\bar \imath} B^c \\
  - \alpha B^{\bar a} && \alpha A^j B^{\bar a} && B^{\bar a c} + \alpha B^{\bar a} B^c \end{array} \right)
  =  \left( \begin{array}{ccccc}
  \alpha  && \alpha \sigma^j/\tau && - \alpha B^{c}  \\  
  \alpha \sigma^{i}/\tau &&  \tau F^{i j} +\alpha (\sigma^i\sigma^j/\tau^2) &&- \alpha \sigma^{i} B^c/\tau \\
  - \alpha B^{\bar a} &&- \alpha \sigma^j B^{\bar a}/\tau && B^{\bar a c} + \alpha B^{\bar a} B^c \end{array} \right) \,,
\ee
where
\be
   \frac{1}{\alpha} =  B_{\ssT\ol\ssT}   - B^{\bar a c} B_{\ssT \bar a} B_{c \ol\ssT} 
\ee
Therefore 
 \be\label{PupAForm}
    \cP^{\ol\ssA} = \cP^{\ol\ssA \ssB} \cP_\ssB =  \left( \begin{array}{c}
  \alpha ( F +B_\ssT - B^a B_a)  \\ \tau F^{ij} F_j + \frac{\alpha}{\tau} \, \sigma^i (F + B_\ssT - B^a B_a) \\ B^{\bar a} - \alpha B^{\bar a} (F + B_\ssT - B^a B_a)  \end{array} \right)  
 \,,
\ee
and so
\bea 
  P^2 &=& \tau F^{ij} F_i F_j + B^a B_a + \alpha | F + B_\ssT - B^a B_a |^2 \nn\\
  &=&  \tau F^{ij} F_i F_j +  \alpha \Bigl[ |F + B_\ssT|^2 + (B_{\ssT \ol\ssT} - 2F - 2 B_\ssT) B^a B_a \Bigr] \,.
\eea

\subsection*{Limiting cases} 

We can check the above against known examples.

\subsubsection*{No $z^a$ sector}

As a check, this should become a no-scale model in the absence of the $z^a$ sector. Setting $B=0$ and dropping the $z^a$ fields, we have $\cP = A$ and so
 \be
    \cP_\ssA =  \left( \begin{array}{c}
  A_\ssT  \\ A_i   \end{array} \right)   =  \left( \begin{array}{c}
  F - (\sigma^i F_i/\tau) \\ F_i   \end{array} \right)  \quad \hbox{and} \quad
\cP^{\ol\ssA\ssB} = \left( \begin{array}{ccc}
  \alpha  && \alpha \sigma^j/\tau   \\  
  \alpha \sigma^{i}/\tau &&  \tau F^{i j} +\alpha (\sigma^i\sigma^j/\tau^2)   \end{array} \right) \,,
\ee
Ignoring the fact that $\alpha \to \infty$ temporarily, it follows that
\be
  \cP^{\ol\ssA} = \cP^{\ol\ssA\ssB} \cP_\ssB =  \left( \begin{array}{c}
  \alpha  F   \\  \frac{\alpha F \sigma^i}{\tau} + \tau F^{ij} F_j  \end{array} \right) \quad \hbox{and so} \quad
  P^2 =  \alpha F^2 + \tau F^{ij} F_i F_j \,.
\ee
Taking now $\alpha \to \infty$ means that $P^2 \to \infty$ as well and so \pref{KKKwrtP} becomes
\be \label{KKKwrtP2}
  K^{\ol\ssA \ssB} K_{\ol\ssA} K_\ssB =  \frac{3P^2}{P^2- \cP} \to 3  \,,
\ee
as is required for a no-scale model. 

\subsubsection*{No $S^i$ fields}
\label{App:CaseNo2}

As a second check, imagine there are no $S^i$ fields, and that $A = \tau$ and so $\cP = \tau + B(\tau, z^a,\bar z^b)$, in which case the above should reduce to the model of the main text (without modulus stabilization). In this case $F = 1$ and so $A_\ssT = 1$ and $A_i = 0$ are to be used in the above formulae, implying in particular that all second derivatives of $A$ vanish. As a consequence 
 \be \label{PdownAForm2}
    \cP_\ssA =  \left( \begin{array}{c}
  1 +B_\ssT  \\ B_a  \end{array} \right)  \quad \hbox{and} \quad
   \cP_{\ssA\bar\ssB} =  \left( \begin{array}{ccc}
  B_{\ssT\ol\ssT}  && B_{\ssT\bar c}  \\  
   B_{a\ol\ssT} && B_{a\bar c}  \end{array} \right) \,.
\ee
and so
\be \label{PupABForm2}
   \cP^{\ol\ssA\ssB} =  \left( \begin{array}{ccc}
  \alpha  &&  - \alpha B^{c}  \\  
  - \alpha B^{\bar a} &&  B^{\bar a c} + \alpha B^{\bar a} B^c \end{array} \right) \quad \hbox{where} \quad
   \frac{1}{\alpha} =  B_{\ssT\ol\ssT}   - B^{\bar a c} B_{\ssT \bar a} B_{c \ol\ssT} \,.
\ee
Therefore 
 \be\label{PupAForm2}
    \cP^{\ol\ssA} = \cP^{\ol\ssA \ssB} \cP_\ssB =  \left( \begin{array}{c}
  \alpha ( 1 +B_\ssT - B^a B_a)  \\ B^{\bar a} - \alpha B^{\bar a} (1 + B_\ssT - B^a B_a)  \end{array} \right)    \,,
\ee
and so \pref{PsqEq} becomes
 \be
   P^2 = \alpha \Bigl[ (1 + B_{\ol\ssT})(
  1+B_\ssT  - B^a B_a) +  B^a B_a ( B_{\ssT\ol\ssT} - 1 - B_\ssT) \Bigr]   \,.
\ee

These expressions imply
\bea \label{KKKwrtP2x2}
  &&K^{\ol\ssA \ssB} K_{\ol\ssA} K_\ssB - 3  =  \frac{3\cP}{P^2- \cP}  \\
  &&\qquad = \frac{3(\tau + B)[B_{\ssT\ol\ssT}   - B^{\bar a c} B_{\ssT \bar a} B_{c \ol\ssT}]}{(1 + B_{\ol\ssT})(
  1+B_\ssT  - B^a B_a) +  B^a B_a ( B_{\ssT\ol\ssT} - 1 - B_\ssT)  - (\tau + B)[B_{\ssT\ol\ssT}   - B^{\bar a c} B_{\ssT \bar a} B_{c \ol\ssT}]}  \,, \nn
\eea
which vanishes if $B$ is independent of $\tau$, since then $P^2 =  \alpha (1 - 2B^a B_a)$ and $\alpha \to \infty$ and so again $K^{\ol\ssA \ssB} K_{\ol\ssA} K_\ssB = {3P^2}/{(P^2- \cP)} \to 3$.

For the rest of the lagrangian we also require $K_\ssA$, $K_{\ssA\ol\ssB}$ and $K^{\ol\ssA\ssB}$ separately, though only to leading order in $1/\tau$ since for these the no-scale cancellations do not occur. For these purposes we can use $B_\ssT \sim \cO(1/\tau)$ and $B_{\ssT\ol\ssT} \sim \cO(1/\tau^2)$. For instance, eq.~\pref{KahlerDownP} becomes
 \be
   K_\ssA = \left( \begin{array}{c}
  K_\ssT  \\  K_a  \end{array} \right) = - \frac{3 }{\cP}  \left( \begin{array}{c}
  1+B_\ssT  \\  B_a  \end{array} \right) \simeq - \frac{3 }{\cP}  \left( \begin{array}{c}
  1 \\  B_a  \end{array} \right)  \,.
\ee

Eq.~\pref{KahlerUpP} similarly becomes
\bea \label{KahlerDownPxx}
   K_{\ssA \ol\ssB} &=& - \frac{3 \cP_{\ssA\ol\ssB}}{\cP} + \frac{3 \cP_\ssA \cP_{\ol\ssB}}{\cP^2} \nn\\
   &=&  - \frac{3}{\cP} \left( \begin{array}{ccc}
  B_{\ssT\ol\ssT}  && B_{\ssT\bar c}  \\  
   B_{a\ol\ssT} && B_{a\bar c}  \end{array} \right)    +  \frac{3}{\cP^2} \left( \begin{array}{ccc}
 |1+B_\ssT|^2   && (1+B_\ssT) B_{\bar c}  \\  
    (1 + B_{\ol\ssT}) B_{a} && B_{a } B_{\bar c} 
   \end{array} \right)  \\
  &\simeq&  - \frac{3}{\cP} \left( \begin{array}{ccc}
 -1/\cP && B_{\ssT\bar c} - B_{\bar c}/\cP \\  
  B_{a\ol\ssT} - B_a/\cP && B_{a\bar c}  \end{array} \right)   =  - \frac{3}{\cP} \left( \begin{array}{ccc}
 -1/\cP && X_{\bar c} \\  
 X_a && B_{a\bar c}  \end{array} \right)     \,.\nn
\eea
where the last equality defines the quantities $X_a$ and $X_{\bar c}$. As is easy to check the inverse of this last matrix is
\be \label{KahlerUpPxx}
  K^{\ol{\ssA}\ssB} \simeq  - \frac{\cP}{3(\cP X^2 + 1)} \left( \begin{array}{ccc}
 -\cP && \cP X^b   \\  
  \cP X^{\bar a} && (\cP X^2 + 1) B^{\bar a b}  - \cP X^{\bar a} X^b \end{array} \right)    
\ee
where $B^{\bar a b}B_{b \bar c} = \delta^{\bar a}_{\bar c}$ and $X^{\bar a} := B^{\bar a b} X_b$ while $X^b = B^{\bar a b} X_{\bar a}$ and $X^2 = X_a X^a = X_{\bar a} X^{\bar a} = B^{\bar a b}X_{\bar a} X_b$ and so on. Keeping only the leading powers of $1/\cP$ and taking $X_a$ and $X_{\bar c}$ to be order $1/\tau$ and $B_{a \bar c}$ to be $\cO(1)$ we have $X^2 = \cO(1/\tau^2)$ and so $\cP X^2 + 1 \simeq 1$. The inverse therefore becomes approximately
\be \label{KahlerUpPxxx}
  K^{\ol{\ssA}\ssB} \simeq  - \frac{\cP}{3} \left( \begin{array}{ccc}
 -\cP && \cP X^b   \\  
  \cP X^{\bar a} &&  B^{\bar a b}   \end{array} \right)    
\ee
and so
\bea \label{KahlerUpVecx}
  K^{\ol{\ssA}\ssB}K_\ssB
  &\simeq&  \left( \begin{array}{ccc}
 -\cP && \cP X^b   \\  
  \cP X^{\bar a} &&  B^{\bar a b}   \end{array} \right) 
  \left( \begin{array}{c}
  1 \\  B_b  \end{array} \right) =  \left( \begin{array}{c}
  \cP(-1 + X^b B_b) \\  \cP X^{\bar a} + B^{\bar a b}B_b  \end{array} \right) \\
  &=&  \left( \begin{array}{c}
  \cP(-1 + B^{\bar a b}B_{\ssT \bar a} B_b) - B^{\bar a b} B_{\bar a}B_b \\  \cP B^{\bar a b} B_{b \ol\ssT} \end{array} \right) \simeq  \cP \left( \begin{array}{c}
 - 1  \\   B^{\bar a b} B_{b \ol\ssT} \end{array} \right) \,. \nn
\eea

With these same approximations eq.~\pref{KKKwrtP2x2} becomes
\be \label{KKKwrtP2x2y}
  K^{\ol\ssA \ssB} K_{\ol\ssA} K_\ssB - 3    \simeq \frac{3\cP (B_{\ssT\ol\ssT}   - B^{\bar a c} B_{\ssT \bar a} B_{c \ol\ssT})}{  1  -2 B^{\bar a b} B_{\bar a} B_b } \sim \cO(\tau^{-1}) \,, \nn
\ee

All that matters for the scalar potential is the superpotential and its first derivative, and so we take the superpotential to have the form
\be
  W = w_0 + w_a z^a
\ee
and evaluate the result at $z^a = 0$ (this is particularly appropriate for $X \in \{ z^a \}$). Both $w_0$ and $w_a$ are imagined to depend on $T$ only as functions of $\ln T$. The ordinary derivatives of $W$ then are $W_\ssT = (w'_0 + w'_a z^a)/T$ and $W_a = w_a$ (where primes denote differentiation with respect to $\ln T$). The K\"ahler covariant derivatives are $D_\ssA W := W_\ssA + K_\ssA W$, and so (evaluated at $z^a = 0$)
\bea
   D_a W &=& W_a + K_a W = w_a - \frac{3B_a w_0}{\cP} \nn\\
    \hbox{and} \quad D_\ssT W &=& W_\ssT + K_\ssT W = \frac{w_0'}{T} - \frac{3w_0}{\cP}  (1 + B_\ssT) \simeq \frac{2w_0'}{\tau} - \frac{3w_0}{\tau} + \cO(1/\tau^2) \,.
\eea
The combination appearing in the $F$-term potential then is
\bea
   K^{\ol\ssA\ssB} \ol{D_\ssA W} D_\ssB W - 3 |W|^2 &=& K^{\ol\ssA\ssB}\Bigl[ \ol{W_\ssA} W_\ssB + K_{\ol \ssA} \ol{W} W_\ssB + K_\ssB \ol{W_\ssA} W  \Bigr] + ( K^{\ol\ssA\ssB} K_{\ol\ssA} K_\ssB - 3) |W|^2 \nn\\
   &=& K^{\ol\ssT \ssT} \ol{W_\ssT} W_\ssT + K^{\ol\ssT a} W_a \ol{W}_\ssT + K^{\bar a \ssT} \ol{W}_a W_\ssT +  K^{\ol\ssT \ssB}  K_\ssB W \ol{W_\ssT} + K^{\ol\ssB \ssT}  K_{\ol\ssB} \ol{W} W_\ssT   \nn\\
   && \quad + K^{\bar a b} \ol{W_a} W_b + K^{\bar a \ssB} K_\ssB W \ol{W_a} + K^{\ol\ssB a} K_{\ol\ssB} \ol{W} W_a + ( K^{\ol\ssA\ssB} K_{\ol\ssA} K_\ssB - 3) |W|^2  \nn\\
   &\simeq&  \frac{\cP^2}{3} \left| \frac{w_0'}{T} \right|^2 - \frac{\cP^2 }{3}\left[ \frac{B^{\bar a b} (B_{b \ol\ssT} - B_b/\cP) \ol{w_a} w_0'}{T} + \hbox{c.c.}\right] - \cP \left( \frac{w_0' \, \ol{w_0}}{T} + \hbox{c.c.} \right) \nn\\
   && \qquad\qquad - \frac{\cP}{3} B^{\bar a b} \ol{w_a} w_b + \cP B^{\bar a b} B_{b \ol\ssT} w_0 \ol{w_a}  + \cP B^{\bar a b} B_{\ssT \bar a} w_b \ol{w_0} \\
   &&\qquad\qquad\qquad\qquad + \frac{B_{\ssT\ol\ssT}   - B^{\bar a c} B_{\ssT \bar a} B_{c \ol\ssT}}{  1  -2  B^{\bar a b} B_{\bar a} B_b} \Bigl( 3\cP |w_0|^2\Bigr) \,, \nn
\eea
and so $V_\ssF$ is given by 
\be 
  V_\ssF = e^K \Bigl[ K^{\ol\ssA\ssB} \ol{D_\ssA W} D_\ssB W - 3 |W|^2 \Bigr] = \frac{1}{\cP^3} \Bigl[ K^{\ol\ssA\ssB} \ol{D_\ssA W} D_\ssB W - 3 |W|^2 \Bigr] 
  = V_2 + V_3 + V_4 + \cdots \,,
\ee
where $V_n$ is of order $\tau^{-n}$ and explicitly
\bea \label{VFBasicFormz}
  V_2 &\simeq&  \frac{1}{\cP^2} \left[ - \frac13 \, B^{\bar a b} \ol{w_a} w_b  \right] \nn\\
 V_3 &\simeq&  \frac{1}{\cP^2} \left[   B^{\bar a b} B_{b \ol\ssT} w_0 \ol{w_a}  + B^{\bar a b} B_{\ssT \bar a} w_b \ol{w_0} \phantom{\frac12} \right. \nn\\
 && \qquad \qquad\left.  +  \frac{\cP}{3} \left| \frac{w_0'}{T} \right|^2 - \frac{\cP }{3}\left[ \frac{B^{\bar a b} (B_{b \ol\ssT} - B_b/\cP) \ol{w_a} w_0'}{T} + \hbox{c.c.}\right] -  \left( \frac{w_0' \, \ol{w_0}}{T} + \hbox{c.c.} \right) \right] \nn\\
 V_4 &\simeq&  \frac{1}{\cP^2} \left[  \frac{B_{\ssT\ol\ssT}   - B^{\bar a c} B_{\ssT \bar a} B_{c \ol\ssT}}{  1  -2  B^{\bar a b} B_{\bar a} B_b } \Bigl( 3 |w_0|^2\Bigr)
 \right] \nn\,,
\eea
and so on. These show that the $\ol{w_a} w_b$ term arises at order $\tau^{-2}$ while the $\ol{w_0} w_b$ terms are at order $\tau^{-3}$ and the $|w_0|^2$ term is order $\tau^{-4}$.

Specializing to the case where $\{z^a\} = X$ and writing $B = - k$ so that $\cP = \tau - k$ the above potential becomes
\be \label{VFBasicFormzz}
  V_\ssF  \simeq  \frac{1}{\cP^2} \left[ \frac13 \, k^{\ol\ssX \ssX} \ol{w_\ssX} w_\ssX +  k^{\ol\ssX \ssX} k_{\ssX \ol\ssT} w_0 \ol{w_\ssX}  + k^{\ol\ssX \ssX} k_{\ssT \ol\ssX} w_\ssX \ol{w_0} - \frac{k_{\ssT\ol\ssT}  - k^{\ol\ssX \ssX} k_{\ssT \ol\ssX} k_{\ssX \ol\ssT}}{  1  + 2 k^{\ssX\ol\ssX}  k_\ssX k_{\ol\ssX} } \Bigl( 3 |w_0|^2\Bigr)
 \right] \,, 
\ee
which has the no-scale form that ensures $T$ is a flat direction when $k_\ssT = 0$. This potential also has no mixing between $w_0$ and $w_\ssX$ whenever $k_{\ssX \ol \ssT} = 0$, and this is why in our simple models stabilization with respect to variations of $w_\ssX$ always led to $w_\ssX = 0$.

In a scenario where $w_\ssX$ is a function of $\Phi$ and so can be minimized, we see (as mentioned above) that the minimization occurs at $w_\ssX = 0$ when $k_{\ssT\ol\ssX} = 0$. More generally extremizing with respect to $\ol{w_\ssX}$ gives the saddle point
\be
   w_\ssX  = -3   k_{\ssX \ol\ssT} w_0   \,,
\ee
and evaluated at this point the potential becomes
\be \label{VFBasicFormzzmin}
  V_\ssF  \simeq  -\frac{3|w_0|^2}{\tau^2} \left[  k^{\ol\ssX \ssX} k_{\ssX\ol \ssT} k_{\ssT\ol\ssX}   + \frac{k_{\ssT\ol\ssT}  - k^{\ol\ssX \ssX} k_{\ssT \ol\ssX} k_{\ssX \ol\ssT}}{  1  +2 k^{\ssX\ol\ssX}  k_\ssX k_{\ol\ssX} }
 \right] \,, 
\ee
where both terms are order $1/\tau^4$. Notice that the kinetic term for $T$ is controlled by $K_{\ssT\ol\ssT}$ and this is independent of $k_{\ssT\ol\ssT}$ once corrections in powers of $1/\tau$ are dropped, and as a result $k_{\ssT\ol\ssT}$ can have either sign, allowing the potential to be positive.

\subsection*{No-scale K\"ahler geometry}
\label{appssec:NoScaleGeo}

In this appendix we compute some of the geometrical quantities that arise in the main text for a general two-field no-scale geometry. To this end consider a general no-scale system with two fields $S$ and $T$ and K\"ahler potential
\be
    K = -3 M_p^2 \ln \Big[ \tau \, \cF(\sigma/\tau) \Bigr] = -3 M_p^2 \Bigl[ \ln \tau + F(\sigma/\tau) \Bigr]
\ee
where $\tau = T + \ol T$ and $\sigma = S + \ol S$ and $\cF = e^F$ is otherwise arbitrary. For such a choice we have
\be
   A^2 = e^{K/(3M_p^2)} = \frac{1}{\tau \, \cF(\sigma/\tau)}
\ee
and the first derivatives of $K$ are
\be \label{K1stD}
   K_\ssT  
   = - 3M_p^2 \left[\frac{1}{\tau} +  F_\ssT \right] =- \frac{3M_p^2}{\tau} \Bigl[1 - x \, F'(x) \Bigr] \quad \hbox{and} \quad
   K_\ssS  
   = - 3M_p^2   F_\ssS = - \frac{3M_p^2 F'(x)}{\tau} \,,
\ee
where $x := \sigma/\tau$. 

The components of the K\"ahler metric then are
\bea
   K_{\ssS\ol\ssS} &=& -3 M_p^2 F_{\ssS\ol\ssS} =  - \frac{3M_p^2 F''(x)}{\tau^2}\nn\\
   K_{\ssS\ol\ssT} &=&  -3 M_p^2 F_{\ssS\ol\ssT} =  \frac{3M_p^2}{\tau^2} \Bigl[ F'(x) + x F''(x) \Bigr] \\
   K_{\ssT\ol\ssT} &=& 3M_p^2 \left[ \frac{1}{\tau^2} - F_{\ssT\ol\ssT} \right] =  \frac{3M_p^2}{\tau^2} \Bigl[ 1 - 2x F'(x) - x^2 F''(x) \Bigr]  \nn
\eea
and so the inverse metric has components 
\bea
   K^{\ol\ssS \ssS} &=& \frac{\tau^2}{3M_p^2[ F'' + (F')^2]} \Bigl[ -1 +2xF' + x^2 F'' \Bigr] \nn\\
   K^{\ol\ssT \ssS} &=&   \frac{\tau^2}{3M_p^2[ F'' + (F')^2]} \Bigl[F' + x F'' \Bigr] \\
   K^{\ol\ssT \ssT} &=&  \frac{\tau^2  F'' }{3M_p^2[ F'' + (F')^2]}   \,.\nn
\eea
Notice that these imply that the combination $\eta^a$ that controls the coupling to matter is
\be\label{Kupper1}
   K^{\ol \ssA \ssT} K_{\ol\ssA} = - \tau \quad \hbox{and} \quad K^{\ol \ssA \ssS} K_{\ol\ssA} = - \sigma \,,
\ee
{\it completely independent} of the function $F(x)$ (in agreement with the general scaling proof given for more general models in \cite{Burgess:2008ir, Burgess:2020qsc}). As a check notice that \pref{K1stD} and \pref{Kupper1} together imply $K^{\ol\ssA \ssB} K_{\ol\ssA} K_\ssB =  3 M_p^2$ as required for a no-scale model. 

The target-space Christoffel symbols that appear in the scalar field equations require the third derivatives:
\bea
   K_{\ssS\ssS\ol\ssS} &=& -3 M_p^2 F_{\ssS\ssS\ol\ssS} =  - \frac{3M_p^2 F'''(x)}{\tau^3}\nn\\
   K_{\ssS\ssS\ol\ssT} &=&  -3 M_p^2 F_{\ssS\ssS\ol\ssT} =  \frac{3M_p^2}{\tau^3} \Bigl[ 2F''(x) + x F'''(x) \Bigr] \\
   K_{\ssS\ssT\ol\ssT} &=& -3M_p^2  F_{\ssS\ssT\ol\ssT} = - \frac{3M_p^2}{\tau^3} \Bigl[ 2 F'(x) + 4x F''(x) + x^2 F'''(x) \Bigr]  \nn
\eea
and
\bea
   K_{\ssT\ssS\ol\ssS} &=& - 3 M_p^2 F_{\ssT\ssS\ol\ssS} =  \frac{3M_p^2}{\tau^3} \Bigl[ 2F''(x) + xF'''(x) \Bigr] \nn\\
   K_{\ssT\ssS\ol\ssT} &=&  -3 M_p^2 F_{\ssS\ssT\ol\ssT} =  -\frac{3M_p^2}{\tau^3} \Bigl[ 2F'(x) +4 x F''(x) + x^2 F'''(x) \Bigr] \\
   K_{\ssT\ssT\ol\ssT} &=& -3M_p^2  \left[ \frac{2}{\tau^3} + F_{\ssT\ssT\ol\ssT} \right] = - \frac{3M_p^2}{\tau^3} \Bigl[2 - 6x F'(x) - 6x^2 F''(x) - x^3 F'''(x) \Bigr]  \nn
\eea
and so  
\bea
  \Gamma^\ssT_{\ssS\ssT} &=& \frac{-2x (F'')^2+x F' F'''}{\tau[F'' + (F')^2]} \nn\\  
  \Gamma^\ssT_{\ssS\ssS} &=&  \frac{2 (F'')^2 - F' F'''}{\tau[F'' + (F')^2]} \\  
  \Gamma^\ssS_{\ssS\ssT} &=& \frac{-2(F')^2-2F''-2xF'F''-xF'''-2x^2(F'')^2+x^2 F'F'''}{\tau[F''+(F')^2]} \nn\\  
  \Gamma^\ssS_{\ssS\ssS} &=&  \frac{2F' F''+F'''+2x(F'')^2-x F' F'''}{\tau[F''+(F')^2]} \nn
\eea
and
\bea
  \Gamma^\ssT_{\ssT\ssT} &=& \frac{-2(F')^2-2F''+2x^2(F'')^2-x^2F'F'''}{\tau[F'' + (F')^2]} \nn\\  
  \Gamma^\ssT_{\ssT\ssS} &=&  \frac{-2x(F'')^2+xF'F''' }{\tau[F'' + (F')^2]} \\  
  \Gamma^\ssS_{\ssT\ssT} &=& \frac{2x(F')^2+2xF''+2x^2F'F''+x^2F'''+2x^3(F'')^2-x^3F'F''' }{\tau[F''+(F')^2]} \nn\\  
  \Gamma^\ssS_{\ssT\ssS} &=&  \frac{-2(F')^2-2F''-2xF'F''-xF'''-2x^2(F'')^2+x^2F'F''' }{\tau[F''+(F')^2]} \,. \nn
\eea

\section{Goldstino production}
\label{AppC}

These notes estimate the size of gravitino production and show how the dilaton couplings appear in the gravitino/goldstino equivalence theorem. To this end we use the component lagrangian describing the coupings of the gravitino to the goldstino $G \in X$ and a proxy for a SM fermion $\chi \in Y$, using constrained fields satisfying $X^2 = XY = 0$.

The K\"ahler potential $K = -3\ln(\tau - k)$ where $k$ is 
\bea\label{kform}
   k &=& \ol X X + \ol Y Y  + \frac{\mfb}2 \Bigl( Y \ol Y^2 + \ol Y Y^2 \Bigr) + \frac{\mfc}4 \, Y^2 \ol Y^2 + \frac{\mfe}2 \Bigl( X \ol Y^2 + \ol X Y^2 \Bigr) \nn\\
   &&\qquad\qquad\qquad\qquad\qquad\qquad + \mu (X + \ol X) +\mfk Y + \frac{\mfm}{2} \, Y^2 \,,
\eea
while the superpotential is as before:
\be \label{WforXY}
  W = W_0 + \mff \, X + \mfg \, Y + \frac{\mfh}{2} \,( Y^2 + \ol Y^2) \,.
\ee

The arguments of the main text show that the gravitino mass is given by
\be \label{m3halves}
  m_{3/2} = e^{K/2} \frac{|W|}{M_p^2} = \frac{\mfF}{\sqrt3\, M_p} \,,
\ee
where 
\be
   \mfF = \Bigl[ \cG_{\ssJ \ol \ssI} F^\ssJ \ol{F}^\ssI \Bigr]^{1/2} = \Bigl[ e^K K^{\ssJ \ol \ssI} D_\ssJ W \ol{D_\ssI W} 
  \Bigr]^{1/2} \sim \frac{ |W_0|}{\tau^{3/2} M_p} \sim \frac{M_p^2}{\tau} \,.
\ee

\subsubsection*{Jordan frame component action}

The component form of the lagrangian is again given by expressions similar to those given in the gauge $G = 0$ by \cite{DallAgata:2015zxp}. The terms that come from the $F$ term include the masses
\be\label{fermmixFz}
   \frac{\cL_\ssF}{\sqrt{-\tilde g}} \ni  - \frac12\left[  \mfh \, \ol \chi \gamma_\ssL \chi + \frac{W_0}{M_g^2} \, \ol \psi_\mu \tilde\gamma^{\mu\nu}\gamma_\ssL \psi_\nu  + \hbox{h.c.} \right] 
\ee
where the tilde on $\sqrt{-\tilde g}$ and $\tilde \gamma^{\mu\nu} = \gamma^{ab} {\tilde e_a}^\mu \, {\tilde e_b}^\nu$ emphasize their dependence on the Jordan-frame metric $\tilde g_{\mu\nu}$. The scalar potential vanishes, $V = 0$, when $k$ is independent of $\tau$, because the system is a no-scale model.

The kinetic and 4-fermi terms similarly arise from the $D$-terms and so come with the prefactor $e^{-K/3} =  \tau - k + \cdots$:
\bea\label{fermmixzDz}
   \frac{\cL_\ssD}{\sqrt{- \tilde g}} &\ni& - e^{-K/3} \left[ \frac{M_g^2}2 \, \widehat{\tilde R} + \frac{i}2 \, \tilde \epsilon^{\mu\nu\lambda\rho} \ol \psi_\mu \gamma_5 \tilde \gamma_\nu D_\lambda \psi_\rho + \frac12 \, K_{\ssY\ol \ssY} \ol \chi  \tilde\Dsl \chi \right]  + \hbox{(4-fermi terms)} \nn\\
   &\simeq& - \tau \left[ \frac{M_g^2}2 \, \widehat{\tilde R} + \frac{i}2 \, \tilde \epsilon^{\mu\nu\lambda\rho} \ol \psi_\mu \gamma_5 \tilde \gamma_\nu D_\lambda \psi_\rho + \frac3{2\tau} \, \ol \chi  \tilde\Dsl \chi \right]  + \hbox{(4-fermi terms)} 
\eea
which shows that the canonically normalized gravitino is given in Jordan frame by $\hat \psi_\mu \simeq \sqrt\tau \; \psi_\mu$. The 4-fermion interactions similarly have the schematic form 
\bea \label{4-fermiz}
  \frac{\cL_{\rm 4-fermi}}{\sqrt{- \tilde g}} &=& e^{-K/3} \left[ \left(-\frac{K_{\ssY\ol\ssY}K_{\ssY\ol\ssY}}{8M_g^2} + K_{\ssY\ssY\ol\ssY\ol\ssY} - K^{\ssY\ol\ssY} K_{\ssY\ol\ssY\ol\ssY} K_{\ssY\ssY\ol\ssY}  - K^{\ssX\ol\ssX} K_{\ssX\ol\ssY\ol\ssY} K_{\ssY\ssY\ol\ssX} \right) (\ol \chi \gamma_\ssL \chi)(\ol \chi \gamma_\ssR \chi) \right.\nn \\
  && \qquad\qquad \left. + \frac{ K_{\ssY\ol\ssY}}{4M_g^2} \Bigl( \tilde \epsilon^{\mu\nu\lambda\rho} \ol \psi_\mu \gamma_\ssL \tilde \gamma_\nu \psi_\lambda + \ol \psi_\mu  \gamma_\ssL \tilde \gamma^\rho \psi^\mu \Bigr)  (\ol\chi  \gamma_\ssL \tilde \gamma_\rho \chi) \right]   \nn\\
  &\simeq& \tau \left[ \left(-\frac9{8M_g^2\tau^2} + \frac{3\mfc}{\tau} - \frac{3( |\mfe|^2 + |\mfb|^2)}{\tau} \right) (\ol \chi \gamma_\ssL \chi)(\ol \chi \gamma_\ssR \chi) \right. \\
  && \qquad\qquad \left. + \frac3{4M_g^2\tau}  \Bigl( \tilde \epsilon^{\mu\nu\lambda\rho} \ol \psi_\mu \gamma_\ssL \tilde \gamma_\nu \psi_\lambda + \ol \psi_\mu  \gamma_\ssL \tilde \gamma^\rho \psi^\mu \Bigr)  (\ol\chi  \gamma_\ssL \tilde \gamma_\rho \chi) \right]  \,, \nn
\eea
and all we need use is that the $\psi^2 \chi^2$ terms have a $\tau$-independent coupling proportional to $1/M_g^2$. Notice that both of the $1/M_g^2$ couplings end up proportional to $1/(M_g^2\tau) \simeq 1/M_p^2$ when written in terms of the canonically normalized fermions, where the coefficient of the Einstein-Hilbert term is taken to define the Planck mass, $M_g^2 \tau = M_p^2$, at least at present day where $\tau = \tau_0$. 

\subsubsection*{Einstein frame component action}

Next we repeat the above estimate directly in Einstein frame, being careful to not change the units of the present-day metric:
\be
    \tilde g_{\mu\nu} = e^{(K-K_0)/3} g_{\mu\nu} = \frac{\tau_0}{\tau} \, g_{\mu\nu} \,.
\ee

The component form of the lagrangian is then obtained by making this metric substitution, which also implies
\be
    \sqrt{-\tilde g} = \sqrt{-g} \; e^{2(K-K_0)/3}  = \sqrt{-g} \; \left( \frac{\tau_0}{\tau} \right)^2
    \quad \hbox{and} \quad
    {\tilde e_a}^\mu = {e_a}^\mu \sqrt{\frac{\tau}{\tau_0}}\,,
\ee
and so
\be\label{fermmixFzapp}
   \frac{\cL_\ssF}{\sqrt{-g}} \ni  - \frac12 \left( \frac{\tau_0}{\tau} \right)^2 \left[ \left( \mfh - \mfe\mff   \right)\, \ol \chi \gamma_\ssL \chi + \frac{W_0}{M_g^2} \left( \frac{\tau}{\tau_0} \right) \ol \psi_\mu \gamma^{\mu\nu}\gamma_\ssL \psi_\nu  + \hbox{h.c.} \right] \,.
\ee
Similarly
\be\label{fermmixzDzapp}
   \frac{\cL_\ssD}{\sqrt{- g}} \ni - \tau_0 \left[ \frac{M_g^2}2 \, \widehat{R} + \frac{i}2 \left( \frac{\tau}{\tau_0} \right)^{1/2} \epsilon^{\mu\nu\lambda\rho} \ol \psi_\mu \gamma_5 \gamma_\nu D_\lambda \psi_\rho \right] - \frac3{2} \left(\frac{\tau_0}{\tau}\right)^{3/2} \ol \chi \Dsl \chi   + \hbox{(4-fermi terms)} \,,
\ee
which shows that $M_p^2 = M_g^2 \tau_0$ is the physical Planck scale (as expected). 

The canonically normalized gravitino is given in Einstein frame by $\hat \psi_\mu \simeq (\tau \tau_0)^{1/4} \; \psi_\mu$ and the canonically normalized spin-half particle is $\hat \chi \sim (\tau_0/\tau)^{3/4} \chi$.  Using these in \pref{fermmixFzapp} verifies that the Einstein-frame fermion masses are $m_\chi \propto \mfh/\sqrt\tau$ and $m_{3/2} \propto W_0/\tau^{3/2}$, as expected. The 4-fermion interactions of \pref{4-fermiz} similarly become 
\bea \label{4-fermizef}
  \frac{\cL_{\rm 4-fermi}}{\sqrt{- \tilde g}} &\simeq&  \left( \frac{\tau_0}{\tau} \right)^2\left[ \left(-\frac9{8M_g^2\tau} + 3\mfc - 3( |\mfe|^2 + |\mfb|^2) \right) (\ol \chi \gamma_\ssL \chi)(\ol \chi \gamma_\ssR \chi) \right. \\
  && \qquad\qquad \left. + \frac3{4M_g^2}  \left( \frac{\tau}{\tau_0} \right) \Bigl( \tilde \epsilon^{\mu\nu\lambda\rho} \ol \psi_\mu \gamma_\ssL \tilde \gamma_\nu \psi_\lambda + \ol \psi_\mu  \gamma_\ssL \tilde \gamma^\rho \psi^\mu \Bigr)  (\ol\chi  \gamma_\ssL \tilde \gamma_\rho \chi) \right]  \,, \nn
\eea
and so once written in terms of canonically normalized fermions the $\psi^2 \chi^2$ terms have a $\tau$-dependent coupling proportional to 
\be
   \frac{1}{M_g^2} \left( \frac{\tau_0}{\tau} \right)\frac{1}{\sqrt{\tau \tau_0}} \left( \frac{\tau}{\tau_0} \right)^{3/2} = \frac{1}{M_g^2\tau} \left( \frac{ \tau}{\tau_0} \right) = \frac{1}{M_p^2}   \left( \frac{ \tau}{\tau_0} \right) 
\ee
and so is Planck-suppressed (as expected).

\subsubsection*{Gravitino/goldstino production}

The estimate for the cross section for gravitino production, $\chi \chi \to \psi \psi$, using the above Einstein-frame action is given in the regime $E \gg m_{3/2}$ by the result 
\be \label{xsectionestz}
  \frac{\exd \sigma}{\exd \Omega} \sim E^2 \left( \frac{1}{M_p^2}  \right)^2 \left(  \frac{E^2}{m_{3/2}^2} \right)^2 
   \sim \frac{E^6}{ (M_p m_{3/2})^4}  = \frac{E^6}{\mfF^4} \,,
\ee
where the factors $(E/m_{3/2})^2$ come from the gravitino spin sums $\cS_{\mu\nu}(k) = \sum_\lambda u_\mu(k) \bar v_\nu(k)$. This shows how the equivalence theorem gives the production rate of longitudinal graviniti in terms of the effective goldstino decay constant $\mfF \sim M_p/\sqrt \tau$. The production rate is acceptably small because this is at the TeV scale.

\end{document}